\documentclass[11pt]{article}
\pdfoutput=1
\usepackage{jcapmod}
\usepackage{booktabs}
\usepackage[english]{babel}
\usepackage{amsmath,amssymb,amsbsy,amstext, amsthm, simplewick}
\usepackage{wrapfig}
\usepackage{graphicx}
\usepackage{amsfonts}
\usepackage{amssymb}
\usepackage{upgreek}
 \usepackage{exscale,relsize}
 \usepackage[makeroom]{cancel}
\usepackage{soul}
\usepackage{bbold}

\RequirePackage{color}
\usepackage{colortbl}
\definecolor{rp}{cmyk}{0.2, 1, 0.6, 0}
\definecolor{green2}{cmyk}{0, 1, 0.5, 0}
\definecolor{lightgreen}{cmyk}{0.2, 0, 0.2, 0.2}
\definecolor{lightgray}{cmyk}{0.1,0.2,0,0.1}
\definecolor{lightgray2}{cmyk}{0.4,0.4,0,0.8}
\definecolor{black}{cmyk}{1.0,1.0,1.0,1.0}

\allowdisplaybreaks[1]

\DeclareMathAlphabet{\mathpzc}{OT1}{pzc}{m}{it}

\usepackage{colortbl}
\definecolor{lightgreen}{cmyk}{0.2, 0, 0.2, 0.2}
\definecolor{lightgray}{cmyk}{0.1,0.2,0,0.1}
\definecolor{lightgray2}{cmyk}{0.1,0.1,0,0.1}

\makeatletter
\newlength{\apb@width}
\newcommand{\autoparbox}[2][c]{\settowidth{\apb@width}{#2}\parbox[#1]{\apb@width}{#2}}

\makeatother

\newcommand{\lsim}{\mathrel{\hbox{\rlap{\lower.55ex\hbox{$\sim$}} \kern-.3em \raise.4ex \hbox{$<$}}}}
\newcommand{\gsim}{\mathrel{\hbox{\rlap{\lower.55ex\hbox{$\sim$}} \kern-.3em \raise.4ex \hbox{$>$}}}}

\newcommand{\vp}{\varphi}

\newcommand{\mpl}{m_{\mbox{\tiny{Pl}}}}
\newcommand{\Beq}{\begin{equation}\begin{aligned}}
\newcommand{\Eeq}{\end{aligned}\end{equation}}

\newcommand{\bq}{{\textbf{\textit{q}}}}
\newcommand{\bk}{{\textbf{\textit{k}}}}
\newcommand{\bx}{{\textbf{\textit{x}}}}

\makeatletter
\def\l@subsubsection#1#2{}
\makeatother

\begin{document}
\begin{titlepage}

\setcounter{page}{1} \baselineskip=15.5pt \thispagestyle{empty}

\bigskip\

\vspace{1cm}
\begin{center}

{\fontsize{20}{24}\selectfont  \sffamily \bfseries  The charged inflaton and its gauge fields: \\
Preheating and initial conditions for reheating}

\end{center}

\vspace{0.2cm}

\begin{center}
{\fontsize{13}{30}\selectfont  Kaloian D.~Lozanov$^{^\spadesuit, ^\bigstar}$ and Mustafa A.~Amin$^{^\spadesuit, ^\bigstar, ^\clubsuit}$}
\end{center}

\begin{center}

\vskip 8pt

\textsl{$^\spadesuit$ Institute of Astronomy, University of Cambridge,
Madingly Road, Cambridge, U.K.}
\vskip 7pt

\textsl{$^\bigstar$ Kavli Institute for Cosmology, University of Cambridge,
Madingly Road, Cambridge, U.K.}
\vskip 7pt

\textsl{$^\clubsuit$ Physics \& Astronomy Department, Rice University,
6100 Main Street, Houston, U.S.A.}
\vskip 7pt

\end{center}

\vspace{1.2cm}
\hrule \vspace{0.3cm}
\noindent {\sffamily \bfseries Abstract} \\[0.1cm]
We calculate particle production during inflation and in the early stages of reheating after inflation in models with a charged scalar field coupled to Abelian and non-Abelian gauge fields. A detailed analysis of the power spectra of primordial electric fields, magnetic fields and charge fluctuations at the end of inflation and preheating is provided. We carefully account for the Gauss constraints during inflation and preheating, and clarify the role of the longitudinal components of the electric field. We calculate the timescale for the back-reaction of the produced gauge fields on the inflaton condensate, marking the onset of non-linear evolution of the fields.  We provide a prescription for initial conditions for lattice simulations necessary to capture the subsequent nonlinear dynamics. On the observational side, we find that the primordial magnetic fields generated are too small to explain the origin of magnetic fields on galactic scales and the charge fluctuations are well within observational bounds for the models considered in this paper.

\vskip 10pt
\hrule
\vskip 10pt

\vspace{0.6cm}
 \end{titlepage}

\tableofcontents

\newpage
\section{Introduction}

Particle  production during inflation \cite{PhysRevD.23.347,LINDE1982389,Starobinsky:1980te,PhysRevLett.48.1220} and reheating \cite{Dolgov:1989us,Traschen:1990sw,Kofman:1994rk,Shtanov:1994ce,Kofman1997,Bassett:2005xm,Allahverdi:2010xz,Amin2014} sets up the initial conditions for the formation of observed structure and the beginning of the hot big bang. Particle production in models with gauge fields is particularly interesting because of the ubiquity of gauge fields in the Standard Model (SM) and their natural appearance in extensions beyond the SM. Gauge fields coupled to scalar fields can have important consequences for the generation of curvature \cite{Yokoyama:2008xw,Dimopoulos:2008yv,ValenzuelaToledo:2009af,ValenzuelaToledo:2009nq,Bartolo:2009pa,Bartolo:2009kg,Karciauskas:2011fp,Dimopoulos:2011ws,Dimopoulos:2011pe,ValenzuelaToledo:2011fj,BeltranAlmeida:2011db,Emami:2011yi,Dimopoulos:2012av,Lyth:2012br,Lyth:2012vn,Jain:2012vm,Nurmi:2013gpa,Barnaby:2010vf,Barnaby:2011vw,Anber:2012du,Barnaby:2012tk,Namba:2012gg,D'Onofrio:2012qy,Nieto:2016gnp,Almeida:2014ava,Rodriguez:2013cj,Enckell:2016xse} and charge perturbations \cite{Goolsby-Cole:2015chd,D'Onofrio:2012qy,Giovannini:2000dj,Dolgov:2006bc,Hertzberg2013,Hertzberg:2013jba,LozAmin}, gravitational waves \cite{Barnaby:2010vf,Anber:2012du,Sorbo:2011rz,Cook:2011hg,Barnaby:2011qe,Barnaby:2012xt,Figueroa:2016ojl,Ferreira:2014zia,Ferreira:2015omg,Domcke:2016bkh,Namba:2015gja} as well as seeding primordial magnetic fields \cite{PhysRevD.37.2743,Ratra:1991bn,PhysRevD.69.043507,Bamba:2006ga,Martin:2007ue,Anber:2006xt,Dimopoulos:2008em,Emami2009,Byrnes:2011aa,Demozzi:2009fu,Suyama:2012wh,Fujita:2012rb,Kandus:2010nw,Giovannini:2000dj,Ferreira:2014hma,Ferreira:2013sqa} during inflation. Such gauge fields can significantly affect the transition to a radiation dominated universe  after inflation \cite{Deskins2013,Adshead2015} and can lead to novel non-perturbative phenomena \cite{GarciaBellido2003,Graham2006,DiazGil2007,Graham2007,DiazGil2008,Figueroa2015,Enqvist:2015sua,Yang2015}. 

An analysis of particle production with gauge fields in the early universe has been undertaken in many previous studies. For example, gauge field production during inflation and its consequences was reviewed in \cite{Maleknejad:2012fw}. Non-perturbative gauge field production during and after inflation was explored in \cite{Finelli:2000sh,Davis:2000zp,PhysRevD.63.103515,D'Onofrio:2012qy,McDonough:2016xvu,HiggsPreheat,Bezrukov2008,Kari,Graham:2015rva,Fujita:2016qab,Kobayashi:2014sga}, whereas  nonlinear dynamics of gauge fields after inflation was considered in (for e.g.) \cite{Rajantie:2000fd,Fujita:2015iga,Deskins2013,Adshead2015,Cheng:2015oqa,Dufaux2010,Figueroa2015} for Abelian fields and \cite{GarciaBellido2003,DiazGil2007,DiazGil2008,Graham2006,Graham2007,Enqvist:2015sua} for non-Abelian ones. 

In this paper we re-visit particle production in locally gauge invariant models with Abelian and non-Abelian fields coupled to charged scalar fields. In these models we assume that a component of the  charged scalar field plays the  role of the inflaton condensate. Care has to be taken with gauge fields because they have to satisfy certain constraint equations (along with the usual evolution equations). The natural gauge redundancy can lead to complications in quantization or to spurious gauge modes during numerical evolution. The non-zero vacuum expectation value (vev) of the inflaton condesate during inflation and reheating changes some of the  common results for gauge fields coupled to scalar fields with zero vevs. Moreover, certain common gauge choices become ill-defined when the inflaton condensate starts oscillating at the end of inflation. Finally, if significant particle production occurs, back-reaction becomes important and classical lattice simulations are needed to fully explore the nonlinear dynamics of the fields. Initial conditions for such lattice simulations can be nontrivial because of the constraints on the different variables that must be satisfied. We pay special attention to all of these issues in this work. While we restrict ourselves to ``minimal" models consistent with local gauge invariance, our techniques and results should carry over to more complicated scenarios such as \cite{HiggsPreheat,Bezrukov2008}.

We calculate particle production during inflation using gauge invariant variables and with proper accounting for the constraints (Section \ref{sec:inflation}). The use of gauge invariant variables naturally avoids issues with spurious gauge degrees of freedom and makes the quantization and subsequent evolution of perturbations particularly transparent. We provide power spectra for the electric and magnetic fields at the end of inflation, along with simple analytic estimate for their shape. In Appendix \ref{sec:dS} we explain their shape via approximate analytic calculations.

While useful during inflation, gauge invariant variables become ill-defined when the inflaton starts oscillating. We argue for the use of well defined Coulumb gauge variables for analysing non-perturbative particle production during preheating at the end of inflation. In Section \ref{sec:preheating}, we carry out a Floquet like analysis for the resonant production of the gauge fields. Here, we point out some minor discrepancies in the literature regarding the productions of the gauge invariant longitudinal component of the electric field. We then estimate the end of preheating by calculating the time when back-reaction of the resonantly produced gauge fields becomes important (Section~\ref{sec:backreaction}).  In Appendix \ref{sec:LSCG} we provide technical details for quantizing and calculating the back-reaction in the Coulomb gauge. 

Once nonlinear effects become important, simulations become essential. Nonlinear simulations with gauge fields (especially non-Abelian fields) can be challenging because of the large number of components and the necessity of satisfying constraint equations. In Section \ref{sec:App} of this work we provide a simple prescription for setting up initial conditions for such lattice simulations which can be applied in any gauge. In our prescription, the lattice initial conditions naturally satisfy the necessary gauge constraints.  We note that our initial conditions accurately account for metric perturbations, field interactions and gauge constraints to linear order. We provide an example of lattice initial conditions in temporal gauge which is a common choice for simulations. We arrive at these initial conditions via gauge invariant variables; this serves as a model to set up initial conditions in any gauge. We will carry out actual lattice simulations in future work. 

We have tried to make the paper self-contained, so that it can be used for future reference easily. To this end, we provide the necessary equations for the perturbations of metric and matter fields (scalar and gauge fields) in position and Fourier space for gauge invariant variables. While we work with gauge invariant variables as far as possible, we also provide a dictionary to translate our results to other popular gauges.

For pedagogical purposes, we carry out the analysis for Abelian fields first (Section~\ref{sec:Amodels}). We show in the later half of the paper (Section~\ref{sec:nAmodels}) how the non-Abelian analysis can be reduced to an analysis of multiple copies of the Abelian case in the linear regime. Hence, the analysis for the Abelian case, including the setting up of the lattice initial conditions, can be easily carried over to the non-Abelian case. We show how to apply the developed techniques to a $SU(2)$ model and its extension: a $SU(2)\times U(1)$ model.

We discuss the observational consequences of a charged inflaton and its gauge fields in Section \ref{sec:observationalconsequences}. We summarize our results in the Conclusions section and discuss future directions.


\section{The Abelian model}
\label{sec:Amodels}
We consider an action with matter minimally coupled to gravity
\Beq
\label{eq:Action0}
S&=-\frac{\mpl^2}{2}\int d^4x\sqrt{-g}R+S_{\rm m}\,,
\Eeq
where $R$ is the Ricci scalar, $g$ is the determinant of the metric and $\mpl$ is the reduced Planck mass. The matter action contains a complex scalar field $\varphi$ and the gauge field $A_\mu$:
\Beq
\label{eq:Action}
&S_{\rm m}=\int \text{d}^4x \sqrt{-g} \mathcal{L}_{\rm m}=\int \text{d}^4x \sqrt{-g}\left[\left(D_{\mu } \varphi \right){}^* \left(D^{\mu} \varphi \right)-\mathcal{V} (\left| \varphi \right|) -\frac{1}{4}F_{\mu \nu } F^{\mu \nu }\right]\,,
\Eeq
where the field tensor $F_{\mu\nu}$ for the gauge fields and the gauge-covariant derivative $D_\mu\varphi$ are given by
\Beq
\label{eq:FTCD}
F_{\mu \nu }(A)=\nabla_{\mu}A_{\nu}-\nabla_{\nu}A_{\mu}\,,\qquad D_{\mu }\varphi=\left(\nabla_{\mu}+ i \frac{g_{\!_A}}{2} A_{\mu }\right)\varphi\,.
\Eeq
In the above equations, $\nabla_\mu$ is the usual Levi-Civita connection. 
The action $S_{\rm m}$ is invariant under local $U(1)$ gauge transformations
\Beq
\label{eq:PhiU1}
\varphi \to e^{- i \frac{g_{\!_A}}{2} \alpha(x^{\nu}) } \varphi\,,\qquad A_{\mu }\to A_{\mu }+\nabla_{\mu}\alpha(x^{\nu})\,,
\Eeq
where $\alpha(x^{\nu})$ is an arbitrary real function of space and time. The total action is also invariant under space-time differomorphisms. 
The gauge symmetry implies that not all of the components of the $4$-vector $A_{\mu}$ and the real and imaginary parts of the scalar field are physical degrees of freedom (dof). We remedy this redundancy by working in the appropriate set of gauge invariant variables or by fixing the gauge. The redundancy due to space-time diffeomorphisms is handled in a similar fashion. 

We will present our answers as power spectra of the electric and magnetic fields, which are defined  in the usual way \cite{Misner:1974qy}:\footnote{$\epsilon^{ijk}$ is the antisymmetric Levi-Civita tensor. It does not raise or lower (spatial) indices.}
\Beq
\label{eq:EBdefs1}
E_{i}\equiv F_{0i}=\nabla_{0}A_{i}-\nabla_iA_{0}\,,\qquad B_{i}\equiv \frac{1}{2}\epsilon^{ilm}F_{lm}=\frac{1}{2}\epsilon^{ilm}\left(\nabla_lA_{m}-\nabla_mA_{l}\right)\,.
\Eeq

\subsection{$U(1)$ gauge invariants}
\label{sec:U(1)in}
When the field $\varphi\neq0$, it can be written in polar co-ordinates as
\Beq
\label{eq:Polars}
\varphi(x^{\mu})=\frac{1}{\sqrt{2}}\rho(x^{\mu})e^{i\frac{g_{\!_A}}{2}\Omega(x^{\mu})}\,.
\Eeq
Under the local $U(1)$ gauge transformation (see eq.~\eqref{eq:PhiU1}) $\rho\rightarrow\rho$ and $\Omega \rightarrow \Omega-\alpha$.
It is convenient to work in local $U(1)$ gauge invariant variables given by the following five fields:
\Beq
\label{eq:defG}
\rho(x^{\nu}) \qquad\text{and}\qquad G_{\mu}(x^{\nu})\equiv A_{\mu}(x^{\nu})+\nabla_{\mu}\Omega(x^{\nu})\,.
\Eeq
In these variables the matter action becomes
\Beq
\label{eq:ActionInfl}
S_{\rm m}=\int \text{d}^4x \sqrt{-g}\Big[\frac{1}{2}\nabla_{\mu}\rho \nabla^{\mu}\rho +\frac{1}{2}\left(\frac{g_{\!_A}\rho}{2}\right)^{\!\!2}G_{\mu}G^{\mu}-V (\rho) -\frac{1}{4}F_{\mu \nu }\left(G\right) F^{\mu \nu }\left(G\right)\Big]\,,
\Eeq
where $V(\rho)=\mathcal{V}(|\varphi|)$. It is worth noting that the variable $\Omega$ appearing in eq.~\eqref{eq:Polars} does not make an appearance in the above action. There is no need to worry about the $U(1)$ gauge redundancy when working with gauge invariant variables. 

Note that the $E$ and $B$ fields are  invariant with respect to $U(1)$ transformations and their expressions in terms of $G_\mu$ are identical to those in terms of $A_\mu$:
\Beq
\label{eq:EBdefs}
E_{i}=\nabla_0G_{i}-\nabla_iG_{0}\,,\qquad B_{i}=\frac{1}{2}\epsilon^{ilm}\left(\nabla_lG_{m}-\nabla_mG_{l}\right)\,.
\Eeq

\subsection{Diffeomorphism invariants}

We will work in a perturbed Friedmann-Roberston-Walker space-time with the metric:
\Beq
\label{eq:Metric}
ds^2&=(\bar{g}_{\mu\nu}+\delta g_{\mu\nu})dx^{\mu}dx^{\nu}\\
    &=\left(1+2\phi\right)a^2(\tau)d\tau^2+2\left(\partial_i\mathcal{B}+\mathcal{S}_{i}\right)a^2(\tau)dx^id\tau\\
    &\qquad-\big[\left(1-2\psi\right)\delta_{ij}-2\partial_i\partial_j\mathcal{E}-\partial_j\mathcal{K}_{i}-\partial_i\mathcal{K}_{j}-\tilde{h}_{ij}\big]a^2(\tau)dx^idx^j\,,
\Eeq
where $\phi(x^\sigma)$, $\mathcal{B}(x^{\sigma})$, $\psi(x^{\sigma})$, $\mathcal{E}(x^{\sigma})$ are scalar perturbations, $\mathcal{S}_{i}(x^{\sigma})$, $\mathcal{K}_i(x^{\sigma})$ are divergence-free $3$-vector perturbations, and $\tilde{h}_{ij}(x^{\sigma})$ is a traceless transverse $3$-tensor perturbation.

In this perturbed space-time, we define the perturbations of the following $U(1)$ invariant variables:
\Beq
\label{eq:rho}
   \rho(x^{\mu})&=\bar{\rho}(\tau)+\delta\rho(x^{\mu})\,,\\
G_{\mu}(x^{\mu})&=\left[G_0(x^{\mu}),\partial_iG^{\parallel}(x^{\mu})+G^{\perp}_i(x^{\mu})\right]\,,
\Eeq
where $G_{0}(x^{\sigma})$ and $G^{\parallel}(x^{\sigma})$ are scalars and $G^{\perp}_i(x^{\sigma})$ is a divergence-free $3$-vector. Note that the gauge fields vanish at the background level: the spatial components are zero from isotropy of the Friedmann-Robertson-Walker (FRW) background and the equations of motion will set the background temporal component to zero. Hence we will work at the linear level in $G_\mu$. 

For our scenario, there are two physical scalar metric perturbations, with scalar field perturbation and scalar parts of the gauge field adding three more. We choose to work with the following five diffeomorphism invariant combinations:
\Beq
\label{eq:diffinv}
\Phi &= \phi - \frac{1}{a}\partial_{\tau}\left[a\left(\mathcal{B}-\partial_{\tau}\mathcal{E}\right)\right]\,,\\
\Psi &= \psi + \mathcal{H}\left(\mathcal{B}-\partial_{\tau}\mathcal{E}\right)\,,\\
\delta \tilde{\rho}& =\delta \rho - \left(\partial_{\tau}\bar{\rho}\right)\left(\mathcal{B}-\partial_{\tau}\mathcal{E}\right)\,,\\
G_0&\,,\\
G^{\parallel}&\,,\\
\Eeq
where $\mathcal{H}=\partial_\tau\ln a$. The first two are the standard Bardeen variables. $G_0$ and $G^{\parallel}$ are diffeomorphism invariant since the gauge field vanishes at the background level. 

Similarly, for the vectors we chose to work with the following diffeomorphism invariant combinations
\Beq
&\tilde{V}_{i}\equiv \mathcal{S}_{i}-\partial_{\tau}\mathcal{K}_{i}\,,\\
&G^{\perp}_i\,,
\Eeq
where $\tilde{V}$ and $G^{\perp}$ are divergence free.

The traceless, transverse $3$-tensor perturbation $\tilde{h}_{ij}$ is already diffeomorphism invariant. Similarly, the electric and magnetic fields defined in eq.~\eqref{eq:EBdefs} are already diffeomorphism invariant. It is convenient to split the electric and magnetic fields into divergence-free and curl-free parts
\Beq
E_i(x^\mu)=\partial_i E^{\parallel}(x^\mu)+E_{\perp i}(x^\mu)\,,\qquad B_i(x^\mu)=B^{\perp}_i(x^\mu)\,.
\Eeq
Note that $B^{\parallel} =0$ from the definition of $B_i$ in eq. \eqref{eq:EBdefs}. 

\subsection{Equations of motion}
\label{sec:FieldDynamics}
The general equations of motion for the matter and metric fields in curved space-time take the following form:
\Beq
\label{eq:PhiEoM}
&D_{\mu } D^{\mu } \varphi +\frac{\partial \mathcal{V}}{\partial\varphi ^*}=0\,,\\
&\nabla_{\mu}F^{\mu \sigma }+ i \frac{g_{\!_A}}{2}\left(\varphi \left(D^{\sigma } \varphi \right)^*-\varphi ^* D^{\sigma }\varphi \right)=0\,,\\
&\mathcal{G}_{\mu\nu}=\frac{1}{\mpl^2}T_{\mu\nu}\,,
\Eeq
where $\mathcal{G}_{\mu\nu}$ is the Einstein tensor and the energy momentum tensor is given by
\Beq
\label{eq:EMT}
T_{\mu \nu }=2 \left(D_{(\mu } \varphi \right)^* D_{\nu) } \varphi - F_{\mu \alpha } F_{\nu}{}^{\alpha}-g_{\mu \nu } \left[\left(D^{\alpha } \varphi \right)^* D_{\alpha } \varphi -\mathcal{V}-\frac{1}{4} F_{\alpha \beta } F^{\alpha \beta }\right].
\Eeq

In terms of the $U(1)$ gauge invariant variables defined in Section \ref{sec:U(1)in}, the above equations become
\Beq
\label{eq:RhoEoM}
\nabla_{\mu}\nabla^{\mu}\rho+\frac{\partial V}{\partial \rho}-\rho\left(\frac{g_{\!_A}}{2}\right)^{\!\!2}G_{\mu}G^{\mu}=0\,,
\Eeq
\Beq
\label{eq:GEoM}
\nabla_{\mu}F^{\mu \sigma }\left(G\right)+\left(\frac{\rho g_{\!_A}}{2}\right)^{\!\!2}G^{\sigma}=0\,,
\Eeq
\Beq
\label{eq:EMTd}
T_{\mu \nu }&=\nabla_{\mu}\rho \nabla_{\nu}\rho-\left(\frac{g_{\!_A}\rho}{2}\right)^{\!\!2}G_{\mu}G_{\nu} - F_{\mu \alpha }(G) F_{\nu}{}^{\alpha}(G)\\
&\qquad-g_{\mu \nu } \left[\frac{1}{2}\nabla_{\alpha}\rho \nabla^{\alpha}\rho-\frac{1}{2}\left(\frac{g_{\!_A}\rho}{2}\right)^{\!\!2}G_{\alpha}G^{\alpha}-V(\rho)-\frac{1}{4} F_{\alpha \beta }(G) F^{\alpha \beta }(G)\right].
\Eeq
Equation \eqref{eq:GEoM} implies the following definition of the conserved 4-current:
\Beq
\label{eq:jmudef}
j^\mu=\left(\frac{\rho g_{\!_A}}{2}\right)^{\!\!2}G^{\mu}\,,\qquad \nabla_\mu j^\mu=0\,.
\Eeq
The equivalent of Maxwell's equations for the electric and magnetic fields are 
\Beq
\nabla^i E_i=\left(\frac{\rho g_{\!_A}}{2}\right)^{\!\!2}G_0= j_0\,,\qquad
\epsilon^{ilm}\nabla^{l} B_{m}-\nabla^{0} E_{i}=\left(\frac{\rho g_{\!_A}}{2}\right)^{\!\!2} G_{i}= j_{i}\,.
\Eeq
Next, we write down the equations of motion for the background (space independent) fields and linearized perturbations around these background fields in terms of gauge invariant variables. 

\subsubsection{Background}
Assuming the scalar field plays the role of the inflaton and treating the gauge fields as perturbations, the evolution of $\bar{\rho}(\tau)$ can be determined from eq.~(\ref{eq:RhoEoM}): 
\Beq
\label{eq:HomEvol}
\partial_{\tau}^{2}\bar{\rho}+2\mathcal{H}\partial_{\tau}\bar{\rho}+a^2\frac{\partial V}{\partial \bar{\rho}}=0\,,
\Eeq
where $\mathcal{H}$ is given by the $00$ background Einstein equation
\Beq
\label{eq:HubbleInflation}
\mathcal{H}^2=\frac{a^2}{3m_{\text{pl}}^2}\left(\frac{\left(\partial_{\tau}\bar{\rho}\right)^{\!2}}{2a^2}+V\right)\,.
\Eeq
The electric and magnetic fields vanish at the background level.

\subsubsection{Linearized perturbations in position space}

From eq.~\eqref{eq:RhoEoM} and eq.~\eqref{eq:GEoM} we get the equations of motion for diffeomorphism and $U(1)$ gauge invariant  scalar perturbations:
\Beq
\label{eq:dRhoEvol}
\partial_{\tau}^2\delta \tilde{\rho}&+2\mathcal{H}\partial_{\tau}\delta\tilde{\rho}-\Delta \delta\tilde{\rho}+a^2\frac{\partial^2V}{\partial\bar{\rho}^2}\delta\tilde{\rho}-\partial_{\tau}\bar{\rho}\left(3\partial_{\tau}\Psi+\partial_{\tau}\Phi\right)+2a^2\frac{\partial V}{\partial \bar{\rho}}\Phi = 0\,,
\Eeq
\Beq
\label{eq:AparlEvol}
\partial_{\tau}^2 G^{\parallel}+a^2\left(\frac{\bar{\rho}g_{\!_A}}{2}\right)^{\!\!2} G^{\parallel}-\partial_{\tau} G_0 = 0\,,
\Eeq
\Beq
\label{eq:LinearGauss}
\partial_{\tau}\Delta G^{\parallel} -\Delta G_0 + a^2\left(\frac{\bar{\rho}g_{\!_A}}{2}\right)^{\!\!2} G_0=0\,,
\Eeq
where eq.~(\ref{eq:LinearGauss}) is the linearized version of the Gauss constraint. Note that the scalar components of the gauge fields $G_0$ and $G^\parallel$ do not depend on the metric perturbations.

The evolution of the $U(1)$ gauge and diffeomorphism invariant {\it vector} perturbations involving matter fields can be obtained from
\Beq
\label{eq:Aperp}
\partial_{\tau}^2 G^{\perp}_i-\Delta G^{\perp}_i + a^2 \left(\frac{\bar{\rho}g_{\!_A}}{2}\right)^{\!\!2}  G^{\perp}_i=0\,.
\Eeq
The perturbations $G^{\perp}_i$ do not couple to metric perturbations. As mentioned earlier, the tensor perturbations are also decoupled from the matter fields at the linear level. We shall not consider tensor perturbations any further in this paper. 

We now turn to the Einstein equations. The energy-momentum tensor in eq.~(\ref{eq:EMTd}) is quadratic in the $G_{\mu}$ (with $G_\mu=0$ at the background level). Hence, to linear order in the perturbations, the energy-momentum tensor depends only on the perturbations in $\rho$ and the metric. Also $T^{i}_j=0$ for $i\ne j$; there is no anisotropic stress. The linearised Einstein equations (for scalar perturbations) yield 
\Beq
\label{eq:Bardeen}
&\Phi=\Psi\,,\\
&\left(\partial_{\tau}\mathcal{H}-\mathcal{H}^2-\Delta\right)\Psi=\frac{1}{2m_{\text{pl}}^2}\left[-\partial_{\tau}\bar{\rho}\left(\partial_{\tau}\delta\tilde{\rho}+\mathcal{H}\delta\tilde{\rho}\right)+\delta\tilde{\rho}\partial_{\tau}^2\bar{\rho}\right]\,,\\
&\partial_{\tau}\Psi+\mathcal{H}\Psi=\frac{1}{2m_{\text{pl}}^2}\delta\tilde{\rho}\partial_{\tau}\bar{\rho}\,.
\Eeq
Note that the gauge fields are completely decoupled from the scalar metric perturbations.

The picture for vector perturbations is even simpler - the linearized Einstein equations for the vector perturbations involve only metric perturbations:
\Beq
\Delta\tilde{V}_i=0\,\qquad{\rm{and}}\qquad \partial_{\tau}\left(\partial_j\tilde{V}_{i}+\partial_i\tilde{V}_{j}\right)+2\mathcal{H}\left(\partial_{j}\tilde{V}_{i}+\partial_{i}\tilde{V}_{j}\right)=0\,,
\Eeq
which do not affect the matter vector perturbations. 

At the linearized level, the equations of motion for the electric and magnetic fields defined in eq.~\eqref{eq:EBdefs} are simple\footnote{Note that in FRW background with conformal time $\nabla_\mu F^{\mu\nu}=g^{\nu\alpha}\nabla^\mu F_{\mu\alpha}=g^{\nu\alpha}\partial^\mu F_{\mu\alpha}$. }
\Beq
\label{eq:j0def}
&-\Delta E^{\parallel}=a^2\left(\frac{\bar{\rho}g_{\!_A}}{2}\right)^{\!\!2}G_{0}=a^2j_{0}\,,\\
&-\partial_{\tau}E^{\parallel}=a^2\left(\frac{\bar{\rho}g_{\!_A}}{2}\right)^{\!\!2} G^\parallel\equiv a^2j^{\parallel}\,,\\
&\epsilon^{ilm}\partial^{l} B^{\perp}_m-\partial_{\tau} E^{\perp}_i=a^2\left(\frac{\bar{\rho}g_{\!_A}}{2}\right)^{\!\!2} G^{\perp}_i\equiv a^2j^{\perp}_i\,.
\Eeq
On the right-hand sides of the last two equations, we have defined the scalar perturbations and divergence-free vector perturbations in the 3-current density respectively.

\subsubsection{Linearized perturbations in Fourier space}
For calculational purposes, we move to Fourier space. Fourier space is also particularly convenient for solving the various constraint equations, which essentially become algebraic in the relevant variables. Our Fourier convention is $f({\bx},\tau)=\int f_{\bk}(\tau)\text{e}^{i{\bk}\cdot{\bx}} \text{d}^\text{3}\bk$.

We begin with the equations of motion for the scalar perturbations. From the (Fourier space version of the) constraints, eqns. \eqref{eq:Bardeen}, we can substitute the gravitational potential $\Psi_{\bk}$ and its derivative into the evolution equation for $\delta\tilde{\rho}_\bk$ (cf. eq.~\eqref{eq:dRhoEvol}) to obtain an equation of motion which only involves $\delta\tilde{\rho}_{\bk}$:
\Beq
\label{eq:dPhi1Quant}
&\partial_{\tau}^2\delta\tilde{\rho}_{\bk}+2\mathcal{H}\partial_{\tau}\delta\tilde{\rho}_{\bk}+k^2\delta\tilde{\rho}_{\bk}+a^2\frac{\partial^2 V}{\partial \bar{\rho}^2}\delta\tilde{\rho}_{\bk}\\
&\qquad\qquad\quad+\frac{2}{\mpl^2}\left[\left(\mathcal{H}\partial_{\tau}\bar{\rho}+\frac{a^2}{2}\frac{\partial V}{\partial \bar{\rho}}\right)\frac{\delta\tilde{\rho}_{\bk}\partial_{\tau}^2\bar{\rho}-\partial_{\tau}\bar{\rho}\left(\partial_{\tau}\delta\tilde{\rho}_{\bk}+\mathcal{H}\delta\tilde{\rho}_{\bk}\right)  }{\partial_{\tau}\mathcal{H}-\mathcal{H}^2+k^2} - \left(\partial_{\tau}\bar{\rho}\right)^{\!2}\delta\tilde{\rho}_{\bk} \right]=0\,.
\Eeq
The remaining scalar perturbations, $G_0$ and $G^{\parallel}$, are governed again by an evolution equation, eq.~(\ref{eq:AparlEvol}), and a constraint, eq.~(\ref{eq:LinearGauss}).  Before moving to the Fourier transformed versions of these equations we define the longitudinal (i.e. curl-free) component of the space-like part of $G_{\mu}$:
\Beq
\left(\boldsymbol{G}^L\right)_{i}=\partial_{i}G^{\parallel}\,.
\Eeq
The Fourier transform of  $\boldsymbol{G}^{L}$ can be expressed in terms of a longitudinal polarisation vector, $\boldsymbol{\epsilon}^L_\bk$, as follows:
\Beq
\boldsymbol{G}^{L}_{\bk}=\boldsymbol{\epsilon}^L_\bk G^L_{\bk}\,,
\Eeq
where we shall call the scalar $G^L_{\bk}$, the {\it longitudinal mode}. The polarisation vector has the following properties:
\Beq
\label{eq:epsL}
&\boldsymbol{\epsilon}^L_\bk=\boldsymbol{\epsilon}^{L*}_{-\bk}\,,\qquad \boldsymbol{\epsilon}^{L*}_{\bk}\cdot\boldsymbol{\epsilon}^L_\bk=1\,,\qquad i\bk\cdot\boldsymbol{\epsilon}^L_\bk=k \,,\qquad i\bk\times\boldsymbol{\epsilon}^L_\bk=0\,.
\Eeq
The Fourier transformed equations, eqns.~\eqref{eq:AparlEvol} and \eqref{eq:LinearGauss}, then take the form
\Beq
\label{eq:GaussL}
\left(\frac{g_{\!_A}\bar{\rho}a}{2}\right)^{\!\!2} G_{0\bk}=-k^2 G_{0\bk}-k\partial_{\tau} G^L_{\bk}\,,
\Eeq
\Beq
\label{eq:dGL}
\partial_{\tau}^2 G^L_{\bk}+k\partial_{\tau} G_{0\bk}+\left(\frac{g_{\!_A}\bar{\rho}a}{2}\right)^{\!\!2} G^L_{\bk}=0\,.
\Eeq
There is a similarity between the pairs $\left( G^L_{\bk}, G_{0\bk}\right)$ and $\left(\delta \tilde{\rho}_{\bk}, {\Psi}_{\bk}\right)$. $G^L_\bk$ and $\delta \tilde{\rho}_{\bk}$ are both dynamical fields, evolved according to second order in time equations of motion, eq.~(\ref{eq:dGL}) and eq.~(\ref{eq:dRhoEvol}), respectively. Each of these perturbations has its own auxiliary field, $G_{0\bk}$ and $\Psi_{\bk}$ respectively, determined by a constraint equation. Substituting the auxiliary field (i.e. $G_{0\bk}$) into the equation of motion we obtain the following expression in terms of $G^L_\bk$ only:
\Beq
\label{eq:dGLfull}
\partial_{\tau}^2 G^L_{\bk}+2\left(\mathcal{H}+\frac{\partial_{\tau}\bar{\rho}}{\bar{\rho}}\right)\frac{\partial_{\tau} G^L_{\bk}}{1+\left(\frac{g_{\!_A}\bar{\rho}a}{2k}\right)^{\!\!2}}+\left[k^2+\left(\frac{g_{\!_A}\bar{\rho}a}{2}\right)^{\!\!2}\right] G^L_{\bk}=0\,.
\Eeq

We now turn to vector perturbations. The matter vector perturbations are decoupled from the metric vector perturbations. Similarly to the longitudinal case, we introduce a pair of transverse polarisation vectors $\boldsymbol{\epsilon}^{T\pm}_\bk$ which satisfy
\Beq
\label{eq:epsT}
\boldsymbol{\epsilon}^{T\pm}_\bk=\boldsymbol{\epsilon}^{T\pm*}_{-\bk} \text{,}\qquad \boldsymbol{\epsilon}^{T\lambda'*}_\bk\cdot\boldsymbol{\epsilon}^{T\lambda}_{\bk}=\delta^{\lambda'\lambda}\,,\qquad
i\bk\cdot\boldsymbol{\epsilon}^{T\pm}_\bk=0 \text{,}\qquad i\bk\times\boldsymbol{\epsilon}^{T\pm}_\bk=\pm k\boldsymbol{\epsilon}^{T\pm}_\bk\,.
\Eeq
The divergence-free perturbations in terms of these polarization vectors are
\Beq
\boldsymbol{G}^{\perp}_{\bk}(\tau)=\sum_{\lambda=\pm}\boldsymbol{\epsilon}^{T\lambda}_\bk G^{T\lambda}_{\bk}(\tau)\,,
\Eeq
with the equations of motion
\Beq
\label{eq:GT}
\partial_{\tau}^2 {G}^{T\pm}_{\bk}+\left[k^2+\left(\frac{g_{\!_A}\bar{\rho}a}{2}\right)^{\!\!2}\right]{G}^{T\pm}_{\bk}=0\,.
\Eeq
The fact that Hubble friction does not appear in the evolution equations for the transverse modes is because of conformal invariance of massless gauge fields. However, the longitudinal components (which exist when the gauge field is effectively massive) do feel Hubble friction.

One can also rewrite the electric and magnetic fields, and the 3-current current, in terms of longitudinal and transverse modes, e.g.
\Beq
E_i=\partial_i E^{\parallel}+E^{\perp}_i=(\boldsymbol{E}^L)_i+E^{\perp}_i\,,
\Eeq
which in terms of the polarization vectors in Fourier space becomes
\Beq
\boldsymbol{E}_\bk=\boldsymbol{\epsilon}^L_\bk E^L_\bk+\sum_{\lambda=\pm}\boldsymbol{\epsilon}^{T\lambda}_\bk E^{T\lambda}_\bk\,.
\Eeq
Similar expansions hold for $j_i$ and $B_i$ with the exception $B^L_\bk=0$. Below we give the expressions used in the subsequent sections to calculate the primordial power spectra of the longitudinal and transverse modes of $E$ and $B$
\Beq
\label{eq:EGBG}
&E^L_\bk=kG_{0\bk}+\partial_\tau G^L_\bk=\frac{\left(\frac{\bar{\rho}g_{\!_A}}{2}\right)^{\!\!2}}{\frac{k^2}{a^2}+\left(\frac{\bar{\rho}g_{\!_A}}{2}\right)^{\!\!2}}\partial_\tau G^L_\bk\,,\\
&E^{T\pm}_\bk=\partial_\tau G^{T\pm}_\bk\,, \qquad B^{T\pm}_\bk=\pm k G^{T\pm}_\bk\,.
\Eeq
The charge and current densities can also be expressed in terms of the $G$ field's longitudinal and transverse modes
\Beq
\label{eq:ChargeAndCurrent}
j_{0\bk}=\left(\frac{\bar{\rho}g_{\!_A}}{2}\right)^{\!\!2}G_{0\bk}=\frac{-\left(\frac{\bar{\rho}g_{\!_A}}{2}\right)^{\!\!2}}{\frac{k^2}{a^2}+\left(\frac{\bar{\rho}g_{\!_A}}{2}\right)^{\!\!2}}\frac{k}{a^2}\partial_\tau G^L_\bk\,,\qquad j^L_\bk=\left(\frac{\bar{\rho}g_{\!_A}}{2}\right)^{\!\!2}G^L_\bk\,,\qquad j^{T\pm}_\bk=\left(\frac{\bar{\rho}g_{\!_A}}{2}\right)^{\!\!2}G^{T\pm}_\bk\,.
\Eeq
The first identity is simply $-kE^L_\bk=a^2j_{0\bk}$, i.e. the Gauss constraint. In a consistent quantum analysis of the perturbations during and after inflation, one cannot set $j_{0\bk}$ and $j^L_{\bk}$ to zero by hand (this differs from the treatment in~\cite{Emami2009},~\cite{Emami2010}). This is because the vacuum fluctuations in $G^L_{\bk}$ can be enhanced due to horizon crossing during inflation or non-adiabatic particle production during preheating. We shall see this aspect in detail in the subsequent sections.

\subsection{Gauge transformations}
\label{sec:GaugeTransf}
In the upcoming sections we will use either the gauge invariant variables discussed above or work in some particular gauge depending on which approach is best for the problem at hand. In this short section we provide the relationships between variables in some of the popular gauges used in the literature and the gauge invariant ones. These relationships can also be used to move from one gauge to another. When the variables are well defined, we can use the equations of motion and the solutions in the gauge invariant case to recover the corresponding expressions in our gauge of choice. Occasionally, some of the variables become ill-defined or the relationships between variables require a patching up of co-ordinate maps (for example during reheating). Such cases can be dealt with on an individual basis, or one simply derives the equations of motion and solutions directly from the equations of motion themselves.

In different gauges (but not the gauge invariant case) the complex scalar field is represented as $\varphi=(\varphi^0+i\varphi^1)/\sqrt{2}$, and the gauge fields by $A^\mu$. For perturbations, we will always work around an FRW background. Using the global $U(1)$ invariance, we set the homogeneous, imaginary part of $\varphi$, $\bar{\varphi}^1=0$. That is $\varphi=(\bar{\varphi}^0+\delta\varphi^0+i\delta\varphi^1)/\sqrt{2}$. For the homogeneous part $\bar{\varphi}^0=\pm\rho$ with the $\pm$ sign accounting for the change in sign during an oscillation through zero. In Fourier space, the transverse modes and the scalar perturbations along the direction of motion of the homogeneous field (in all the gauges discussed below) are related to the gauge invariant variables as follows:
\Beq
\label{eq:GaugeTransfrhoAT}
\delta\vp^0_\bk=\delta\rho_\bk\,,\qquad A^{T\pm}_\bk=G^{T\pm}_\bk\,,
\Eeq
with the equations of motion for $\delta\varphi^0_\bk$ and $A_\bk^{T\pm}$ being identical to the gauge invariant case with $\rho \rightarrow \bar{\varphi}^0$.\\ \\
\noindent {\it  Coulomb} gauge:\\
In this gauge $\partial_i  A^i=0$. In Fourier space we get
\Beq
\label{eq:GaugeTransfCoulomb}
\delta\vp^1_\bk=-\frac{\bar{\rho}g_{\!_A}}{2k}G^L_\bk\,,\qquad A_{0\bk}=G_{0\bk}+\frac{1}{k}\partial_\tau G^L_\bk\,,\qquad A^L_\bk=0\,.
\Eeq\\
\noindent {\it Unitary} gauge:\\
In this gauge  $\varphi^1=0$, which yields
\Beq
 \delta\varphi^1_\bk=0\,,\qquad A_{0\bk}=G_{0\bk}\,,\qquad A^L_{\bk}=G^L_{\bk}\,.
 \Eeq
The equations in the Unitary gauge are identical to those in the gauge invariant one.  \\ \\
\noindent {\it Temporal} gauge:\\
In this gauge $A_0=0$. However, the theory is still invariant under the time-independent transformation $A_i\rightarrow A_i+\partial_i \alpha({\bf{x}})$ and $\delta\vp^1\rightarrow \delta\vp^1-\bar{\vp}^0g_{\!_A}\alpha({\bf{x}})/2$, which in Fourier space translates to $A^L_\bk\rightarrow A^L_\bk-k\alpha_\bk$ and $\delta\vp^1_\bk\rightarrow \delta\vp^1_\bk-\bar{\vp}^0g_{\!_A}\alpha_\bk/2$. Hence, we completely fix the gauge by choosing an $\alpha$ such that at some moment of time, $\tau=\tau_\text{in}$, $\delta\vp^1_\bk(\tau_\text{in})=0$. With this condition, we have
\Beq
\label{eq:GaugeTransfTemporal}
\delta\vp^1_\bk=\frac{\bar{\rho}g_{\!_A}}{2}\int_{\tau_\text{in}}^\tau\, G_{0\bk}(\eta)d\eta\,,\qquad A_{0\bk}=0\,,\qquad A^L_\bk=G^L_\bk+k\int_{\tau_\text{in}}^\tau\, G_{0\bk}(\eta)d\eta\,.
\Eeq
\noindent  {\it Lorenz} gauge: \\
In this gauge $\nabla_\mu A^{\mu}=0$. In this case
\Beq
\label{eq:GaugeTransfLorenz}
\delta\vp^1_\bk&=\bar{\rho}g_{\!_A}(A^L_\bk-G^L_\bk)/(2k)\,,\qquad A_{0\bk}=G_{0\bk}+\partial_\tau(G^L_{\bk}-A^L_{\bk})/k\,,\\
A^L_\bk&=\int\,\mathcal{G}(\tau,\eta)\left(\partial_\eta+2\mathcal{H}(\eta)\right)\left(k G_{0\bk}(\eta)+\partial_\eta G^L_\bk(\eta)\right)d\eta\,,
\Eeq 
where $\mathcal{G}(\tau,\eta)$ is the Greens function of the linear operator $\mathcal{L}_\tau \equiv \partial^2_\tau+2\mathcal{H}(\tau)\partial_\tau+k^2$. Arriving at the above form of the relationship between variables requires a bit of explanation. 
In Fourier space the Lorenz gauge condition translates to $\partial_\tau A_{0\bk}+2\mathcal{H}A_{0\bk}-kA^L_{\bk}=0$, and the equation governing $A^L_\bk$ yields $\mathcal{L}_\tau A^L_\bk =(\partial_\tau+2\mathcal{H})(k G_{0\bk}+\partial_\tau G^L_\bk)$. This equation yields the particular solution above only if we can set the complementary solution to zero.
This can always be done since there is a residual degree of freedom $\chi$ such that under the transformations $A_\mu\rightarrow A_\mu +\nabla_\mu \chi$ and $\delta\vp^1\rightarrow \delta\vp^1-\bar{\vp}^0g_{\!_A}\chi/2$, the theory remains invariant; provided $\chi$ obeys $\nabla_\mu\nabla^\mu\chi=0$. In Fourier space we get $\mathcal{L}_\tau \chi_\bk=0$. Since the operator evolving $\chi_\bk$ is identical to the one evolving  $A^L_\bk$, we can always choose $\chi_\bk$ such that the complementary part of $A^L_\bk$ vanishes, thus arriving at the particular solution provided above.


\section{Inflationary dynamics}
\label{sec:inflation}
The background dynamics are relatively straightforward during inflation. At the phenomenological level, with an appropriate choice of the potential $V$ and initial conditions we can easily arrange for $-\partial_t{H}/H^2=-a\partial_\tau(\mathcal H/a)/\mathcal{H}^2\ll 1$ for sufficient number of $e$-folds.  For simple models, this corresponds to $H=(\mathcal H/a)\approx $ const and $\bar{\rho}\approx$ const during inflation. Inflation ends when $\partial_\tau{\mathcal H}=0$, when accelerated expansion stops and the field starts rolling quickly. Assuming such a background solution has been found, we focus on the quantum fluctuations around this classical background. Quantization of constrained systems, like the problem at hand, can be tricky. We find that by working in Fourier space with gauge invariant variables and substituting the constraints before quantizing, the process becomes straightforward. Once the appropriate quantized solutions for the scalar and gauge fields are available, we construct the power spectra of the electric and magnetic fields at the end of inflation.

\subsection{Quantized scalar and vector perturbations}
\label{sec:QuantPerturb}
For the purposes of quantization, it is convenient to write down the action for the Fourier components of the dynamical perturbation variables left after imposing the constraints. The equations of motion for these variables $\delta\tilde{\rho}_\bk, G^L_\bk$ and  $G^{T\pm}_\bk$ were provided in the previous section (see eqns.~\eqref{eq:dPhi1Quant}, \eqref{eq:dGLfull} and \eqref{eq:GT}). The total quadratic action of these variables  naturally splits into four parts:
\Beq
\label{eq:S2}
S_{\rm m}^{(2)}=S^{\rho}+S^{L}+S^{T+}+S^{T-}=\sum_I S^I\,,
\Eeq
where $S^I$ are the quadratic actions for the perturbations in the $I$-th variable $f^I$ with
\Beq
\label{eq:S_I}
S^I=\int \text{d}\tau L^I(\tau)=\int d\tau{\int d^3k\,\,b_{I}(k,\tau) \left[\frac{1}{2}|\partial_\tau f^I_{\bk}|^2-\frac{1}{2}\omega^2_{I}(k,\tau)|f^I_{\bk}|^2\right]}\,.
\Eeq
The explicit forms of $b_I(k,\tau)$ and $\omega_I(k,\tau)$ will be provided below for the different components. We first outline the general quantization procedure common to all of the components. The conjugate momentum density of the $I$-th variable
\Beq
\label{eq:pi}
\pi^{I}_{\bk}(\tau)=\frac{\delta\left(L^I(\tau)\right)}{\delta\left(\partial_\tau f^I_{-\bk}\right)}=b_I\partial_\tau f^I_{\bk}\,,
\Eeq
where we have made use of $f_{\bk}^{I*}=f^I_{-\bk}$. The field operators $\hat{f}^I_\bk$ along with their conjugate momenta operators $\pi^I_\bk$  must obey the standard equal time commutators:
\Beq
\label{eq:commutators}
\left[\hat{f}^I_\bk(\tau),\hat{f}^J_{\bq}(\tau)\right]=0\,,\quad\left[\hat{\pi}^{I}_{\bk}(\tau),\hat{\pi}^{J}_{\bq}(\tau)\right]=0\,,\quad\left[\hat{f}^I_\bk(\tau),\hat{\pi}^{J}_{\bq}(\tau)\right]=i\left(2\pi\right)^{-3}\delta^{IJ}\delta(\bk+\bq)\,.
\Eeq
Note that our fields are not canonically normalized. The non-vanishing commutator and eq. \eqref{eq:pi} imply
\Beq
\label{eq:Comm}
\left[\hat{f}^I_\bk(\tau),\partial_\tau\hat{f}^{J}_{\bq}(\tau)\right]=i\left(2\pi\right)^{-3}\left(b_I(k,\tau)\right)^{-1}\delta^{IJ}\delta(\bk+\bq)\,.
\Eeq
The field operators $\hat{f}^I_\bk$ can be expanded in terms of operators $\hat{a}^I_\bk$ and mode functions $u^I_k(\tau)$ as 
\Beq
\label{eq:ftou}
\hat{f}^I_\bk(\tau)=\hat{a}^{I}_{\bk}u^{I}_{k}(\tau)+\hat{a}^{I\dagger}_{-\bk} u^{I*}_{k}(\tau)\,,
\Eeq
where each of the mode functions $u^I_{k}(\tau)$ satisfies the corresponding field equations of motion obtained by varying the action $S^I$  (same equations as those satisfied by $f^I_\bk$)
\Beq
\label{eq:uIEoM}
\partial_\tau^2u^I_k+\left(\partial_\tau \ln b_I\right)\partial_\tau u^I_k+\omega_I^2u^I_k=0\,.
\Eeq

The final ingredient needed for evolving the mode function $u^{I}_{k}(\tau)$ (and hence the field operators), are the initial conditions for the mode functions.  Their normalization will in turn also determine the commutation relation for the operators $\hat{a}^{I}_{\bk}$. We will determine the initial conditions by constructing WKB solutions for the mode functions satisfying eq.~\eqref{eq:uIEoM} at very early times. 

To proceed to the WKB solutions for the initial conditions we need the explicit forms of $b_I(k,\tau)$ and $\omega_I(k,\tau)$, which are provided below. For $I=\rho$, i.e. when $f^\rho_{\bk}=\delta\tilde{\rho}_\bk$ we have
\Beq
\label{eq:frho}
b_\rho(k,\tau)&=a^2\exp\left[\frac{1}{2\mpl^2}\int_{\tau_i}^\tau d\tau \left(\frac{\partial_\tau (\partial_\tau\bar{\rho})^2}{\partial_\tau{\mathcal{H}}-\mathcal{H}^2+k^2}\right)\right]\,,\\
\omega_\rho^2(k,\tau)&=k^2+a^2\partial_{\bar{\rho}}^2V-\frac{1}{\mpl^2}\left[\partial_\tau^2\bar{\rho}\left(\frac{\partial_\tau^2\bar{\rho}-\mathcal{H}\partial_\tau\bar{\rho}}{\partial_\tau \mathcal{H}-\mathcal{H}^2+k^2}\right)+2(\partial_\tau\bar{\rho})^2\right]\,.
\Eeq
Similarly, for $I=L, T\pm$, i.e. when $f^I_\bk=G^L_\bk, G^{T\pm}_\bk$ we have
\Beq
\label{eq:fLT}
&b_{L}(k,\tau)=\left[1+\left(\frac{2 k}{\bar{\rho}g_{\!_A}a}\right)^{\!\!2}\right]^{-1}\,,\qquad \omega_L^2(k,\tau)=k^2+\left(\frac{\bar{\rho}g_{\!_A}a}{2}\right)^{\!\!2}\,,\\
&b_{T\pm}(k,\tau)=1\,,\qquad \omega_{T\pm}^2(k,\tau)=k^2+\left(\frac{\bar{\rho}g_{\!_A}a}{2}\right)^{\!\!2}\,.
\Eeq
One can check that extremising the action in eq. \eqref{eq:S2} with respect to each of the field perturbations gives the corresponding equations of motion from the previous section, namely eqns. \eqref{eq:dPhi1Quant}, \eqref{eq:dGLfull} and \eqref{eq:GT}.

At early enough times during inflation as $a\rightarrow 0$, a given $k$ mode of interest will be deep inside the horizon $k\gg \mathcal{H}$ and will dominate all other physical scales (for example, $k\gg (\bar{\rho} g_{\!_A} a)$) as $a\rightarrow 0$. At such early enough times during inflation $\partial_\tau^2\ln b_I,\,\left(\partial_\tau\ln b_I\right)^{\!2}\ll \omega_I^2\,$. 
This hierarchy can be verified by noting that for each component $\omega_I^2\rightarrow k^2$ and $\left(\partial_\tau\ln b_{\rho,L}\right)^{\!2},\partial^2_\tau\ln b_{\rho,L}\rightarrow \mathcal{O}[\mathcal{H}^2],\,\partial_\tau\ln(b_{T\pm})=0$. With this information at hand, the WKB solution of eq.~\eqref{eq:uIEoM} at early enough times during inflation is\footnote{This can be obtained by making the usual transformation to eliminate the ``friction term" from eq.~\eqref{eq:uIEoM} and then using the general form of the 1st order WKB solution. The final form also assumes that $\omega_I^2$ dominates over other terms in the WKB solution.}
\Beq
\label{eq:uearly}
u^I_k(\tau)\rightarrow  \frac{1}{(2\pi)^{3/2}\sqrt{2}}\frac{1}{\sqrt{b_I(k,\tau)\omega_I(k,\tau)}}\exp{\left(-i\int_{\tau_\text{in}}^\tau \text{d}\tau' \omega_I(k,\tau')\right)}\,, 
\Eeq
where for the WKB solution to be valid 
\Beq
\left|\partial_\tau\left[\omega^2_I-\frac{1}{2}\partial_\tau^2 \ln b_I-\frac{1}{4}(\partial_\tau\ln b_I)^2\right]^{-1/2}\right|\ll 1\,.
\Eeq
The time independent normalization of the mode functions ($\left[(2\pi)^{3/2}\sqrt{2}\right]^{-1}$) was chosen so that the usual commutation relation for the creation and annihilation operators is satisfied. That is, 
using the above mode functions in eqns. \eqref{eq:Comm} and \eqref{eq:ftou} implies that at early times
\Beq
\left[\hat{a}^I_{\bk},\hat{a}^J_{-\bk}\right]=0\,,\qquad\left[\hat{a}^I_{\bk},\hat{a}^{J\dagger}_{-\bq}\right]=\delta^{IJ}\delta(\bk+\bq)\,.
\Eeq
Since these operators are time-independent, these relationships remain true at all times. Thus by starting with the initial conditions for the mode functions $u^I_k$ in \eqref{eq:uearly}, we can evolve them using eq.~\eqref{eq:uIEoM} to any later time and obtain the necessary power spectra. The forward time evolution will take the initially sub-horizon and/or ultra-relativistic ($k\gg \mathcal H$ and/or $k\gg \bar{\rho}g_{\!_A}a$) solutions to superhorizon and non-relativistic ones. 

Before ending this section we make some general comments about the early time solutions for the mode functions $u^I_k$ during inflation written explicitly below:
\Beq
\label{eq:uWKB}
&u^\rho_k(\tau)\rightarrow \frac{\text{e}^{-ik\tau}}{\left(2\pi\right)^{3/2}a(\tau)\sqrt{2k}}\,,\\
&u^L_k(\tau)\rightarrow \frac{\left(\frac{2k}{\bar{\rho}g_{\!_A}}\right)\text{e}^{-ik\tau}}{\left(2\pi\right)^{3/2}a(\tau)\sqrt{2k}}\,,\\
&u^{T\pm}_k(\tau)\rightarrow \frac{\text{e}^{-ik\tau}}{\left(2\pi\right)^{3/2}\sqrt{2k}}\,.
\Eeq
The early time solution for the mode function $u^\rho_k$ of $\delta\hat{\tilde{\rho}}_\bk$ reflects the fact that at early enough times, we are simply dealing with an effectively massless ($k^2\gg a^2\partial_\rho^2 V$) field on subhorizon $k\gg \mathcal{H}$ scales where metric perturbations are negligible. The early time solution is identical to the mode functions for the Minkowski vacuum apart from the trivial $a(\tau)$ scaling.
The mode function $u^L_k$ of the longitudinal component of the gauge fields $\hat{G}^L_\bk$ is related to $u^\rho_k$ in a simple way: $u^L_k= \left(\frac{2k}{\bar{\rho}g_{\!_A}}\right) u^\rho_k$. The rescaled field $(\bar{\rho}g_{\!_A}/2k)\hat{G}^L_\bk$ behaves as a massless scalar field, which is a manifestation of the Goldstone Boson Equivalence Theorem. The scalefactor does not appear in the early time mode functions $u^{T\pm}_k$ of the transverse gauge field modes $\hat{G}^{T\pm}_\bk$. This reflects the fact that massless transverse modes of gauge fields are conformally invariant. 

Finally, we note that the above early solutions were constructed assuming slow roll inflation. However, for large enough $k$ these solutions remain valid even at the end of inflation as is to be expected, albeit with slightly different conditions on $k$.  The conditions for these solutions to be valid are $k^2\gg \mathcal{H}^2, a^2\partial^2_\rho V$ for $u^\rho_k$; $k\gg \partial_\tau(a\bar{\rho})/(a\bar{\rho}), a\bar{\rho}g_{\!_A}/2$ and $k^2\gg |\partial_\tau^2(a\bar{\rho})/(a\bar{\rho})|$ for $u^L_k$; and $k\gg a\bar{\rho}g_{\!_A}/2$ for $u^{T\pm}_k$. These are useful for checking the ultra-relativistic solutions at the end of inflation.

\subsection{Inflationary power spectra}
\label{sec:Inflps}

Let us now use the developed formalism to compute the power spectra of the matter fields at the end of inflation - the first moment when $\partial_\tau\mathcal{H}=0$.\footnote{Which is equivalent to the standard expression $\ddot{a}(t)=0$, where {\it t} is cosmic time.} We will calculate the power spectra of the gauge invariant electric and magnetic fields. 

\begin{figure*}[t] 
   \centering
   \includegraphics[width=6.4in]{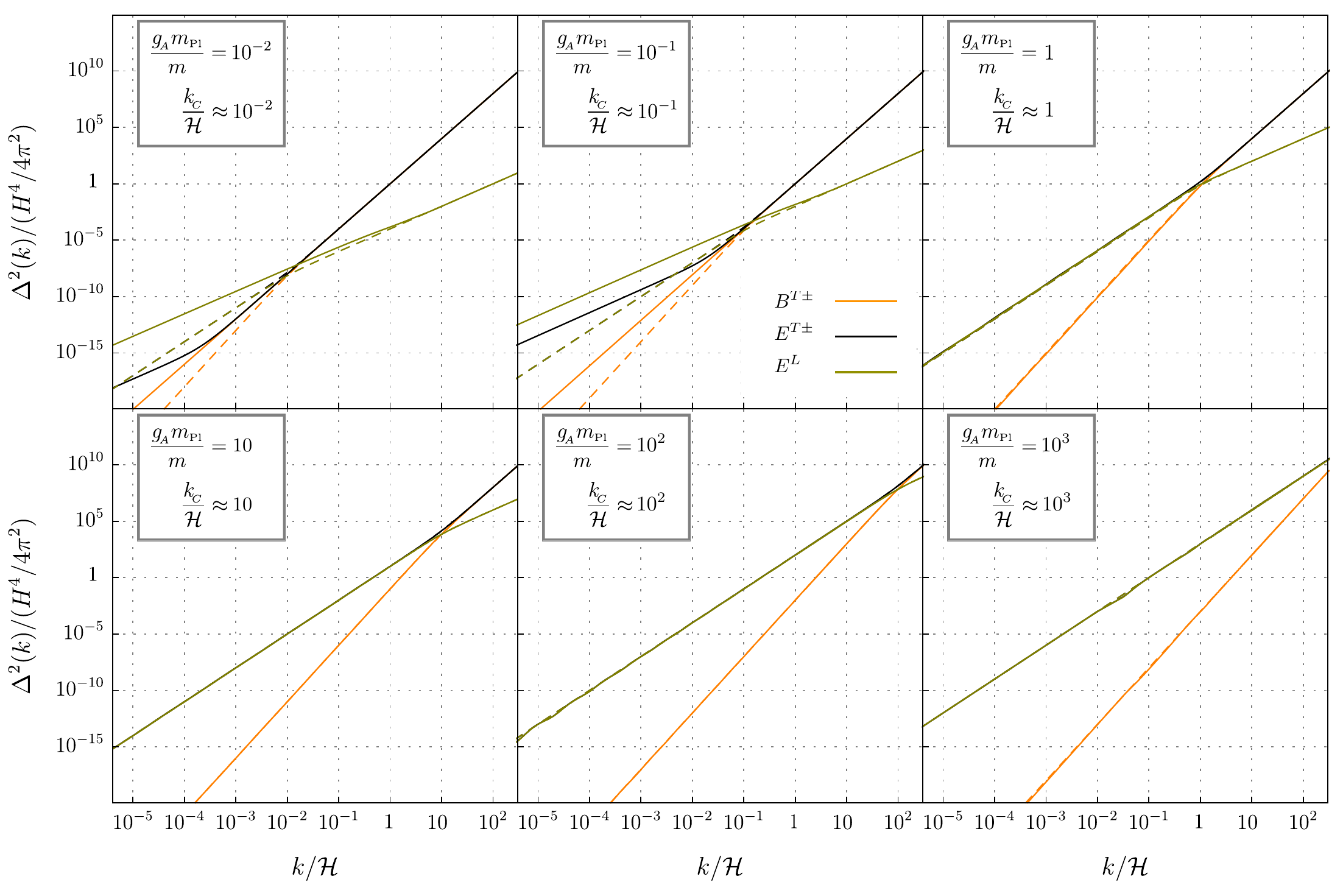}
   \caption{The power spectra of the transverse and longitudinal electric and magnetic fields at the end of inflation for $m^2|\varphi|^2$ inflation ({solid lines}) for different strengths of the couplings $g_{\!_A}$ between gauge fields and the inflaton condensate. Power spectra at the end of inflation ignoring the effects of expansion ({dashed lines}) are shown for comparison. We show three different spectra: the longitudinal electric field, $\Delta_{E^L}^2$ ({green}), the transverse electric field, $\Delta_{E^{T\pm}}^2$ ({black}), and the magnetic field, $\Delta_{B^{T\pm}}^2$ ({orange}). The mode functions of the fields responsible for these spectra can have two distinct forms depending on the magnitude of the scale $k_{\!_C}\equiv a\bar{\rho}g_{\!_A}/2$ where $\bar{\rho}$ is the magnitude of the inflaton field at the end of inflation. For the upper row with $k_{\!_C}\lesssim\mathcal{H}$, $\Delta_{E^L}^2$ (green) and $\Delta_{B^{T\pm}}^2$ (orange) are nearly single power-laws, whilst $\Delta_{E^{T\pm}}^2$ (black) is a double power-law with a break $k\sim k_{\!_C}(k_{\!_C}/\mathcal{H})$. For the lower row with $k_{\!_C}\gtrsim\mathcal{H}$, all three power spectra have double power-law forms, with a break at $k\sim k_{\!_C}$.  Note that one can also use these plots to read-off the power spectrum of the charge density, $\Delta_{j_{0}}^2=(k/a)^2\Delta_{E^L}^2$. More detailed expressions for the spectra are provided in eqns. \eqref{eq:subhCompton} and \eqref{eq:suphCompton}. An analysis of these spectra in de Sitter space (which explains the different power-laws) is provided in Appendix~\ref{sec:dS}. Note that the Hubble parameter $H$ (and its conformal counterpart $\mathcal{H}$) {\it at the end of inflation} was used to make the relevant quantities dimensionless on the horizontal and vertical axes.}   
   \label{fig:PsEndOfInflation}
\end{figure*}

The correlation functions of the electric and magnetic fields can be written in terms of the longitudinal and transverse mode functions as follows:
\Beq
\left<0\right|\hat{E}_{j\bq}\hat{E}^{\dagger}_{j\bk}\left|0\right>&=\delta\left(\bq-\bk\right)\left({\left|u^{E^L}_k\right|^2}+\sum_{\lambda=\pm}{\left|u^{E^{T\lambda}}_k\right|^2}\right)\,,\\
\left<0\right|\hat{B}_{j\bq}\hat{B}^{\dagger}_{j\bk}\left|0\right>&=\delta\left(\bq-\bk\right)\sum_{\lambda=\pm}\left| u^{B^{T\lambda}}_k\right|^2\,,\\                                                                                           
\Eeq
where from \eqref{eq:EGBG}
\Beq
\label{eq:EBmodefnsGmodefns}
u^{E^L}_k=\left[1+\left(\frac{2 k}{\bar{\rho}g_{\!_A}a}\right)^{\!\!2}\right]^{-1}\partial_\tau u^L_k\,,\qquad u^{E^{T\lambda}}_k=\partial_\tau u^{T\lambda}_k\,, \qquad u^{B^{T\pm}}_k=\pm k u^{T\pm}_k\,.
\Eeq
The power spectra of the electric and magnetic fields are defined as\footnote{The two transverse modes would have different power spectra if there was axion coupling which we do not consider here. The way we have split the transverse modes into states $\pm$ with well-defined helisities makes our analysis easily extendible to the case of charged Higgs with an additional axion-like interaction.}
\Beq
\label{eq:EBps}
\Delta^2_{E^{T\pm}}=4\pi k^3\frac{|u^{E^{T\pm}}_k|^2}{a^4}\,,\qquad
\Delta^2_{B^{T\pm}}=4\pi k^3\frac{|u^{B^{T\pm}}_k|^2}{a^4}\,,\qquad
\Delta^2_{E^L}=4\pi k^3\frac{|u^{E^{L}}_k|^2}{a^4}\,.
\Eeq
For concreteness, we compute these power spectra at the end of inflation for the chaotic inflation scenario with $V(\rho)=(1/2)m^2\rho^2$. However, our results hold more generally. For an analytic understanding of the features in the power spectra we solve the equations of motion in de Sitter space to zeroth order in slow-roll (i.e. we ignore time derivatives of the inflaton) in the Appendix~\ref{sec:dS}. We assume that inflation lasts $60$ e-folds and take the inflaton mass to be $m=10^{-6}\,\mpl$. The longitudinal and transverse  fields, $\hat{G}^L_{\bk}$ and $\hat{G}^{T\pm}_{\bk}$ are evolved numerically from very early times when they are deep inside the horizon with $k\gg\mathcal{H}$, $k\gg a\bar{\rho}g_{\!_A}/2$ and $k^2\gg \left|\partial_{\tau}^2(a\bar{\rho})/a\bar{\rho})\right|$, and have corresponding electric and magnetic WKB power spectra (cf. eqns. \eqref{eq:uWKB}, \eqref{eq:EBmodefnsGmodefns}, \eqref{eq:EBps}) 
\Beq
\Delta_{E^{T\pm}}^2=\frac{H^4}{4\pi^2}\left(\frac{k}{\mathcal{H}}\right)^{\!\!4}\,,\qquad
\Delta_{B^{T\pm}}^2=\frac{H^4}{4\pi^2}\left(\frac{k}{\mathcal{H}}\right)^{\!\!4}\,,\qquad
\Delta_{E^L}^2=\frac{H^4}{4\pi^2}\left(\frac{k}{\mathcal{H}}\right)^{\!\!2}\left(\frac{k_{\!_C}}{\mathcal{H}}\right)^{\!\!2}\,.
\Eeq
The power spectra of the electric and magnetic fields at the end of inflation are shown in Fig.~\ref{fig:PsEndOfInflation}. To understand these plots, a `Compton wavenumber' $k_{\!_C}$ corresponding to the effective mass is particularly important:
\Beq
k_{\!_C}\equiv a\bar{\rho}g_{\!_A}/2\,,
\Eeq
where the time dependent quantities on the right-hand side are all evaluated at the time of interest (usually at the end of inflation). The spectra behave differently based on the relative size of $k_{\!_C}$ and the Hubble scale $\mathcal{H}$. When the Compton wavenumber of $\hat{G}^L_{\bk}$ and $\hat{G}^{T\pm}_{\bk}$ is subhorizon, i.e. $k_{\!_C}\gtrsim \mathcal{H}$, we have
\Beq
\label{eq:subhCompton}
\Delta^2_{B^{T\pm}}&\approx\frac{H^4}{4\pi^2}\times\begin{cases}
                     \left(k/\mathcal{H}\right)^{\!4}(k/k_{\!_C})\,, & \text{if } k\ll k_{\!_C}\,,\\
                     \left(k/\mathcal{H}\right)^{\!4}\,,        & \text{if } k\gg k_{\!_C}\,,
                   \end{cases}\\
\Delta^2_{E^{T\pm}}&\approx \frac{H^4}{4\pi^2}\times\begin{cases}
                     \left(k/\mathcal{H}\right)^{\!3}\left(k_{\!_C}/\mathcal{H}\right)\,, & \text{if } k\ll k_{\!_C}\,,\\
                     \left(k/\mathcal{H}\right)^{\!4}\,,        & \text{if } k\gg k_{\!_C}\,,
                   \end{cases}\\
\Delta^2_{E^L}&\approx\frac{H^4}{4\pi^2}\times\begin{cases}
                     \left(k/\mathcal{H}\right)^{\!3}\left(k_{\!_C}/\mathcal{H}\right)\,, & \text{if } k\ll k_{\!_C}\,,\\
                     \left(k/\mathcal{H}\right)^{\!2}\left(k_{\!_C}/\mathcal{H}\right)^{\!2}\,, & \text{if } k\gg k_{\!_C}\,.
                   \end{cases}
\Eeq
As is evident from the above scalings, the magnetic field and both the transverse and longitudinal electric field power spectra have double power-law forms, with $k_{\!_C}$ setting the break in all three of them. This is indeed expected for $\Delta^2_{E^{T\pm}}$, $\Delta^2_{B^{T\pm}}$, since $\hat{G}^{T\pm}_\bk$ cares only about $k_{\!_C}$ (cf. eq. \eqref{eq:GT}). However, $\hat{G}^{L}_\bk$ is affected by the expansion rate as well. From its equation of motion we would expect to see something in $\Delta^2_{E^L}$ near $k_{\!_C}$ and the Hubble scale, $\mathcal{H}$. The reason why there are no features in $\Delta^2_{E^L}$ near $\mathcal{H}$ in the strong coupling regime is given in Appendix~\ref{sec:dS}. 

On the other hand when $k_{\!_C}\lesssim {\mathcal{H}}$, we find that
\Beq
\label{eq:suphCompton}
\Delta^2_{B^{T\pm}}&\approx \frac{H^4}{4\pi^2}\left(\frac{k}{\mathcal{H}}\right)^{\!4}\,, \\
\Delta^2_{E^{T\pm}}&\approx \frac{H^4}{4\pi^2}\times\begin{cases}
                     \left(k/\mathcal{H}\right)^{\!2}\left(k_{\!_C}/\mathcal{H}\right)^{\!2}\,, & \text{if $k\ll k_{\!_C}^2/\mathcal{H}$}\,,\\
                     \left(k/\mathcal{H}\right)^{\!4}\,,        & \text{if $k\gg k_{\!_C}^2/\mathcal{H}$}\,,
                   \end{cases}\\
\Delta^2_{E^L}&\approx\frac{H^4}{4\pi^2}\times\begin{cases}
                     \mathcal{T}_k\left(k/\mathcal{H}\right)^{\!2}\left(k_{\!_C}/\mathcal{H}\right)^{\!2}\,, & \text{if $k\ll \mathcal{H}$}\,,\\
                     \left(k/\mathcal{H}\right)^{\!2}\left(k_{\!_C}/\mathcal{H}\right)^{\!2}\,, & \text{if $k\gg \mathcal{H}$}\,.
                   \end{cases}
\Eeq
In this regime, the transverse modes are still unaffected upon crossing the Hubble horizon. $\Delta^2_{E^{T\pm}}$ is again a double power-law. However, this time the break is at $k=k_{\!_C} (k_{\!_C}/\mathcal{H})$. The additional suppression of $k_{\!_C}/\mathcal{H}$ is unexpected, at least if one looks at the equation of motion of $\hat{G}^{T\pm}_\bk$, eq. \eqref{eq:GT}. For the interested reader, we explain this in Appendix~\ref{sec:dS}. Also note that in this regime $\Delta^2_{B^{T\pm}}$ is a single power-law, given by deep subhorizon Minkowskian power spectrum. For $0.1<k_{\!_C}/\mathcal{H}<1$ there is a crossover behaviour from a double power-law of the $k_{\!_C}\gg \mathcal{H}$ regime to the single power-law of $k_{\!_C}\ll \mathcal{H}$. The power spectrum for the longitudinal component of the electric field $\Delta^2_{E^L}$ exhibits a power excess on super-Hubble scales, with the correction factor found numerically to be $1<\mathcal{T}_k<2$ at the end of inflation. On super-Compton scales, $k<k_{\!_C}$, $\mathcal{T}_k\approx2$. We also do not see any features in the range $k_{\!_C}<k<\mathcal{H}$ in $\Delta^2_{E^L}$ because the {\it k}-dependent pre-factor multiplying $\partial_\tau \hat{G}^L_\bk$ in the definition of $\hat{E}^L_\bk$, eq. \eqref{eq:EGBG}, cancels the additional {\it k}-dependence. Note that $\mathcal{T}_k$ is time dependent and $\mathcal{T}_k\rightarrow 1$ in de Sitter space-time (see Appendix~\ref{sec:dS} for the evaluation of the power spectra in de Sitter space-time). The deviation of $\mathcal{T}_k$ from $1$ is observed towards the end of inflation. It is thus important to solve for the mode functions in a background that deviates from de Sitter in order to capture this effect and obtain the correct power spectra at the end of inflation.

We also give the power spectra calculated under the assumption that $H=0$ and $\bar{\rho}=\text{const}$ (hence $k_{\!_C}=\text{const}$), cf. dashed lines in Fig.~\ref{fig:PsEndOfInflation}. These are `quasi-Minkowski' conditions in the sense that the effects of expansion are ignored. We can then calculate the evolution of the mode functions and obtain the following expressions for the power for electric and magnetic fields:
\Beq
\label{eq:MinkowskiPs}
\Delta_{E^{T\pm}}^2=\frac{H^4}{4\pi^2}\left(\frac{k}{\mathcal{H}}\right)^{\!\!4}\left(1+\frac{k_{\!_C}^2}{k^2}\right)^{\!1/2}\,,&\qquad
\Delta_{B^{T\pm}}^2=\frac{H^4}{4\pi^2}\left(\frac{k}{\mathcal{H}}\right)^{\!\!4}\left(1+\frac{k_{\!_C}^2}{k^2}\right)^{\!-1/2}\,,\\
\Delta_{E^L}^2=\frac{H^4}{4\pi^2}\left(\frac{k}{\mathcal{H}}\right)^{\!\!2}&\left(\frac{k_{\!_C}}{\mathcal{H}}\right)^{\!\!2}\left(1+\frac{k_{\!_C}^2}{k^2}\right)^{\!-1/2}\,.
\Eeq
When $k_{\!_C}\ll\mathcal{H}$ expansion is important for superhorizon modes $k<\mathcal{H}$. However, if $k_{\!_C}\geq\mathcal{H}$ the effects due to expansion are negligible and eq. \eqref{eq:MinkowskiPs} is a good approximation to the power spectrum in the electric and magnetic fields.


\section{Preheating dynamics}
\label{sec:preheating}

At the end of inflation the inflaton field begins to oscillate around the minimum of its potential. The linearised equations describing the evolution of the fluctuations in the inflaton and gauge fields will then contain oscillating terms. Such oscillating terms can lead to an exponential growth of matter fluctuations. This {\it preheating} period begins soon after the end of inflation and its end is marked by the back-reaction of the fluctuations on the inflaton condensate and/or the moment when the linearised analysis of the fluctuations stops being a good approximation \cite{Kofman1997}. 

\begin{figure*}[h] 
   \centering
   \includegraphics[width=6.4in]{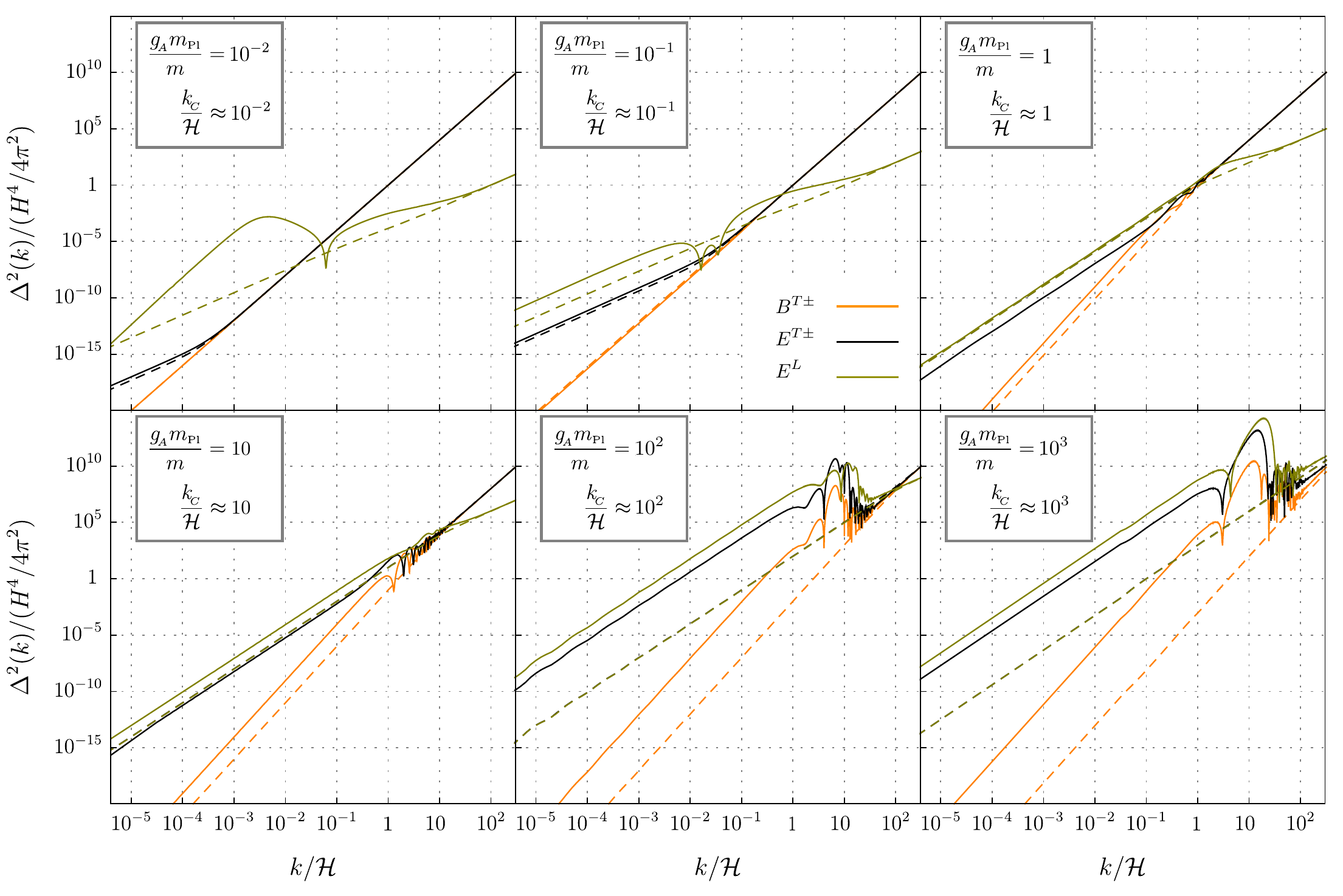}
   \caption{The power spectra of the transverse and longitudinal electric and magnetic fields {\it after} a few inflaton oscillations at the end of inflation: the longitudinal electric field, $\Delta_{E^{L}}^2$ (green), the transverse electric field, $\Delta_{E^{T\pm}}^2$ ({black}), and the magnetic field, $\Delta_{B^{T\pm}}^2$ (orange).  Each plot was evaluated for a different coupling $g_{\!_A}$, and the result is presented at a fixed time $t_{\rm br}\approx 10^2\,m^{-1}$ after inflation. $t_{\rm br}$ is the time of back-reaction for the specific coupling $g_{\!_A}\mpl/m=10^{3}$ (lower right corner). The effects of non-adiabatic particle production during preheating is clearly visible in the lower row (large coupling) for $k/\mathcal{H}\lesssim \sqrt{m/H} \sqrt{k_{\!_C}/\mathcal{H}}\sim\sqrt{k_{\!_C}/\mathcal{H}}$. The boundary between small and large coupling $g_{\!_A}\mpl/m\approx 1$, can be understood via a Floquet analysis of the instabilities. A requirement of {\it broad resonance} \cite{Kofman1997}, leads to the upper bound on $k$. The features in the top row (small couplings) can be also understood in terms of the inflaton oscillations. We have rescaled $\Delta_{E^{T\pm}}^2$ and $\Delta_{B^{T\pm}}^2$ by $a^4$, and rescaled $\Delta_{E^{L}}^2$ by $a^5$ (with $a=1$ at the end of inflation) to roughly separate the effects of resonant particle production from the expected red-shifting of the fields. We also plot the power spectra before resonant particle production begins, i.e. right at the end of inflation ({dashed lines}) for comparison. For ease of comparison with Fig. \ref{fig:PsEndOfInflation}, $H$ (and its conformal counterpart $\mathcal{H}$) at the end of inflation was used for constructing the relevant dimensionless ratios on the horizontal and vertical axes. Note that $\mathcal{H}_{\rm br}\approx 0.25 \mathcal{H}$, that is $k/\mathcal{H}_{\rm br}\approx 4 k/\mathcal{H}$, so the main features in the lower row are all significantly subhorizon at the time of evaluation.}
   \label{fig:Ps}
\end{figure*}

The power spectra of electric and magnetic field at the end of the preheating era are shown in Fig.~\ref{fig:Ps}. These spectra are calculated numerically (for quadratic inflation) and properly account for the quantum nature of the fields, effects of expansion and metric perturbations. A detailed understanding of these spectra is the main goal of this section. As we will see, calculating these spectra using gauge invariant variables (and {\it Unitary gauge} variables) becomes somewhat unwieldy (in particular for the longitudinal components of the gauge fields). The {\it Coulomb gauge} turns out to be well suited for this calculation. 

In Section \ref{sec:Floquet}, we will use Floquet theory to explore the instabilities in matter perturbations using gauge invariant variables and gauge dependent variables. In Section \ref{sec:backreaction}, a Hartree approximation is used to understand the effects of back-reaction on the homogeneous condensate and quantify a time when preheating ends.


\subsection{Floquet analysis}
\label{sec:Floquet}
We ignore metric perturbations for the Floquet analysis of the instabilities in the matter perturbations. This is a plausible approximation since the vector and tensor metric perturbations are decoupled from matter anyway, while the scalar metric perturbations are suppressed on subhorizon scales. We also ignore expansion at the background level since the universe does not expand much during the short period of preheating. For notational simplicity we drop quantum operators from our expressions while carrying out Floquet analysis.\footnote{More explicitly, we could have done the calculation with the mode functions with operators coming along for the ride rather than notationally using classical Fourier modes of the fields. However, this gets rather cumbersome.} The linearized equations of motion for the Fourier modes of the matter fields have the general form (cf. Section~\ref{sec:QuantPerturb}, but now ignoring expansion, quantum operators and metric perturbations)
\Beq
\label{eq:fIEoM}
\partial_\tau^2f^I_\bk+\left(\partial_\tau \ln b_I\right)\partial_\tau f^I_\bk+\omega^2_If^I_\bk=0\,,
\Eeq
where $\omega_I(k,\tau)$ and $b_I(k,\tau)$ are periodic because of the oscillating inflaton. Floquet theory tells us that eq.~(\ref{eq:fIEoM}) has solutions of the form
\Beq
\label{eq:fISol}
f^I_\bk(\tau)=\mathcal{P}^I_{\bk+}(\tau)\exp(\mu^I_k\tau)+\mathcal{P}^I_{\bk-}(\tau)\exp(-\mu^I_k\tau)\,,
\Eeq
where $\mathcal{P}^{I}_{\bk\pm}(\tau)$ are periodic functions having the same period $T$ as the inflaton, and $\mu_{k}^I$ is the Floquet exponent. The Floquet exponents and the periodic functions can be calculated by solving the equation of motion for $f^I_\bk$ twice in the interval $\tau_0\le \tau\le\tau_0+T$  with $\left(f^{I(1)}_\bk,\partial_\tau f^{I(1)}_\bk\right)=(1,0)$ and $\left(f^{I(2)}_\bk,\partial_\tau f^{I(2)}_\bk\right)=(0,1)$ as initial conditions. The real part of the Floquet exponents is given by (cf. e.g. \cite{Amin2014} for details)
\Beq
\Re[\mu_k]=\frac{1}{T}\ln\left|\frac{1}{2}\left(f^{I(1)}_\bk+\partial_\tau f^{I(2)}_\bk+\sqrt{\left\{f^{I(1)}_\bk-\partial_\tau f^{I(2)}_\bk\right\}^2+4f^{I(2)}_\bk \partial_\tau f^{I(1)}_\bk} \right)\right|
\Eeq
where $f^{I(i)}_\bk$ and $\partial_\tau f^{I(i)}_\bk$ are evaluated at $\tau=\tau_0+T$. The effects of the more general initial conditions (for example, those from the end of inflation) can then be incorporated by appropriately scaling the periodic solutions.  

\subsubsection{Gauge invariant analysis}
\begin{figure*}[t] 
   \centering
   \includegraphics[width=2.42in]{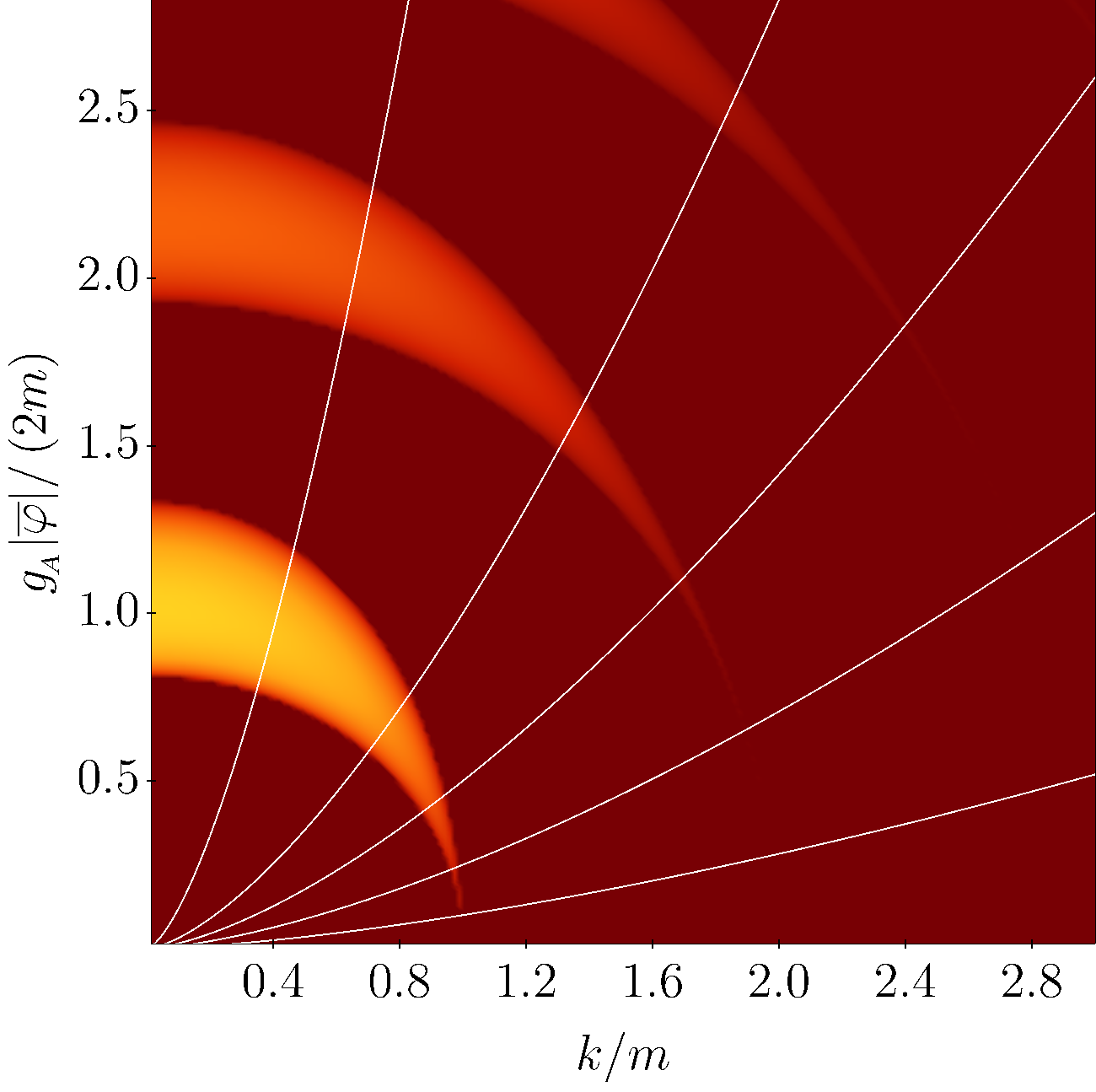} 
     \raisebox{0.058\height}{\includegraphics[width=0.64in]{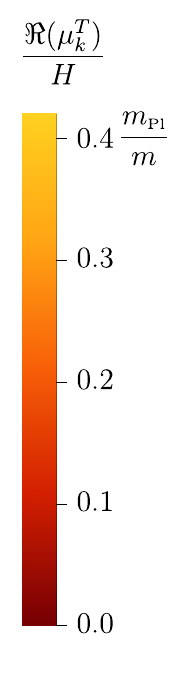}}
   \includegraphics[width=2.42in]{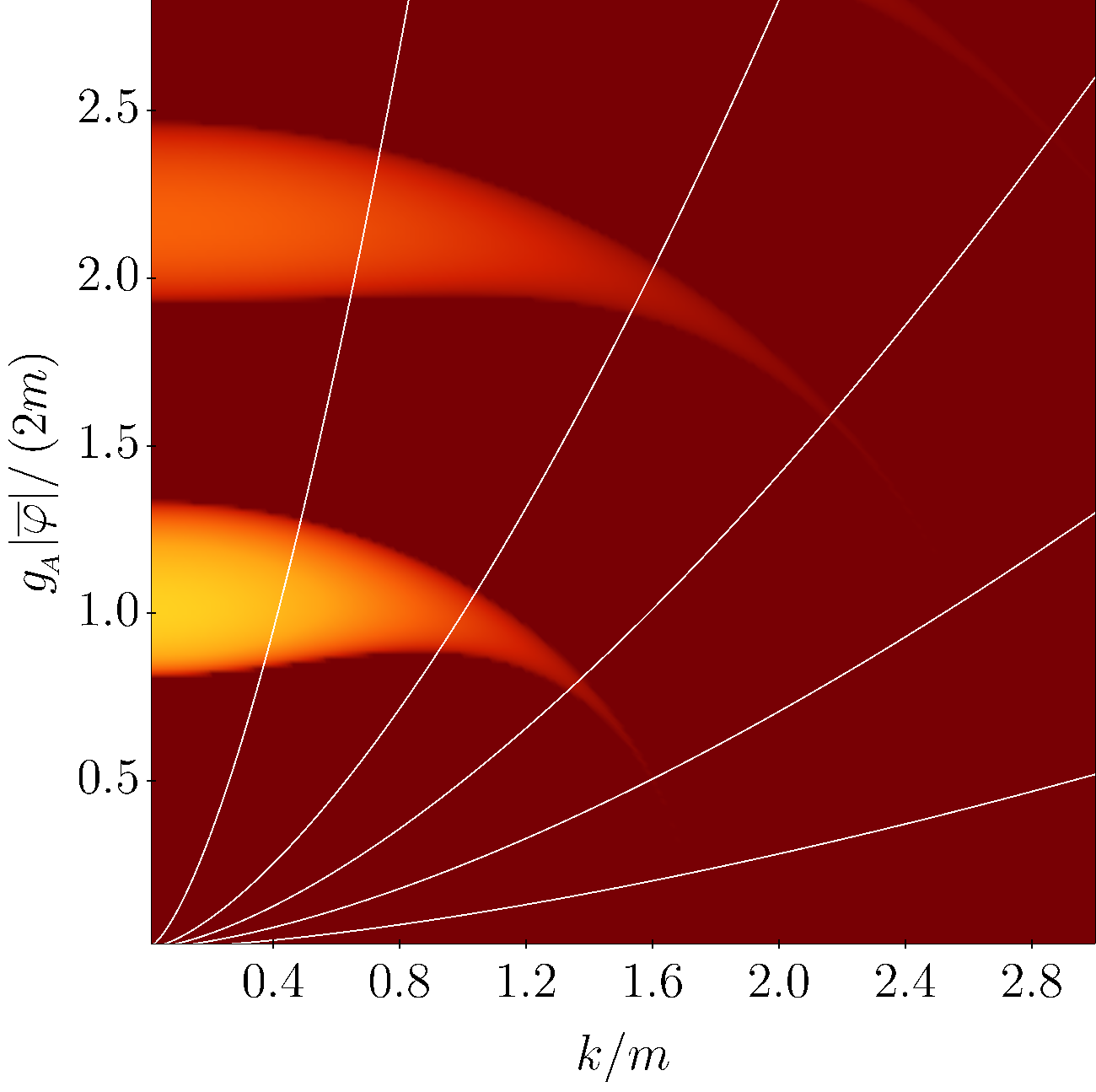} 
     \raisebox{0.058\height}{\includegraphics[width=0.64in]{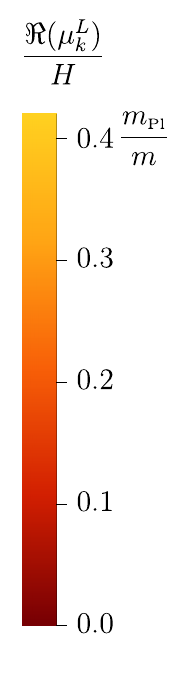}}
   \caption{The above figures show Floquet exponents $\mu_k$ of the transverse (left) and longitudinal (right) modes of the gauge fields during preheating as a function of the physical wavenumber $k$ and the scaled amplitude of the inflaton field oscillations. The gauge fields are coupled to an inflaton with a coupling strength $g_{\!_A}$, and the inflaton potential is $m^2|\varphi|^2$.  Note that there is no qualitative difference between the longitudinal and transverse modes. As the universe expands, a rough estimate of the amount of particle production in a given Fourier mode can be made by flowing across this plot towards the origin along the white lines (momentum $k$ on the horizontal axis redshifts as $a^{-1}$ and the inflaton field amplitude decays on the vertical axis as $\sim a^{-3/2}$). Note that the ``local" Floquet exponent $\mu_k$ is large compared to the instantaneous expansion rate $H$: $\Re[\mu_k]/H \propto \mpl/m$ in the light colored bands,  with $\mpl/m\sim10^{6}$ from CMB observations. However, whether a given Fourier mode passes through these bands depends on $g_{\!_A}$. To see this note that at the beginning of preheating $|\bar{\vp}|\approx 0.1\mpl$, hence $g_{\!_A}|\bar{\varphi}|/m \lesssim  10^{5} g_{\!_A}$. Hence, for $g_{\!_A}\lesssim 10^{-5}$, most of the low momentum modes start ``below" the lowest resonance bands and never experience significant growth. For $g_{\!_A}\gg 10^{-5}$ the unstable momentum modes pass through many Floquet bands as the universe expands and this can lead to significant particle production.}
   \label{fig:2dFloq_m2Phi2}
\end{figure*}

For the gauge invariant variables, the specific forms of the coefficients $b_I$ and $\omega_I$ in eq.~\eqref{eq:fISol} are quite simple in flat spacetime. For the inflaton perturbations $f^\rho_\bk =\delta\rho_\bk,$ (cf. eq.~\eqref{eq:frho}, with $a=1$, $\mpl^2\rightarrow \infty$)
\Beq
b_\rho=1\,,\qquad \omega_\rho^2=k^2+\partial^2_{\bar{\rho}} V(\bar{\rho})\,.
\Eeq 
Similarly, for the transverse components of the gauge fields $f^{T\pm}_\bk =G^{T\pm}_\bk,$
\Beq
\label{eq:omegaT}
b_{T}=1\,,\qquad\omega_{T}^2=k^2+\left(\frac{g_{\!_A}\bar{\rho}}{2}\right)^{\!\!2}\,.
\Eeq
In both of these cases, one can calculate the respective Floquet exponents $\mu^\rho_k$ and $\mu^{T\pm}_k$ once $\bar{\rho}(\tau)$ is obtained from the background dynamics using the usual techniques.\footnote{From the background equation of motion for $\bar{\rho}$ near the minimum of $V(\bar{\rho})$, eq.~\eqref{eq:HomEvol}, one might infer that $\bar{\rho}$ is oscillatory with positive and negative values. This seems apparently at odds with the positive definiteness of $\bar{\rho}$ as implied by its definition as a ``radial" variable. This is of course a minor inconvenience due to our choice of ``radial" gauge invariant variables and can be understood in terms of a discontinuous jump of the angular variable at the origin which we have ignored. There is the other issue of the definition $G_\mu$ when $\bar{\rho}=0$, cf. eq. \eqref{eq:defG}, though it remains well defined away from this point.} For a chaotic inflation potential $V(\rho)=(1/2)m^2\rho^2$, $\Re[{\mu^\rho_k}]=0$ whereas $\Re[\mu^{T}_k]$ is shown in Fig. \ref{fig:2dFloq_m2Phi2}. 
\\ \\
\noindent {\it Longitudinal mode}: The analysis of the longitudinal mode of the gauge field $f^L_\bk=G^L_\bk$ is a bit more subtle. For this case
\Beq
\label{eq:dGLFloq}
b_{L}(k,\tau)=\left[1+\left(\frac{2 k}{\bar{\rho}g_{\!_A}}\right)^{\!\!2}\right]^{-1}\,,\qquad \omega_L^2(k,\tau)=k^2+\left(\frac{g_{\!_A}\bar{\rho}}{2}\right)^{\!\!2}\,.
\Eeq
To deal with the singularity, $\bar{\rho}=0$, in the coefficient $\partial_\tau \ln b_L$  of $\partial_\tau G^L_\bk$ (the ``damping" term) appearing in eq.~\eqref{eq:fIEoM}, we make the following change of variables
\Beq
\label{eq:dGLNewvf}
            G^L_{\bk}(\tau)=\check{h}(\bar{\rho}(\tau))\check{G}^L_{\bk}(\tau)\, \qquad \text{where} \qquad \check{h}(\bar{\rho}(\tau))=\sqrt{1+\left(\frac{2k}{g_{\!_A}\bar{\rho}(\tau)}\right)^{\!\!2}}\,.
\Eeq
Thereby, we lose the singular damping term in eq.~(\ref{eq:fIEoM}) for $f^L_\bk=\check{G}^L_\bk$, at the expense of the singularity in $\check{h}(\bar{\rho}(\tau))$. The explict form of the coefficients after the transformation are
\Beq
\label{eq:vdGL}
\check{b}_L(k,\tau)=1\,,\qquad \check{\omega}_L^2(k,\tau)=\frac{3\left(\frac{g_{\!_A}\partial_{\tau}\bar{\rho}}{2 k}\right)^{\!\!2}}{\left[1+\left(\frac{g_{\!_A} \bar{\rho}}{2 k}\right)^{\!\!2}\right]^{2}}-\frac{\frac{\partial_{\tau}^2\bar{\rho}}{\bar{\rho}}}{1+\left(\frac{g_{\!_A} \bar{\rho}}{2k}\right)^{\!\!2}}+k^2+\left(\frac{g_{\!_A} \bar{\rho}}{2}\right)^{\!\!2}\,.
\Eeq
We note that the second term in $\check{\omega}^2_L$, proportional to $\partial_{\tau}^2\bar{\rho}(\tau)/\bar{\rho}(\tau)$, is never singular provided $\bar{\rho}^{-1}\partial_{\bar{\rho}} V(\bar{\rho})$ is well-defined at the origin $\bar{\rho}=0$\footnote{Recall that $\partial_{\tau}^2\bar{\rho}/\bar{\rho}=-\bar{\rho}^{-1}\partial_{\bar{\rho}} V(\bar{\rho})$.}${}^{,}$\footnote{{If we are in a broken-symmetry, static state with $\partial_{\bar{\rho}} V(\bar{\rho})|_{\bar{\rho}\neq 0}=0$ and $\partial_{\tau}\bar{\rho}=0$, the effective equations of motion $G^L_{\bk}$ (seen more easily with $\check{G}^L_{\bk}$) and $G^{T\pm}_{\bk}$ are equal. Equal effective masses and similar behaviors for $G^L_{\bk}$ and $G^{T\pm}_{\bk}$ were assumed for example in~\cite{HiggsPreheat,Bezrukov2008}. However, since during preheating $\partial_{\bar{\rho}} V(\bar{\rho})|_{\bar{\rho}\neq 0}\ne0$ and $\partial_{\tau}\bar{\rho}\ne0$ and hence $G^L_{\bk}$ and $G^{T\pm}_{\bk}$ have different effective masses (although the masses still remain comparable).}}, which is usually the case. Hence the solutions
\Beq
\label{eq:checkdGLFloqSol}
\check{G}^L_{\bk}(\tau)=\check{\mathcal{P}}_{\bk+}^{L}(\tau)\exp\left(\mu_{k}^{L}\tau\right)+\check{\mathcal{P}}_{\bk-}^{L}(\tau)\exp\left(-\mu_{k}^{L}\tau\right)\,,
\Eeq
are well behaved with no singularities present in the Floquet exponents or in the periodic functions. Reverting back to the original variables, cf. eq.~\eqref{eq:dGLNewvf}, we have
\Beq
\label{eq:dGLFloqSol}
               G^L_{\bk}(\tau)&=\mathcal{P}_{\bk+}^{L}(\tau)\exp\left(\mu_{k}^{L}\tau\right)+\mathcal{P}_{\bk-}^{L}(\tau)\exp\left(-\mu_{k}^{L}\tau\right)\,,\\
\mathcal{P}_{\bk\pm}^{L}(\tau)&=\check{h}(\bar{\rho}(\tau))\check{\mathcal{P}}_{\bk\pm}^{L}(\tau)\,.
\Eeq
The Floquet exponents, $\mu_{k}^L$, remain unaffected and singularity-free and are shown in Fig. \ref{fig:2dFloq_m2Phi2}. Only the periodic functions, $\mathcal{P}_{\bk\pm}^{L}(\tau)$, are singular (or have `spikes') at the zeroes of $\bar{\rho}$, because of the $\check{h}(\bar{\rho}(\tau))$ factor. Importantly, the physical observables: the longitudinal electric field $E^L_\bk$ and the charge and current densities, $j_{0\bk}$ and $j^L_\bk$, are immune to the singularities because of the $\bar{\rho}$-dependent pre-factors in their definitions cancel appropriately (cf. eqns.~\eqref{eq:EGBG} and~\eqref{eq:ChargeAndCurrent}). 

We note that an identical analysis holds for the {\it Unitary gauge} ($\Im[{\varphi}]=0$) as well. In that gauge, the equation of motion for the $A^L_\bk$ has $b_L$ and $\omega_L$ given in eq.~\eqref{eq:dGLFloq} with $\bar{\rho}\rightarrow \Re[{\bar{\varphi}}]$ (cf. Section \ref{sec:GaugeTransf}). Hence $A^L_\bk$ grows exponentially as well with periodic spikes occurring whenever $\Re[{\bar{\varphi}}]=0$.  This still leads to {\it spike-free} exponential growth of the longitudinal electric field and charge and current densities. We also confirm this result below by carrying out the calculation in the Coulumb gauge where no singularities are present in the equations of motion. Our conclusion is different from that of~\cite{Kari}, where the authors argued that the coefficient, $\partial_\tau \ln b_L$, of the $\partial_\tau A^L_\bk$ would drive  $A^L_{\bk}(\tau)\rightarrow 0 $ due to its large amplitude near $\Re[{\bar{\varphi}}(\tau)]=0$. We also note that the authors in \cite{Dimopoulos:2001wx} argued that there should be strong resonance in the longitudinal modes (stronger than the transverse modes), though they did not pursue this issue in detail. We find that the resonance in both longitudinal and transverse modes is comparable.

\subsubsection{Coulomb gauge analysis}
One might be troubled by the presence of singularities in the intermediate steps of the gauge invariant analysis, especially if one wishes to carry out numerical calculations. In the Coulumb gauge the calculation of the physical observables $E^L_{\bk}$, $j^0_{\bk}$ and $j^L_{\bk}$, is `clean' throughout, i.e. no singularities arise whatsoever in the intermediate steps.

For more details on the well defined field content and the equations of motion in the Coulomb gauge see Section \ref{sec:GaugeTransf} and Appendix~\ref{sec:LSCG}.

The equations of motion for perturbations $\delta\vp^0_\bk$ and $A^{T\pm}_\bk$ in the Coulumb gauge are identical to the ones for $\delta\rho_\bk$ and $G^T_\bk$ respectively in the gauge invariant scenario (with $\bar{\rho}\rightarrow \bar{\varphi}^0$). For a chaotic inflation potential $V(\bar{\varphi}^0)=(1/2)m^2(\bar{\varphi}^0)^2$, $\Re[{\mu^0_k}]=0$ whereas $\Re[\mu^{T\pm}_k]$ is shown in Fig. \ref{fig:2dFloq_m2Phi2} and are identical to the related Floquet exponents in the gauge invariant case. 

In the Coulumb gauge we need to analyse $\delta\varphi^1_\bk$ instead of the longitudinal mode $A^L_\bk$. For $f^1_\bk=\delta\vp^1_\bk$ in eq.~\eqref{eq:fISol}, the coefficients are 
\Beq
b_1(k,\tau)=\left[1+\left(\frac{g_{\!_A}\bar{\vp}^0}{2k}\right)^{\!\!2}\right]^{-1}\,,\qquad\omega_1^2(k,\tau)=-\frac{\partial_{\tau}^2\bar{\vp}^0}{\bar{\vp}^0}+\frac{2\left(\frac{g_{\!_A}\partial_{\tau}\bar{\vp}^0}{2k}\right)^{\!\!2}}{1+\left(\frac{g_{\!_A}\bar{\vp}^0}{2k}\right)^{\!\!2}}+k^2+\left(\frac{g_{\!_A}\bar{\vp}^0}{2}\right)^{\!\!2}\,.
\Eeq
Both coefficients are non-singular (assuming $\left(\bar{\vp}^0\right)^{-1}\partial_{\bar{\vp}^0} V(\bar{\vp}^0)$ is well-defined at the origin $\bar{\vp}^0=0$ as is usually the case). We can calculate the Floquet exponents based on this equation, however, we shall change variables and  remove the damping term for reasons that will become clear below. We make the non-singular transformation
\Beq
\label{eq:deltaphi2check}
\delta \vp_{\bk}^1(\tau)=\check{l}(\bar{\vp}^0(\tau))\delta \check{\vp}_{\bk}^1(\tau)\,\qquad {\textrm{where}}\qquad\check{l}(\bar{\vp}^0(\tau))\equiv\sqrt{1+\left(\frac{g_{\!_A}\bar{\vp}^0(\tau)}{2k}\right)^{\!\!2}}\,,
\Eeq
which yields $\check{b}_1=\check{b}_L$ and $\check{\omega}_1=\check{\omega}_L$  with $\bar{\rho}\rightarrow \bar{\varphi}^0$ (cf. eq.~\eqref{eq:vdGL}). This identification immediately yields $\delta\check{\vp}^1_\bk=\check{G}^L_\bk$ where $\check{G}^L_\bk$ was provided explicitly in eq.~\eqref{eq:checkdGLFloqSol}. Finally undoing our transformation of variables in eq. \eqref{eq:deltaphi2check}, we have the solution
\Beq
\delta {\vp}_{\bk}^1(\tau)={\mathcal{P}}_{\bk+}^{1}(\tau)\exp\left(\mu_{k}^{L}\tau\right)+{\mathcal{P}}_{\bk-}^{1}(\tau)\exp\left(-\mu_{k}^{L}\tau\right)\,,
\Eeq
with $\mathcal{P}_{\bk\pm}^{1}(\tau)=\check{l}(\bar{\vp}_0(\tau))\check{\mathcal{P}}_{\bk\pm}^{L}(\tau)$. The Floquet exponent $\mu^L_\bk$ is identical to the one in eq.~\eqref{eq:checkdGLFloqSol}. A numerical computation of this exponent assuming a chaotic inflationary potential is shown in Fig.~\ref{fig:2dFloq_m2Phi2}.  Also note that now the periodic functions, $\mathcal{P}_{\bk\pm}^{1}(\tau)$, have no troublesome spikes since $\check{l}$ is non-singular everywhere. This is an improvement over the gauge invariant analysis. In that case,  $G^L_\bk=\check{h}\check{G}^L_\bk$, had singular spikes because $\check{h}$ was singular.

Hence, the Coulomb gauge allows us to calculate $\delta \vp_{\bk}^1(\tau)$, in a safe, singularity-free way. It is well suited for numerical computations. There is no problem with calculating physical variables such as $E^L_\bk$, $B^{T\pm}_\bk$ and $E^{T\pm}_\bk$. For example the longitudinal electric field (cf. Appendix~\ref{sec:LSCG})
\Beq
E^L_{\bk}=\frac{g_{\!_A} k}{2}\frac{\left[\delta \vp_{\bk}^1\partial_{\tau}\bar{\vp}^0-\bar{\vp}^0\partial_{\tau}\delta \vp_{\bk}^1\right]}{k^2+\left(\frac{g_{\!_A}\bar{\vp}^0}{2}\right)^{\!\!2}}\,,
\Eeq
has no singularities in it.

Before moving on to back-reaction, let us revisit the features seen in Fig. \ref{fig:Ps} in light of what we now understand from our Floquet analysis. The resonance structure of the transverse mode, eq. (\ref{eq:omegaT}), is identical to that of $\chi$ in the popular $\tilde{g}^2\varphi^2\chi^2$ toy model \cite{Finelli:2000sh}. Making use of well-known results about this model (see for example \cite{Mukhanov1}), the parametric excitation of transverse modes is expected to be most efficient in the broad resonance regime $g_{\!_A}|\bar{\varphi}|/m\sim g_{\!_A}\mpl/m\gg1$ for modes in the range $k\lesssim m \sqrt{g_{\!_A}|\bar{\varphi}|/m}\sim m\sqrt{g_{\!_A}\mpl/m}$. This should be true for the longitudinal mode as well based on the Floquet plots shown in Fig. \ref{fig:2dFloq_m2Phi2}. The lower three plots in Fig. \ref{fig:Ps} show how for $g_{\!_A}\mpl/m\gg1$, the transverse and longitudinal modes in the range $k/\mathcal{H}\lesssim \sqrt{m/H} \sqrt{k_{\!_C}/\mathcal{H}}\sim\sqrt{k_{\!_C}/\mathcal{H}}$ are significantly enhanced, in agreement with the analysis of this section (once expansion is appropriately included). 

\subsection{Back-reaction and end of preheating}
\label{sec:backreaction}
So far we have extensively relied on a linear analysis of perturbations. As we have seen, during preheating, the covariant derivative coupling between the gauge fields and the inflaton causes resonant Fourier modes of electric and magnetic fields to grow exponentially fast. This growth cannot proceed forever. The resonance is eventually shut-off by the gauge field's back-reaction on the oscillating inflaton condensate, which ultimately leads to the condensate's fragmentation and our linear analysis stops holding any more.

To estimate the time of back-reaction, we shall investigate the effective equation of motion of the inflaton condensate in the Hartree approximation. We shall work in the non-singular Coulomb gauge, and include the background expansion as well (but ignore metric perturbations). Assuming that the slow-roll inflation happens along the real $\bar{\vp}^0$ axis, the effective condensate equation in the Hartree approximation becomes 
\Beq
\label{eq:EffBack}
&\partial_{\tau}^2\bar{\vp}^{0}+2\mathcal{H}\partial_{\tau}\bar{\vp}^{0}+a^2\left[m^2-\left(\frac{g_{\!_A}}{2}\right)^{\!\!2}\left<0\right|\hat{A}_{\mu}\hat{A}^{\mu}\left|0\right>\right]\bar{\vp}^0=\\
&\qquad \qquad \qquad \qquad a^2\frac{g_{\!_A}}{2}\Re\Big[2\left<0\right|\partial_{\mu}\delta\hat{{\vp}}^{1}\hat{A}^{\mu}\left|0\right>+\left<0\right|\delta \hat{{\vp}}^{1}\partial^{\mu}\hat{A}_{\mu}\left|0\right>+2\mathcal{H}\left<0\right|\delta \hat{{\vp}}^{1}\hat{A}^{0}\left|0\right>\Big]\,.
\Eeq
During preheating the occupation numbers become much greater than one, and the fields are expected to approach the classical limit. In this limit, the commutators of non-commuting variables vanish and the ambiguity regarding operator ordering is not significant. Moreover, note that the expectation values of non-commuting operators above can be complex. However, taking the real par is a reasonable approximation since in the classical limit the imaginary part becomes negligible compared to the real one (we have verified this numerically as well). 

\begin{figure*}[t] 
   \centering
   \includegraphics[width=2.68in]{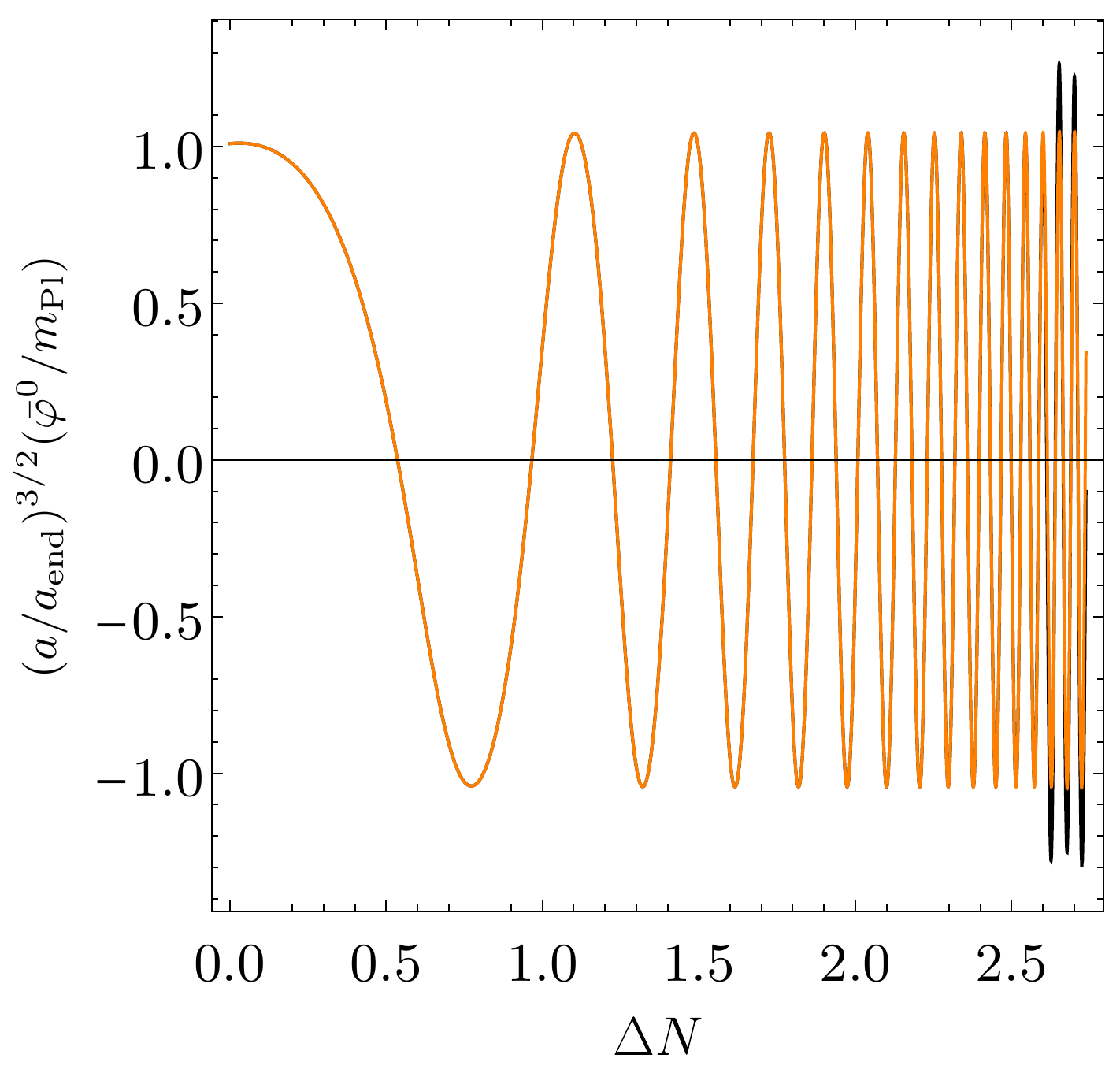} \hfill
   \includegraphics[width=2.5in]{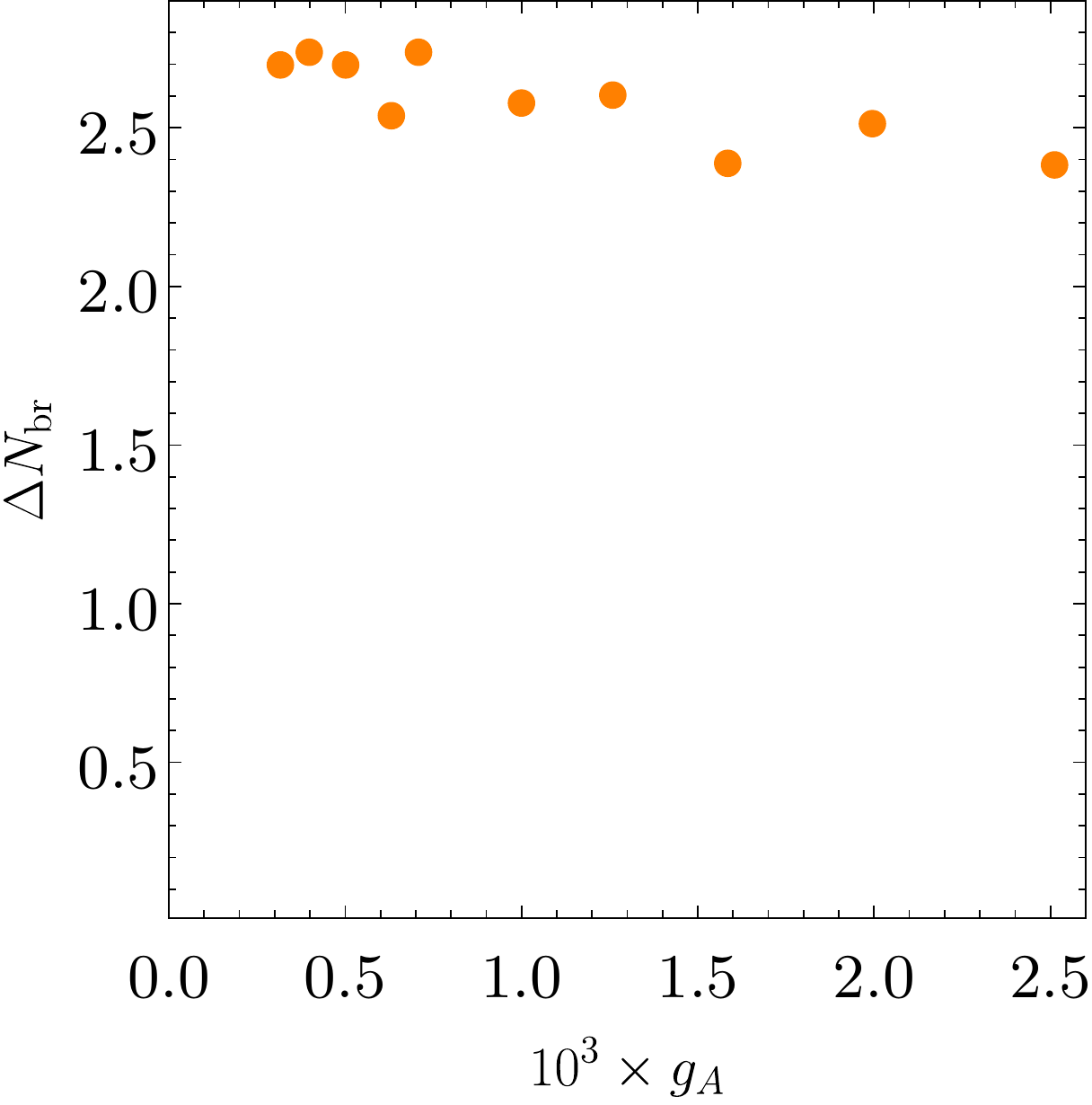} 
\hspace*{\fill} 
  \caption{Left: The evolution of the homogeneous inflaton field as a function of number of e-folds after the end of inflation with no back-reaction (orange) and with back-reaction taken into account (black) for $g_{\!_A}\approx1.26\times10^{-3}$. Right: The number of e-folds after the end of inflation for the back-reaction to become important as a function of $g_{\!_A}$.} 
   \label{fig:BackReactionCriterion}
\end{figure*}

Our next step is to derive expressions for the expectation values in eq.~\eqref{eq:EffBack} in terms of mode functions for the different fields. In the Coulomb gauge, and in Fourier space, the temporal component of the gauge field can be written in terms of $\delta\varphi^1_\bk$ (see Appendix ~\ref{sec:LSCG}):
\Beq
\hat{A}_{0\bk}=\frac{g_{\!_A}}{2}\frac{\left[\delta \hat{\vp}_{\bk}^1\partial_{\tau}\bar{\vp}^{0}-\bar{\vp}^0\partial_{\tau}\delta\hat{\vp}_{\bk}^1\right]}{\left(\frac{k}{a}\right)^{\!\!2}+\left(\frac{g_{\!_A} \bar{\vp}^0}{2}\right)^{\!\!2}}\,,
\Eeq
and its first derivative, after taking into account the equations of motion, conveniently reduces to
\Beq
\partial_{\tau}\hat{A}_{0\bk}=a^2\frac{g_{\!_A}}{2}\bar{\vp}^0\delta \hat{\vp}_{\bk}^1\,.
\Eeq
These then translate to the mode functions, $u_{k}^1(\tau)$ and $u_{k}^{A_0}(\tau)$, of the quantised fluctuations
\Beq
\delta \hat{\vp}^1(\boldsymbol{x},\tau)=\int \textbf{d}^3\bk e^{i\boldsymbol{k.x}} \delta \hat{\vp}_{\bk}^1(\tau)=\int \textbf{d}^3\bk e^{i\boldsymbol{k.x}}\left[\hat{a}_{\bk}^Lu_{k}^1(\tau)+\hat{a}_{-\bk}^{L\dagger}u_{k}^{1*}(\tau)\right]\,,
\Eeq
\Beq
\hat{A}_0(\boldsymbol{x},\tau)=\int \textbf{d}^3\bk e^{i\boldsymbol{k.x}} \hat{A}_{0\bk}(\tau)=\int \textbf{d}^3\bk e^{i\boldsymbol{k.x}}\Big[\hat{a}_{\bk}^L u_{k}^{A_0}(\tau)+\hat{a}_{-\bk}^{L\dagger}u_{k}^{A_0*}(\tau)\Big]\,.
\Eeq
This immediately yields the following expectation values:
\Beq
\left<0\right|\hat{A}_{0}\hat{A}^{0}\left|0\right>
&=\int \frac{dk k^2}{2\pi^2}\left(\frac{g_{\!_A}}{2a}\right)^{\!\!2}\left|\frac{\partial_{\tau}\bar{\vp}^{0}u_{k}^1-\bar{\vp}^0\partial_{\tau}u_{k}^1}{\left(\frac{k}{a}\right)^{\!2}+\left(\frac{g_{\!_A}\bar{\vp}^0}{2}\right)^{\!\!2}}\right|^2\,,\\
\left<0\right|\partial_{\mu}\delta \hat{\vp}^{1}\hat{A}^{\mu}\left|0\right>
&=\left<0\right|\partial_{\tau}\delta \hat{\vp}^{1}\hat{A}^{0}_{}\left|0\right>=\int \frac{dk k^2}{2\pi^2}\frac{g_{\!_A}}{2a^2}\frac{\partial_{\tau}u_{k}^1\left[\partial_{\tau}\bar{\vp}^{0}u_{k}^{1*}-\bar{\vp}^0\partial_{\tau}u_{k}^{1*}\right]}{\left(\frac{k}{a}\right)^{\!2}+\left(\frac{g_{\!_A}\bar{\vp}^0}{2}\right)^{\!\!2}}\,,\\
\left<0\right|\delta \hat{\vp}^{1}\partial^{\mu} \hat{A}_{\mu}\left|0\right>
&=\left<0\right|\delta \hat{\vp}^{1}\partial^{\tau} \hat{A}_{0}\left|0\right>=\int \frac{dk k^2}{2\pi^2}\frac{g_{\!_A}}{2}\bar{\vp}^0\left|u_{k}^1(\tau)\right|^2\,,\\
\left<0\right|\delta \hat{\vp}^{1} \hat{A}^{0}\left|0\right>
&=\int \frac{dk k^2}{2\pi^2}\frac{g_{\!_A}}{2a^2}u_{k}^1\frac{\left[\partial_{\tau}\bar{\vp}^{0}u_{k}^{1*}-\bar{\vp}^0\partial_{\tau}u_{k}^{1*}\right]}{\left(\frac{k}{a}\right)^{\!2}+\left(\frac{g_{\!_A}\bar{\vp}^0}{2}\right)^{\!\!2}}\,.
\Eeq
The only term which involves transverse modes is
\Beq
\left<0\right|\hat{A}_{j}\hat{A}^{j}\left|0\right>=-\int \frac{dk k^2}{2\pi^2}a^{-2}\left[|u^{T+}_{k}|^2+|u^{T-}_{k}|^2\right]\,,
\Eeq
defined in the Coulomb gauge as
\Beq
\hat{A}_j(\boldsymbol{x},\tau)=\int \textbf{d}^3\bk \sum_{\lambda=\pm} e^{i\boldsymbol{k.x}} (\boldsymbol{\epsilon}^{T\lambda}_\bk)_j\hat{A}_{\bk}^{T\lambda}(\tau)=\int \textbf{d}^3\bk \sum_{\lambda=\pm} e^{i\boldsymbol{k.x}}(\boldsymbol{\epsilon}^{T\lambda}_\bk)_j\Big[\hat{a}_{\bk}^{T\lambda}u_{k}^{T\lambda}+\hat{a}_{-\bk}^{T\lambda\dagger}u_{k}^{T\lambda*}\Big]\,.
\Eeq
With these expressions, we can now calculate back-reaction effects iteratively. We first evolve the mode functions from vacuum initial conditions in eq.~\eqref{eq:uWKB}, using the background solution $\bar{\varphi}^0$ of  eq.~(\ref{eq:EffBack}) with the correction terms (quadratic in the perturbations) ignored. These then allow us to get the necessary expectation values explicitly. Next, we re-evaluate the background solution, now including the correction terms (quadratic in the perturbations) in eq.~(\ref{eq:EffBack}). The moment when the new background solution deviates from the uncorrected one (see the left panel in Fig.~\ref{fig:BackReactionCriterion}) back-reaction has become important. For large couplings, $g_{\!_A}>10^{-4}$ in the Chaotic inflation model ($V=m^2\rho^2/2$, $m=10^{-6}\mpl$), we get the following expression for the number of e-folds of expansion after the end of inlation when gauge fields back-react on the condensate $\Delta N_{\rm br}\approx1.72 g_{\!_A}^{-0.1}$.  We caution that this scaling is to be taken as a very rough guide in the range of $3\times10^{-4}< g_{\!_A} < 2.5\times 10^{-3}$ (see the right panel in Fig. \ref{fig:BackReactionCriterion}), the true dependence is likely non monotonic. Below the lowest value of $g_{\!_A}$ we have considered, the Hubble expansion quickly redshifts all modes below the lowest resonance band and we see negligible gauge field production.

\section{Initial conditions for lattice simulations}
\label{sec:App}
In this section we give a concise description of initial conditions for numerical lattice studies of reheating. Our linearlized calculations including expansion, metric perturbations and the quantum nature of the fields were sufficient during inflation and preheating. However, {\it after} preheating the dynamics of matter fields on subhorizon scales  can become highly non-linear. Since (i) the matter field occupation numbers grow rapidly during preheating, (ii) predominantly on subhorizon scales, while (iii) the metric perturbations remain small, one can evolve the classical equations of motion of the matter fields numerically after preheating using subhorizon 3+1d lattice simulations including gravity only at the background level. This standard approximation captures all of the relevant physical phenomena during the non-linear stage of reheating.
 
There are two distinct parts to setting up initial conditions on the lattice. First, for making the transition from quantized fluctuations to classical ones in the continuum limit, a prescription is needed. The classical field fluctuations must also satisfy the necessary constraints. The second is discretizing these initialized continuous fields on the lattice. We first focus on getting from quantized fluctuations to classical ones, with an accounting for the constraints. We will move from gauge invariant variables to variables in the temporal gauge (a popular choice of lattice simulations), but the prescription is applicable in any gauge.

Working in the local $U(1)$ gauge invariant variables, the matter fields with the quantized fluctuations at the end of preheating, written in terms of the mode functions in Fourier space (in the continuum limit) are as follows. In the expressions below, the time $\tau$ is the time of specification of the initial conditions on the lattice (this could be the end of inflation or some time before back-reaction).
\begin{flalign*}
&\delta \hat{\rho}(\boldsymbol{x},\tau)=\int \textbf{d}^3\bk e^{i\boldsymbol{k\cdot x}} \delta \hat{\rho}_{\bk}(\tau)=\int \textbf{d}^3\bk e^{i\boldsymbol{k\cdot x}}\Big[\hat{a}_{\bk}^{\rho}u_{k}^{\rho}(\tau)+\hat{a}_{-\bk}^{\rho\dagger}u_{\rho k}^{\rho*}(\tau)\Big]\,,&
\end{flalign*}
\begin{flalign}
\label{eq:GLInit}
&\hat{G}_j^L(\boldsymbol{x},\tau)=\int \textbf{d}^3\bk e^{i\boldsymbol{k\cdot x}} \left(\boldsymbol{\epsilon}_{\bk}^{L}\right)_j\hat{G}_{\bk}^L(\tau)=\int \textbf{d}^3\bk e^{i\boldsymbol{k\cdot x}}\left(\boldsymbol{\epsilon}_{\bk }^{L}\right)_j\Big[\hat{a}_{\bk}^{L}u_{k}^{L}(\tau)+\hat{a}_{-\bk}^{L\dagger}u_{k}^{L*}(\tau)\Big]\,,&
\end{flalign}
\begin{flalign}
\label{eq:GTInit}
\hat{G}_j^T(\boldsymbol{x},\tau)&=\int \textbf{d}^3\bk \sum_{\lambda=\pm} e^{i\boldsymbol{k\cdot x}} \left(\boldsymbol{\epsilon}_{\bk}^{T\lambda}\right)_j\hat{G}_{\bk}^{T\lambda}(\tau)\\
 &=\int \textbf{d}^3\bk \sum_{\lambda=\pm} e^{i\boldsymbol{k\cdot x}} \left(\boldsymbol{\epsilon}_{\bk}^{T\lambda}\right)_j\Big[\hat{a}_{\bk}^{T\lambda}u_{k}^{T\lambda}(\tau)+\hat{a}_{-\bk}^{T\lambda\dagger}u_{k}^{T\lambda*}(\tau)\Big]\,,&
\end{flalign}
\begin{flalign}
\label{eq:G0Init}
\hat{G}_0(\boldsymbol{x},\tau)&=-\int \textbf{d}^3\bk e^{i\boldsymbol{k\cdot x}} \frac{\partial_{\tau}\hat{G}_{\bk}^L(\tau)}{k^2+\left(\frac{\bar{\rho}(\tau)g_{\!_A}a(\tau)}{2}\right)^{\!\!2}}
                                           =-\!\!\int \textbf{d}^3\bk e^{i\boldsymbol{k\cdot x}} \frac{\Big[\hat{a}_{\bk}^L\partial_{\tau}u_{k}^L(\tau)+\hat{a}_{-\bk}^{L\dagger}\partial_{\tau}u_{k}^{L*}(\tau)\Big]}{k^2+\left(\frac{\bar{\rho}(\tau)g_{\!_A}a(\tau)}{2}\right)^{\!\!2}}\,.&
\end{flalign}
We also need time derivatives of the above expressions. In that case, the mode functions $u_{k}^I$ get differentiated; the creation and annihilation operators $(\hat{a}_{\bk}^I,\hat{a}_{\bk}^{I\dagger})$ are time independent.
By switching from the quantum two point function in real space $\langle 0|\hat{f}^I({\bf x})\hat{f}^J({\bf x+r})|0\rangle$ to the classical 2-point correlation functions in real space $\langle f^I({\bf x})f^J({\bf x+r})\rangle =V^{-1}\int d^3{\bf x} f^I({\bf x})f^J({\bf x+r})$ with $V\rightarrow \infty$, we switch from a quantum to a classical description.\footnote{That is, we replace the creation and annihilation operators with stochastic complex numbers (with Gaussian probability distribution) whose covariance matrix is determined by the mode functions. This holds for fields whose mode functions have increased (are occupied) significantly \cite{Polarski1995}. This is not necessarily true for the gauge fields at the end of inflation. However, if gauge field modes are resonantly excited during preheating, the stochastic approach becomes consistent.} We now treat the creation and annihilation operators as complex numbers 
\Beq
\label{eq:aNoHat}
\hat{a}_{\bk}^I\to a_{\bk}^I \qquad\text{and}\qquad \hat{a}_{\bk}^{I\dagger}\to a_{\bk}^{I*}\,,
\Eeq
with phases distributed uniformly on $[0,2\pi)$, and with a Rayleigh distributed amplitudes set in the following manner \cite{Polarski1995}:
\Beq
\label{eq:aRand}
\left|a_{\bk}^I\right|=\sqrt{-\ln\left(X_{\bk}^I\right)} \qquad\text{and}\qquad \text{arg}\left(a_{\bk}^I\right)=2\pi Y_{\bk}^I\,.
\Eeq
Here $X_{\bk}^I$ is a uniform deviate on $(0,1)$ and $Y_{\bk}^I$ is a uniform deviate on $[0,1)$. The amplitude is governed by the requirement of consistency with the quantum 2-point function. The index $I$ denotes the field under consideration, i.e. $I=\left\{\rho,L,T^+,T^-\right\}$. Note that the complex numbers $a_{\bk}^L$ in eq.~(\ref{eq:GLInit}) and eq.~(\ref{eq:G0Init}) are the same. 

So far the analysis have been carried out in a gauge invariant framework. It is not difficult to find the initial conditions for lattice simulations in a particular gauge. For example, in the $A_0=0$ (temporal) gauge - the common gauge choice for numerical simulations, the conversion between gauge invariant variables and gauge dependent fields is given in eqns. \eqref{eq:GaugeTransfrhoAT} and \eqref{eq:GaugeTransfTemporal}. For convenience one can choose $\tau_\text{in}=\tau$ in eq. \eqref{eq:GaugeTransfTemporal}, so that $\delta\vp^1_{\bk}(\tau)=0$ and $A^L_{\bk}(\tau)=G^L_{\bk}$, but $\partial_\tau \delta\vp^1_{\bk}(\tau)=\bar{\rho}g_{\!_A}G_{0\bk}/2$ and 
\Beq
\label{eq:TemporalInitdtAL}
\partial_\tau A^L_{\bk}(\tau)=\partial_\tau G^L_{\bk}+kG_{0\bk}\,.
\Eeq

We note that despite the random nature of the complex numbers in eq.~(\ref{eq:aRand}), the Gaussian constraint on the lattice is automatically satisfied to linear order in the perturbations, since the expression for $G_0(\bx,\tau)$ in eq.~(\ref{eq:G0Init}), in terms of the complex $a_{\bk}^L$, takes care of the first order terms in eq.~(\ref{eq:GEoM}). 

If one wishes the Gauss constraint to be met to machine precision on the lattice, a simple correction has to be made. After initialising $\delta\rho(\bx,\tau)$, $\partial_\tau\delta\rho(\bx,\tau)$, $G_i(\bx,\tau)$, $\partial_\tau G_i(\bx,\tau)$, $G_0(\bx,\tau)$ we define the corrected time derivative 
\Beq
\partial_\tau G_j^{L,\text{corr}}(\bx,\tau)=\int \textbf{d}^3\bk e^{i\boldsymbol{k\cdot x}} \left(\boldsymbol{\epsilon}_{\bk}^{L}\right)_j\partial_\tau G_{\bk}^{L,\text{corr}}(\tau)\,,
\Eeq
where
\Beq
\partial_\tau G_{\bk}^{L,\text{corr}}=-k G_{0\bk}-\frac{\tilde{j}_{0\bk}}{k}\,,\qquad\tilde{j}_{0\bk}=\int \frac{\textbf{d}^3\bx}{\left(2\pi\right)^{3}} e^{-i\boldsymbol{k\cdot x}}(\bar{\rho}+\delta\rho)\bar{\rho}\left(\frac{g_{\!_A}a}{2}\right)^{\!\!2}G_0\,.
\Eeq
Note the $\delta\rho$ appearing on the right-hand side of the last equation. We then use $\partial_\tau G_j^{L,\text{corr}}(\bx,\tau)$ in place of $\partial_\tau G_j^L(\bx,\tau)$ to generate the necessary gauge dependent variable in eq. \eqref{eq:TemporalInitdtAL}.

We remind the reader that all expressions so far in this section are for fields on a continuous space-time manifold. The reason is transparency of the transition from quantum to classical fields accounting for the constraints. For completeness we outline the discrete lattice counterparts to the continuous variables. Let $\Delta$ be the separation between neighbouring lattice points and {\it L} the length of the lattice. Then the following substitutions should be used to define the fields on the lattice
\Beq
\bx&\to\bx_n=\boldsymbol{n}\Delta\,,\\
\bk&\to\bk_n=\frac{2\pi}{L}\boldsymbol{n}\,,\\
\int \textbf{d}^3\bk&\to\left(\frac{2\pi}{L}\right)^3\sum_{\bk_n}\,,
\Eeq
where $\boldsymbol{n}=[n_x,n_y,n_z]$ with $-L/(2\Delta) \leq n_{x,y,z} \leq L/(2\Delta)$. For example
\Beq
\label{eq:drhoInitDisc}
\delta \rho(\bx_n,\tau)=\left(\frac{2\pi}{L}\right)^3\sum_{\bk_n} e^{i\bk_n\cdot\bx_n} \delta \rho_{\bk_n}(\tau)=\left(\frac{2\pi}{L}\right)^3\sum_{\bk_n} e^{i\bk_n\cdot\bx_n}\Big[a_{\bk_n}^\rho u_{k_n}^\rho(\tau)+a_{-\bk_n}^{\rho*}u_{ k_n}^{\rho*}(\tau)\Big]\,.
\Eeq
There is a minor subtlety in the definition of the polarisation vectors. Consider a finite difference lattice code in which the discrete divergence of a vector is given by\footnote{E.g. in the standard Wilsonian approach to lattice gauge theories, gauge fields live on {\it space-time} links, $G_\mu(x_\mu+\hat{\mu}\frac{\Delta}{2})$ and scalar fields live on the nodes of the space-time lattice, $\rho(x^\mu)$. Here $\hat{\mu}$ is the space-time unit vector and $\Delta$ is the separation between two neighbouring points on the space-time lattice.}
\Beq
\label{eq:DiscreteDiv}
\left(\boldsymbol{\nabla}\cdot\textbf{G}\right)(\bx,\tau)\to\frac{1}{\Delta}\sum_{i=x,y,z} G_i(\bx+\hat{i}\frac{\Delta}{2},\tau)-G_i(\bx-\hat{i}\frac{\Delta}{2},\tau)\,.
\Eeq
Then the continuous directional definitions of the longitudinal and transverse polarisation vectors from eq.~\eqref{eq:epsL} and eq.~\eqref{eq:epsT} should be modified to
\Beq
i\bk\cdot\boldsymbol{\epsilon}^{T\pm}_\bk=0\to 2i\frac{\sin{\left(\bk\frac{\Delta}{2}\right)}}{\Delta}\cdot\boldsymbol{\epsilon}^{T\pm}_\bk=0\,,\\
i\bk\times\boldsymbol{\epsilon}^L_\bk=0\to 2i\frac{\sin{\left(\bk\frac{\Delta}{2}\right)}}{\Delta}\times\boldsymbol{\epsilon}^L_\bk=0\,.
\Eeq
These modifications hold for the spatial discretization stencil in eq.~\eqref{eq:DiscreteDiv}. For other stencils one can derive similar expressions for the orientation of the polarisation vectors. We stress that if the spatial discretization is not accounted for by the polarization vectors, there will be a violation of the Gauss constraint.


\section{The non-Abelian models}
\label{sec:nAmodels}

We now show that the developed techniques in the previous sections can be used for more complicated non-Abelian models. The main purpose here is to reduce calculations of these complicated models (as far as possible) to the ones we have carried out in the Abelian case. We shall first consider the case of $SU(2)$ non-Abelian gauge fields and then extend to $SU(2)\times U(1)$, describing the Electroweak sector of the Standard Model. 
\subsection{$SU(2)$ gauge fields}
\label{sec:SU2}

We consider the following action for a charged scalar doublet
\Beq
\label{eq:NonAbelianActionSU2}
S_{\rm m}=\int \text{d}^4x \sqrt{-g} \mathcal{L}_{\rm m}=\int \text{d}^4x \sqrt{-g}\Big[\left(\textbf{D}_{\mu } \boldsymbol{\varphi} \right){}^{\dagger} \left(\textbf{D}^{\mu} \boldsymbol{\varphi} \right)-\mathcal{V} (\left| \boldsymbol{\varphi} \right|)-\frac{1}{2}\rm{tr}\,\textbf{F}^2(\textbf{A})\Big]\,,
\Eeq
where
\Beq
\label{eq:FSU2M}
\textbf{D}_{\mu }\boldsymbol{\varphi}&=\nabla_{\mu}\boldsymbol{\varphi}+ i g_{\!_A} \textbf{A}_{\mu } \boldsymbol{\varphi}\,,\\
                \textbf{F}_{\mu \nu }&=\nabla_{\mu}\textbf{A}_{\nu}-\nabla_{\nu}\textbf{A}_{\mu}+ig_{\!_A}\left(\textbf{A}_{\mu}\textbf{A}_{\nu}-\textbf{A}_{\nu}\textbf{A}_{\mu}\right)\,.
\Eeq
The action is invariant under the local $SU(2)$ transformation
\Beq
\label{eq:NonAbelianSU2transSU2}
\boldsymbol{\varphi}&\to \textbf{U}\boldsymbol{\varphi}=e^{-ig_{\!_A}\boldsymbol{\beta}(x^{\nu})}\boldsymbol{\varphi}\,,\\
    \textbf{A}_{\mu}&\to \textbf{U}\textbf{A}_{\mu}\textbf{U}^{-1}+\frac{i}{g_{\!_A}}\left(\nabla_{\mu}\textbf{U}\right)\textbf{U}^{-1}\,,
\Eeq
where the non-Abelian gauge fields are $\textbf{A}_{\mu}=A_{\mu}^a\boldsymbol{\sigma}^a/2$, and $\boldsymbol{\beta}(x^{\nu})=\beta^a(x^{\nu})\boldsymbol{\sigma}^a/2$, with $\left\{\boldsymbol{\sigma}^1,\boldsymbol{\sigma}^2,\boldsymbol{\sigma}^3\right\}$ being the three Pauli matrices. The repeated indices are summed over (both for field and spacetime indices).

It is possible to write the above action in terms of local $SU(2)$ invariant fields (analogous to the Abelian case). We will show below that these gauge invariant fields decouple from each other at the linear level and the problem reduces to three identical copies of the Abelian model. The quantisation scheme, preheating analysis as well as setting up of lattice initial conditions discussed in the previous sections then carries over without any difficulty.

We begin by writing the scalar doublet as~\cite{Mukhanov1}
\Beq
\label{eq:Su2phi}
\boldsymbol{\varphi}=\frac{\rho}{\sqrt{2}}\textbf{M}\begin{pmatrix} _{}0_{} \\ _{}1_{} \end{pmatrix}\,,
\Eeq
where $\rho$ will be our real scalar field which forms a condensate during inflation (e.g. the inflaton), and $\textbf{M}$ is a {\it unitary} matrix of {\it unit} determinant
\Beq
\textbf{M}=\exp\left(i\frac{g_{\!_A}}{2}\Omega^a\boldsymbol{\sigma}^a\right)\,.
\Eeq
This $\{\rho,\Omega^a\}$ decomposition is similar to the polar one give in eq.~\eqref{eq:Polars} for the Abelian model.
In a now familiar way, cf. eq.~\eqref{eq:defG}, we proceed to define $SU(2)$ invariant non-Abelian fields
\Beq
\label{eq:Su2G}
\textbf{G}_{\mu}\equiv\textbf{M}^{-1}\textbf{A}_{\mu}\textbf{M}-\frac{i}{g_{\!_A}}\textbf{M}^{-1}\nabla_{\mu}\textbf{M}\,.
\Eeq
More explicitly, $\textbf{G}_\mu=G^a_\mu\boldsymbol{\sigma}^a/2$, which in component form is
\Beq
\label{eq:Su2GPauli}
\textbf{G}_{\mu}=\frac{1}{2}\begin{pmatrix}\quad G^3_{\mu} \qquad G^{1}_{\mu}-iG^{2}_{\mu} \qquad\\ G^{1}_{\mu}+iG^{2}_{\mu}\quad -G^3_{\mu} \qquad\end{pmatrix}\,.
\Eeq
The action in eq.~\eqref{eq:NonAbelianActionSU2} then simplifies to
\Beq
\label{eq:nAbelianGaugeInvSU2}
S_{\rm m}=\int \text{d}^4x \sqrt{-g}\Bigg[\frac{\nabla_{\mu}\rho \nabla^{\mu}\rho}{2}-V (\rho)+\frac{g_{\!_A}^2\rho^2}{8}G^a_{\mu}G^{a\mu}-\frac{1}{4}F^a_{\mu\nu}(G)F^{a\mu\nu}(G)\Bigg]\,,
\Eeq
where the interactions between the gauge bosons are hidden in the definition of the field tensor
\Beq
\label{eq:FSU2}
F^a_{\mu\nu}(G)&\equiv \nabla_{\mu}G^a_{\nu}-\nabla_{\nu}G^a_{\mu}-g_{\!_A}\epsilon^{abc}G_{\mu}^bG^c_{\nu}\,.
\Eeq
We will again work at the linear level in $G^a_{\mu}$, since the arguments for the vanishing backgrounds of the gauge fields used in the Abelian model apply to this case as well. The inflaton fluctuations ($\delta \rho$) are decoupled from the gauge fields, $G^{a}_{\mu}$. Furthermore $\delta \rho$ is the only field coupled to the metric perturbations. At linear order we can ignore the interactions between the gauge bosons, i.e. the last term in~\eqref{eq:FSU2}. Thereby, the three gauge fields, $G^{a}_{\mu}$, are decoupled from each other and each of them is treated as the gauge field from the Abelian model. The quadratic action for the matter perturbations splits into  (cf. eq.~\eqref{eq:S2})
\Beq
\label{eq:S2SU2}
S_{\rm m}^{(2)}=S^{\rho}+\sum_{a=1}^3\left(S^{aL}+S^{aT+}+S^{aT-}\right)=\sum_I S^I\,.
\Eeq
Each $S^I$ is of the general form given in eq.~\eqref{eq:S_I}. The coefficients $b_\rho(k,\tau)$, $\omega_\rho(k,\tau)$ are those in eq.~\eqref{eq:frho}, and $b_{aL}(k,\tau)$, $\omega_{aL}(k,\tau)$ and $b_{aT\pm}(k,\tau)$, $\omega_{aT\pm}(k,\tau)$  are equal to the longitudinal and transverse ones in eq.~\eqref{eq:fLT}, respectively.
Therefore, the inflationary power spectra will be those of three identical copies of longitudinal modes and six identical copies of transverse modes (i.e. one longitudinal and two transverse modes per gauge field). The quantization procedure is identical to the Abelian case with no additional subtleties.

The inflaton oscillations during preheating again present a problem for the gauge invariant fields. The gauge invariant fields, $G^{a}_{\mu}$, are ill-defined when $\Omega^a$ are ill-defined which happens every time $\boldsymbol{\vp}=\boldsymbol{0}$. In analogy with the Abelian case, we will work in the Coulomb gauge. In terms of its ``cartesian" components,
\Beq
\boldsymbol{\vp}(x^{\mu})=\frac{1}{\sqrt{2}}\begin{pmatrix} \delta\varphi^2 + i\delta\varphi^1 \\ \bar{\vp}^0+\delta\varphi^0 - i\delta\varphi^3 \end{pmatrix}\,,
\Eeq
where we have used the global $SU(2)$ invariance of the action to rotate the internal $\boldsymbol{\varphi}$ axes to align with the direction of motion of the homogeneous field. We have taken this direction to be along $\bar{\vp}^0$, with all the other homogeneous components set to zero.\footnote{Note that $\Omega^a\ll 1$, and to linear order in perturbations, the scalar doublet in  eq.~\eqref{eq:Su2phi} can be written as
\Beq
\label{eq:Su2phiOmega}
\boldsymbol{\vp}(x^{\mu})=\frac{\bar{\rho} + \delta\rho}{\sqrt{2}}\begin{pmatrix} \frac{g_{\!_A}}{2}\Omega^2 + i\frac{g_{\!_A}}{2}\Omega^1 \\ 1 - i\frac{g_{\!_A}}{2}\Omega^3 \end{pmatrix}\approx\frac{1}{\sqrt{2}}\begin{pmatrix} \frac{g_{\!_A}}{2}\bar{\rho}\Omega^2 + i\frac{g_{\!_A}}{2}\bar{\rho}\Omega^1 \\ \bar{\rho} + \delta\rho - i\frac{g_{\!_A}}{2}\bar{\rho}\Omega^3 \end{pmatrix}  \,.
\Eeq
This explains the choice of signs and labeling of components in $\boldsymbol{\vp}$. The terms on the right-hand side in eq.~\eqref{eq:Su2G} reduce to
\Beq
\textbf{M}^{-1}\textbf{A}_{\mu}\textbf{M}&=\frac{1}{2}\begin{pmatrix}\quad A^3_{\mu} \qquad A^{1}_{\mu}-iA^{2}_{\mu} \qquad\\ A^{1}_{\mu}+iA^{2}_{\mu}\quad -A^3_{\mu} \qquad\end{pmatrix}\,,\\
-\frac{i}{g_{\!_A}}\textbf{M}^{-1}\nabla_{\mu}\textbf{M}&=\frac{\nabla_{\mu}}{2}\begin{pmatrix}\quad \Omega^3 \qquad \Omega^1-i\Omega^2 \qquad\\ \Omega^1+i\Omega^2\quad -\Omega^3 \qquad\end{pmatrix}\,.
\Eeq 
Recalling eq.~\eqref{eq:Su2GPauli}, one arrives at the familiar expression for the gauge invariant fields at linear order in perturbations (cf. eq.~\eqref{eq:defG}): $G^a_{\mu}=A^a_{\mu}+\nabla_{\mu}\Omega^a$. This relation, along with $\delta\varphi^a=g_{\!_A}\bar{\rho}\Omega^a/2$ and $\bar{\rho}=\bar{\vp}^0$, then allows us to easily derive the equations of motion for $A^a_\mu$ and $\delta \varphi^a$ from the equations for $G^a_\mu$, since $G^a_\mu=A^a_{\mu}+\nabla_{\mu}(2\delta\vp^a/g_{\!_A}\bar{\rho})$.} In the Coulomb gauge for each $a=1,2,3$, the longitudinal modes $A^{aL}_\bk=0$ from the gauge condition $\partial_i A^{ai}=0$. The relevant perturbative degrees of freedom in this gauge are $\{\delta\varphi^0_\bk,\delta\varphi^a_\bk\}$ and the transverse components of the gauge field $A^{aT\pm}_\bk$. The equation of motion for $\bar{\vp}^0$ and its perturbation in Fourier space ($\delta\tilde{\varphi}^0_\bk$) is identical to eq.~\eqref{eq:EoMbckgrndCoulomb} and the first equation in eq. \eqref{eq:EoMdphiCoulomb}, respectively. This field plays the role of the inflaton. The equations of motion for $\delta\varphi^{a}_\bk$ are copies of the second equation in eq.~\eqref{eq:EoMdphiCoulomb}.

\subsection{The ``Electroweak'' sector: $SU(2)\times U(1)$}

Let us consider a more realistic scenario in which the scalar has $SU(2)\times U(1)$ charges
\Beq
\label{eq:NonAbelianAction}
S_{\rm m}=\int \text{d}^4x \sqrt{-g} \mathcal{L}_{\rm m}=\int \text{d}^4x \sqrt{-g}\Big[\left(\textbf{D}_{\mu } \boldsymbol{\varphi} \right){}^{\dagger} \left(\textbf{D}^{\mu} \boldsymbol{\varphi} \right)-\mathcal{V} (\left| \boldsymbol{\varphi} \right|)-\frac{1}{4}\mathcal{F}^2(B)-\frac{1}{2}\rm{tr}\textbf{F}^2(\textbf{A})\Big]\,,
\Eeq
where
\Beq
\textbf{D}_{\mu }\boldsymbol{\varphi}&=\nabla_{\mu}\boldsymbol{\varphi}+ i g_{\!_A} \textbf{A}_{\mu } \boldsymbol{\varphi} - \frac{i}{2}g_{\!_B}B_\mu\boldsymbol{\varphi}\,,
\Eeq
and $\mathcal{F}_{\mu\nu}(B)$ and $\textbf{F}_{\mu\nu}(\textbf{A})$ are the field tensors defined in eqns. \eqref{eq:FTCD} and \eqref{eq:FSU2M}, respectively.
The coefficients in the covariant derivative are defined in this particular way, so that one can refer the scalar doublet to the Standard Model Higgs field, whose hypercharge is $-1/2$. The action is invariant under the local $U(1)$ and $SU(2)$ transformations 
\Beq
\label{eq:U1SU2transf}
\boldsymbol{\varphi}\to \exp[ig_{\!_B}\alpha(x^\nu)/2]\textbf{U}\boldsymbol{\varphi}\,,\qquad B_\mu\rightarrow B_\mu+\nabla_\mu\alpha(x^\nu) \,,\qquad \textbf{A}_{\mu}\to \textbf{U}\textbf{A}_{\mu}\textbf{U}^{-1}+\frac{i}{g_{\!_A}}\left(\nabla_{\mu}\textbf{U}\right)\textbf{U}^{-1}\,,
\Eeq 
where $\textbf{U}$ was defined in eq. \eqref{eq:NonAbelianSU2transSU2}.

We can write the $SU(2)$ sector (but not the $U(1)$ sector) in terms of gauge invariant variables, $\rho$ and $G^a_\mu$, by repeating the initial steps (eqns.~\eqref{eq:Su2phi}-\eqref{eq:Su2GPauli}) from Section \ref{sec:SU2}. We could have defined $SU(2)$ {\it and} $U(1)$ invariant fields. However, following the standard treatment of the Electroweak sector of the Standard Model of Particle Physics, we will fix the gauge for the $U(1)$ sector. With an eye towards the preheating analysis, we will choose Coulomb gauge for the $U(1)$ fields, where $\partial_iB^i=0$ (in Fourier space, $B_\bk^L=0$). In addition, instead of proceeding to quantization and then to a preheating analysis, we will work with certain linear combinations of $G$ and $B$ fields to make contact with the Standard Model. To this end, we define the charged {\it W} bosons as (cf. eq.~\eqref{eq:Su2GPauli}) $W^{\pm}_{\mu}=({G^1_{\mu}\mp iG^2_{\mu}})/{\sqrt{2}}$. The {\it Z} boson and the massless photon, $\mathcal{A}$, emerge after rotating in the $G^3B$ space
\Beq
\label{eq:WeinbergRot}
\begin{pmatrix} \mathcal{A}_{\mu} \\ Z_{\mu} \end{pmatrix}\equiv\begin{pmatrix}\cos{\theta_{\!_W}} \qquad \sin{\theta_{\!_W}} \\ -\sin{\theta_{\!_W}} \qquad \cos{\theta_{\!_W}} \end{pmatrix}\begin{pmatrix} B_{\mu} \\ G^3_{\mu} \end{pmatrix}\,,
\Eeq
through the Weinberg angle
\Beq
\cos{\theta_{\!_W}}\equiv-\frac{g_{\!_A}}{\sqrt{g_{\!_A}^2+g_{\!_B}^2}}\,.
\Eeq
The action from eq.~(\ref{eq:NonAbelianAction}) becomes
\Beq
\label{eq:nAbelianGaugeInv}
S_{\rm m}=\int \text{d}^4x \sqrt{-g}\Bigg[\frac{\nabla_{\mu}\rho \nabla^{\mu}\rho}{2}-&V (\rho)+\frac{\left(g_{\!_A}^2+g_{\!_B}^2\right)\rho^2}{8}Z_{\mu}Z^{\mu}-\frac{1}{4}F^2(\mathcal{A})-\frac{1}{4}F^2(Z)\\
                                                                                                              &\qquad+\frac{g_{\!_A}^2\rho^2}{4}W^+_{\mu}W^{-\mu}-\frac{1}{2}F_{\mu\nu}(W^{+})F^{\mu\nu}(W^{-})\Bigg]\,,
\Eeq
where the interactions between the gauge bosons are included into the definitions of the field tensors
\Beq
\label{eq:FZFAFW}
          F_{\mu\nu}(Z)&\equiv \nabla_{\mu}Z_{\nu}-\nabla_{\nu}Z_{\mu}-ig_{\!_A}\cos\theta_{\!_W}\left(W_{\mu}^-W_{\nu}^+-W_{\nu}^-W_{\mu}^+\right)\,,\\
F_{\mu\nu}(\mathcal{A})&\equiv \nabla_{\mu}\mathcal{A}_{\nu}-\nabla_{\nu}\mathcal{A}_{\mu}-ig_{\!_A}\sin\theta_{\!_W}\left(W_{\mu}^-W_{\nu}^+-W_{\nu}^-W_{\mu}^+\right)\,,\\
    F_{\mu\nu}(W^{\pm})&\equiv \mathcal{D}^{\pm}_{\mu}W_{\nu}^{\pm}-\mathcal{D}^{\pm}_{\nu}W_{\mu}^{\pm}\,,
\Eeq
where
\Beq
\mathcal{D}^{\pm}_{\mu}W^{\pm}_{\nu}\equiv \nabla_{\mu}W^{\pm}_{\nu}\pm ig_{\!_A}\text{sin}\theta_{\!_W}\mathcal{A}_{\mu}W^{\pm}_{\nu}\pm ig_{\!_A}\text{cos}\theta_{\!_W}Z_{\mu}W^{\pm}_{\nu}\,.
\Eeq
Note that $\rho,\mathcal{A}_\mu,Z_\mu$ and $W^{\pm}_\mu$ are invariant under $SU(2)$ transformations, eq. \eqref{eq:NonAbelianSU2transSU2}. However, $\mathcal{A}_\mu$ and $W^{\pm}_\mu$  change under a $U(1)$ transformation, eq.~\eqref{eq:U1SU2transf}, as follows:
\Beq
\mathcal{A}_{\mu}\to \mathcal{A}_{\mu}+\frac{1}{\text{cos}\theta_{\!_W}}\nabla_{\mu}\alpha(x^{\nu})\,,\quad W_{\mu}^{\pm}\to e^{\pm ig_{\!_B}\alpha(x^{\nu})}W^{\pm}_{\mu}\,.
\Eeq
That is why, to remove this redundancy, we choose $B^L_\bk=0$, which implies $\mathcal{A}^L_\bk=Z^L_\bk\tan{\theta_{\!_W}}$, cf. eq. \eqref{eq:WeinbergRot}. This is also a good place to point out that $W^{\pm}_{\mu}$ are a conjugate pair of {\it complex} fields. The `$\pm$' should not be confused with the two transverse polarisation states. 

We work at the linear level in $\delta \rho$, $W^{\pm}_{\mu}$, $Z_{\mu}$, $\mathcal{A}_{\mu}$ (the arguments for the vanishing backgrounds of the gauge fields used in the previous cases apply here as well). Once again the inflaton fluctuations $\delta \rho$ are decoupled from the rest of the matter fields, and the only ones coupled to the metric perturbations. Neglecting the interactions between the gauge bosons, i.e. the second order terms in eq.~\eqref{eq:FZFAFW} and expressing $W^{\pm}_{\mu}$ in terms of $G^1_\mu$, $G^2_\mu$, the second order in perturbations matter action splits into, cf. eqns. \eqref{eq:S2} and \eqref{eq:S2SU2},\footnote{While defined $W^\pm_\mu$ because of our desire to connect to the Standard model, we write the action for the perturbations in terms of $G^1_\mu$ and $G^2_\mu$, because it brings part of the action in a form which looks like multiple copies of the Abelian case.}
\Beq
S_{\rm m}^{(2)}&=S^{\rho}+S^{ZL}+S^{ZT+}+S^{ZT-}+S^{\mathcal{A}T+}+S^{\mathcal{A}T-}+\sum_{a=1}^2\left(S^{aL}+S^{aT+}+S^{aT-}\right)\\
                       &=\sum_I S^I\,.
\Eeq
The $S^I$ have the form shown in eq.~\eqref{eq:S_I}. For  $I=\rho$, $b_\rho(k,\tau)$ and $\omega_\rho(k,\tau)$ are given in eq. \eqref{eq:frho}. For $I=aL,aT\pm$  (with $a=\{1,2\})$, the coefficient pairs $\{b_{aL}(k,\tau)$, $\omega_{aL}(k,\tau)\}$ and $\{b_{aT\pm}(k,\tau)$, $\omega_{aT\pm}(k,\tau)\}$ are the longitudinal and transverse coefficients in eq.~\eqref{eq:fLT}, respectively. The other coefficients are
\Beq
&b_{ZL}(k,\tau)=\left[1+\left(\frac{2 k\cos{\theta_{\!_W}}}{\bar{\rho}g_{\!_A}a}\right)^{\!\!2}\right]^{-1}\,,\qquad \omega_{ZL}^{2}(k,\tau)=k^2+\left(\frac{\bar{\rho}g_{\!_A}a}{2\cos{\theta_{\!_W}}}\right)^{\!\!2}\,,\\
&b_{ZT\pm}(k,\tau)=1\,,\qquad \omega_{ZT\pm}^2(k,\tau)=k^2+\left(\frac{\bar{\rho}g_{\!_A}a}{2\cos{\theta_{\!_W}}}\right)^{\!\!2}\,,\\
&b_{\mathcal{A}T\pm}(k,\tau)=1\,,\qquad \omega_{\mathcal{A}T\pm}^2(k,\tau)=k^2\,.
\Eeq
Notice that only the transverse modes of the photon, $\mathcal{A}$, contribute to the action. The longitudinal modes of $\mathcal{A}$ do not contribute since the photon is {\it massless} - a manifestation of the Ward identity. We can now quantise the transverse modes of $\mathcal{A}$ and the longitudinal and transverse modes of {\it Z}, $G^1$, $G^2$ as done in Section \ref{sec:QuantPerturb}.

We remind the reader that the procedure outlined above is well-defined during inflation, when $\rho^2/2=\boldsymbol{\vp}^{\dagger}\boldsymbol{\vp}\neq0$. Similarly to the Abelian and the $SU(2)$ cases, $\textbf{G}_{\mu}$, are ill-defined when $\rho=0$ and this can generally happen during preheating. Once again the right prescription is to work in well-defined variables in a non-singular gauge. In the Coulomb gauge, $B^L=0$ and $A^{aL}=0$, the longitudinal modes $G^{1L}$, $G^{2L}$ and $Z^{L}$ can be expressed in terms of the well-defined perturbations of $\boldsymbol{\varphi}$: $\delta\vp^1$, $\delta\vp^2$ and $\delta\vp^3$ respectively. In this gauge, the transverse modes are even easier to handle: $G^{1T\pm}$ and $G^{2T\pm}$ reduce to $A^{1T\pm}$ and $A^{2T\pm}$, respectively; similarly $Z^{T\pm}\rightarrow -\sin{\theta_{\!_W}}B^{T\pm}+\cos{\theta_{\!_W}}A^{1T\pm}$ and $\mathcal{A}^{T\pm}\rightarrow\cos{\theta_{\!_W}}B^{T\pm}+\sin{\theta_{\!_W}}A^{1T\pm}$. Hence we can keep the expressions for the transverse modes since they have no singularities in them with $\rho\rightarrow \bar{\varphi}^0$. As usual, using the $SU(2)$ global invariance, we have rotated our internal axes along the direction of motion of the inflaton.

\begin{figure*}[t] 
   \centering
   \includegraphics[width=2.42in]{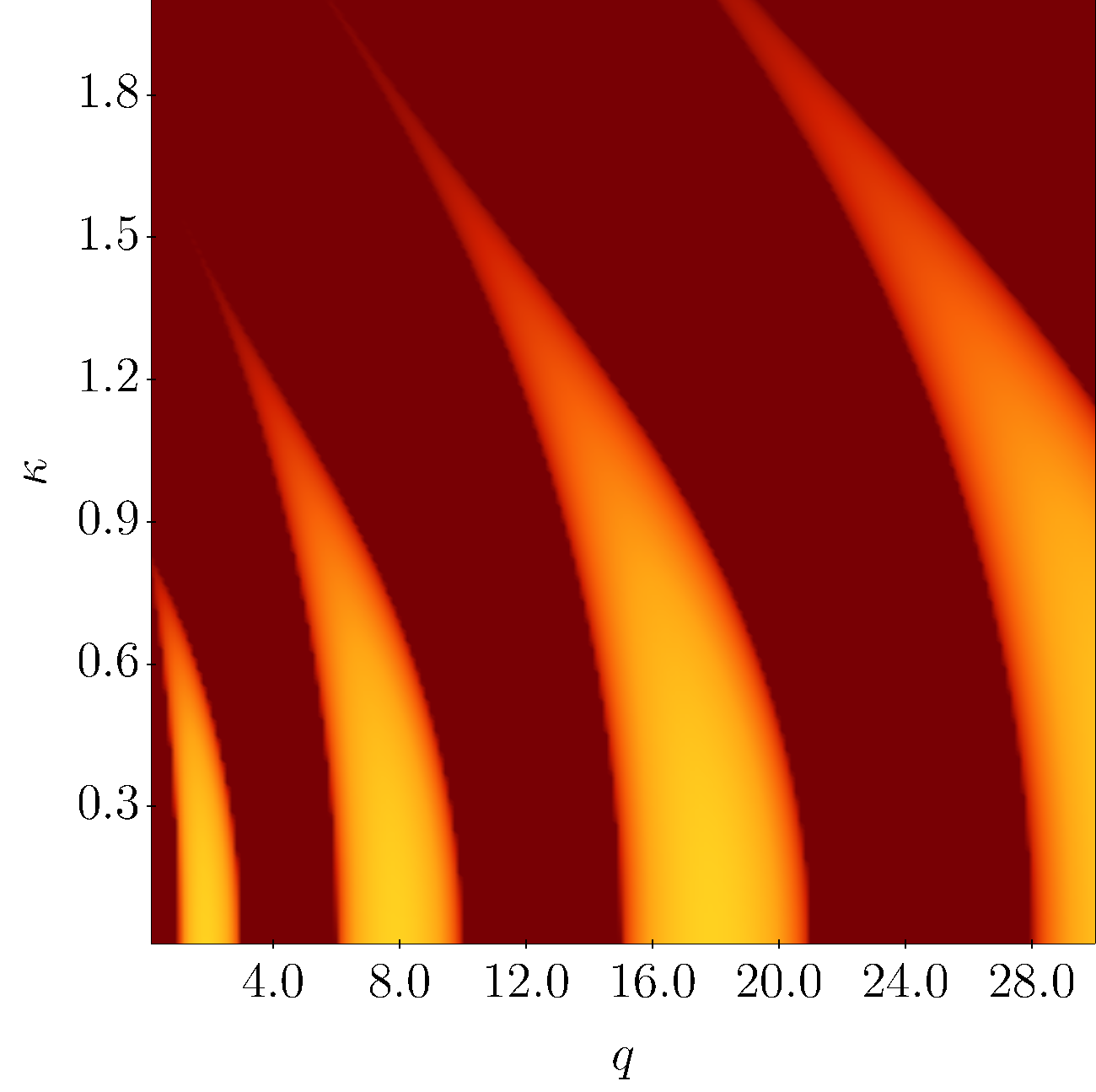} 
     \raisebox{0.054\height}{\includegraphics[width=0.52in]{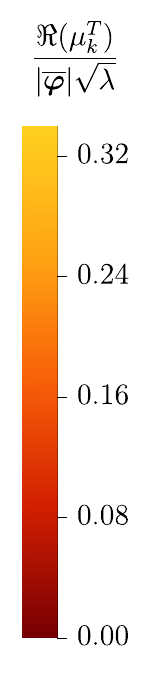}}
   \includegraphics[width=2.42in]{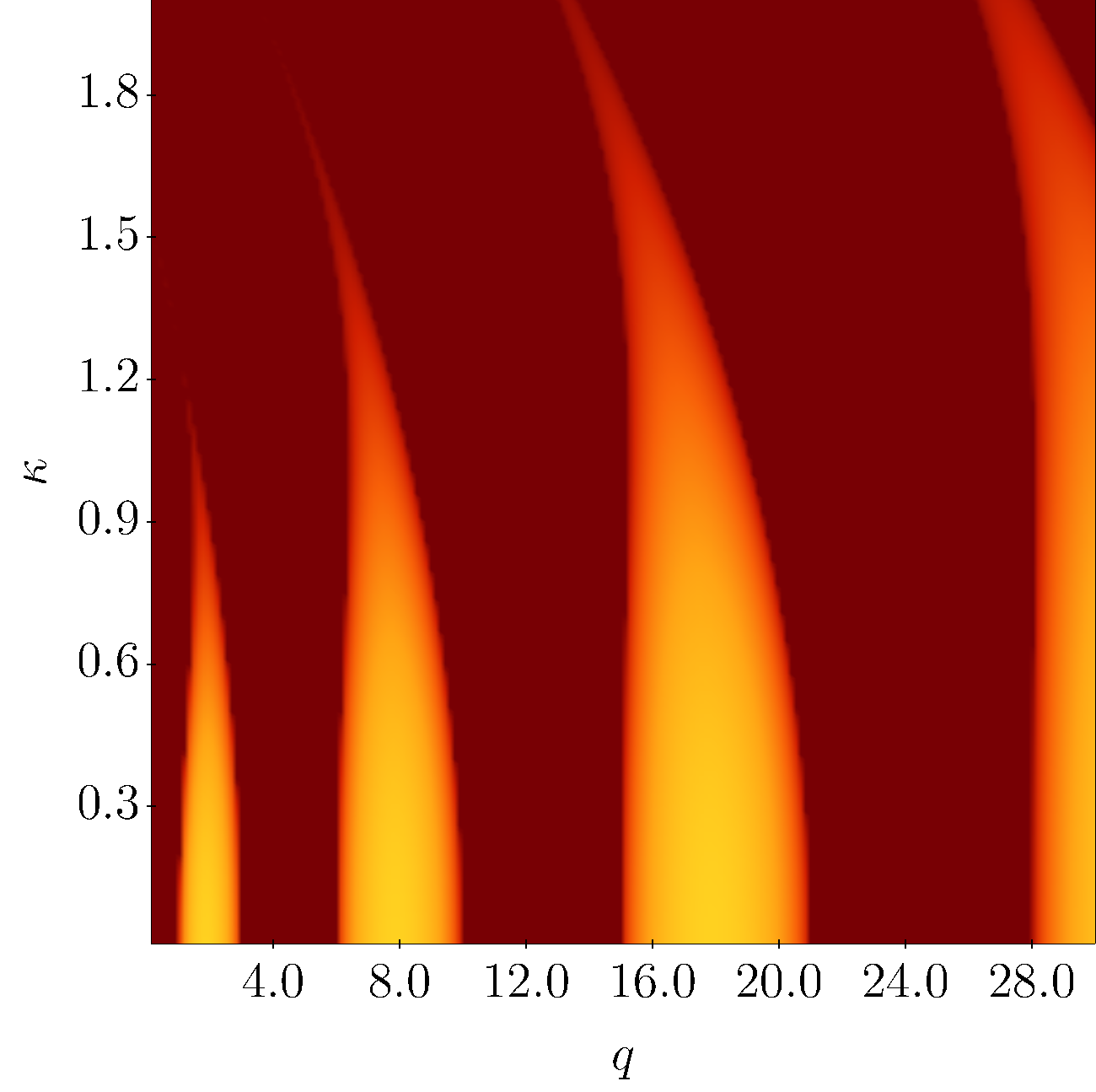} 
     \raisebox{0.054\height}{\includegraphics[width=0.52in]{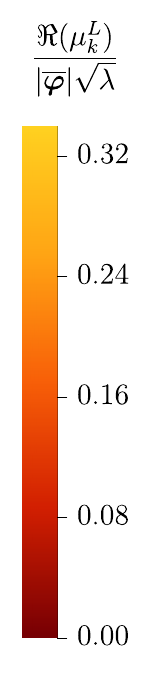}}
   \caption{Floquet exponents of transverse and longitudinal modes of {\it W} and {\it Z} bosons during preheating with non-Abelian fields. The Floquet plot for the transverse modes on the left is the same as Figure 1 in~\cite{Kari}. Our analysis shows a similar qualitative behavior for the longitudinal modes as well (right). For this plot the scalar field potential is taken to be $V(\rho)=\lambda(\boldsymbol{\varphi}^{\dagger}\boldsymbol{\varphi})^2=\lambda\rho^4/4$. We use the same notation as in~\cite{Kari}: $\kappa=k/\sqrt{2\lambda|\bar{\boldsymbol{\varphi}}|^2}=k/\sqrt{\lambda\bar{\rho}^2}$, $q_W=g_{\!_A}^2/4\lambda$ and $q_Z=(g_{\!_A}^2+g_{\!_B}^2)/4\lambda$. The amplitude of the oscillating scalar field and the momentum redshift as $a^{-1}$, rendering $\kappa$ redshift-independent. That is why the flow lines from Fig. \ref{fig:2dFloq_m2Phi2} do not appear here. As the universe expands, Fourier modes do not cross through multiple bands. This is a manifestation of the conformal nature of the quartic potential after the end of inflation.}
   \label{fig:2dFloq_lambdaPhi4}   
\end{figure*}

For purposes of comparison with the literature, we provide the Floquet charts for the longitudinal and transverse modes of the {\it W} and {\it Z} fields which are excited by the oscillations of the Higgs condensate for the case when $V (\left| \boldsymbol{\varphi} \right|)=\lambda\left(\boldsymbol{\vp}^{\dagger}\boldsymbol{\vp}\right)^{\!2}=\lambda\rho^4/4$, see Fig.~\ref{fig:2dFloq_lambdaPhi4}. The longitudinal modes are excited parametrically during reheating. The inflaton decay to transverse modes of the gauge fields mimics the well studied $\vp^2\chi^2$ scalar field model \cite{Felder:2006cc}. 

Without the longitudinal modes, the back-reaction of the transverse gauge fields would be identical to the one in $\vp^2\chi^2$. However, recent numerical experiments~\cite{Figueroa2015} indicate that there are small differences between the behaviour of the inflaton decaying to scalars and to Abelian gauge fields. This might be due to the longitudinal modes being excited and playing a role in the back-reaction process. In general, calculations during (p)reheating and estimation of the effects of non-Abelian interaction terms are convenient in the Coulomb gauge (or some other well-defined gauge during preheating). The traditional unitary gauge is {\it not} a good choice because the inflaton oscillates through the origin. Although as we saw in the Abelian case, the electric and magnetic fields can be well defined even when using gauge invariant variables, the intermediate steps leading up their calculation often involve singular behavior which at the very least leads to numerical unpleasantness. The same comment holds for the Unitary gauge since the equations of motion are identical to those in gauge invariant variables.

\section{Observational consequences}
\label{sec:observationalconsequences}

Particle production during and after inflation in models with a charged inflaton coupled to gauge fields may leave potentially observable signatures such as primordial magnetic fields, charge fluctuations as well as scalar and tensor metric perturbations. We schematically discuss each one in turn.

\subsection{Magnetic fields}
At the end of inflation the magnetic field power spectrum on superhorizon scales is blue and given by a power-law, see eqns. \eqref{eq:subhCompton} and \eqref{eq:suphCompton}. If the gauge fields are `lighter' than the co-moving Hubble scale during inflation, i.e. for $k_{\!_C}\lesssim \mathcal{H}$ we have $\Delta_{B^{T\pm}}\propto k^2$. If the gauge fields are ``massive" enough, $k_{\!_C}\gtrsim \mathcal{H}$, the power spectrum is much steeper: $\Delta_{B^{T\pm}}\propto k^{5/2}$. After inflation, the oscillating inflaton can resonantly amplify the gauge fields. For the $m^2|\varphi|^2$ model considered in Section \ref{sec:Inflps}, the resonance is broad and effective for $k_{\!_C}/a\sim g_{\!_A}|\bar{\varphi}|\gtrsim m$, where $\bar{\varphi}$ is the vev of the inflaton during the early stages of preheating at the end of inflation.\footnote{In the more popular notation, the parameter determining the strength of the resonance, $q$, is given by $q=k_{\!_C}/(am)$. The resonance is broad when $q\gg1$ and narrow or absent when $q\lesssim1$.} 

The above considerations show that the optimal range for production of magnetic fields on superhorizon scales is $k_{\!_C}\lesssim \mathcal{H}$ during inflation and $k_{\!_C}/a\gtrsim m$ during preheating. However, for $m^2|\varphi|^2$ inflation, $k_{\!_C}/(am)\gtrsim 1$ during preheating implies $k_{\!_C}/\mathcal{H}\gtrsim 1$ during inflation, thus excluding the possibility of broad resonance {\it and} a relatively shallow power law for the magnetic field spectrum. We reached the same conclusion via a more detailed calculation whose results were summarized in Fig. \ref{fig:Ps}. We have to settle for two separate regimes:  $\Delta_{B^{T\pm}}\propto k^2$ which receives no resonant amplification during preheating, or $\Delta_{B^{T\pm}}\propto k^{5/2}$ which can be resonantly amplified until it backreacts.

In the contemporary universe, observations indicate a magnetic field strength of $\sim10^{-7} \,\text{G}$ on $1 \,\text{Mpc}$ scales (for example, see \cite{Kandus:2010nw}). If a {\it seed} magnetic field $B_\text{seed}\gtrsim10^{-25} \,\text{G}$ is present on comoving scales corresponding to $1\, \text{Mpc}$ at the time of matter-radiation equality, the galactic dynamo mechanism can potentially amplify $B_{\rm seed}$ to the observed values today  \cite{Kandus:2010nw}. As we show below, the seed field generated in the models under consideration are far too small compared to what is required observationally. 

For the case when $k_{\!_C}\lesssim \mathcal{H}$ during inflation, the magnetic field has a shallow spectrum $\propto k^2$ on superhorizon scales but is not resonantly amplified during preheating. At the time of decoupling we obtain $B_\text{seed}\sim\Delta_{B^{T\pm}}^{\text{dec}}\sim (2\pi)^{-1}\left({k}/{a_\text{dec}}\right)^2\sim 10^{-50}\,\text{G}\,$(for example, see \cite{Emami2009}) where we used $a_\text{dec}\sim 10^{-3}$, $k={\rm few}\,\text{Mpc}^{-1}\sim 10^{-38}\,\text{GeV}$ and $1\,\text{GeV}^2\sim 10^{20}\,\text{G}$ . When $k_{\!_C}\gtrsim\mathcal{H}$, the magnetic field can be resonantly amplified, with an amplification factor that can be as large as $\sim 10^4 - 10^5$. The maximum value is set by back-reaction considerations. However, there is a suppression factor $\sqrt{k/k_{\!_C}}$ (see eq. \eqref{eq:subhCompton}) which turns out to be more important. The net effect is that $B_\text{seed}$ is suppressed compared to the no-resonance scenario. For the case when $k_{\!_C}=10^3\,\mathcal{H}$ at the end of inflation (shown in Fig. \ref{fig:Ps}), $B_\text{seed}\sim 10^{-58}\,\text{G}$. For this estimate we assumed that the back-reaction takes place $2.5$ e-folds after inflation, and subsequently the magnetic field redshifts in the usual way: $a^4\Delta^2_{B^{T\pm}}=\text{const}$ for another $57.5$ e-folds.

We note that while broad resonance does not happen for $k_{\!_C}\lesssim \mathcal{H}$ for the $m^2|\vp|^2$ models, it can occur for $k_{\!_C}<\mathcal{H}$ for steeper potentials (e.g. $\lambda|\vp|^4$). It might be possible to boost the amplitude of the seed field up to $10^{-45}\,\text{G}$ in such cases. However, this is still too small to be observationally relevant. Successful magnetogenesis from reheating has been recently discussed in models different from ours (see for example, \cite{Fujita:2016qab,Kobayashi:2014sga}).

\subsection{Charge fluctuations}

Just like the magnetic field, the charge fluctuations also have a blue power spectrum at the end of inflation. Recalling Gauss law, $-kE^L_{\boldsymbol{k}}=a^2j_{0\boldsymbol{k}}$, we arrive at the useful expression $\Delta_{j_0}=(k/a)\Delta_{E^L}$. Indeed eqns. \eqref{eq:subhCompton} and \eqref{eq:suphCompton} imply that on superhorizon scales $\Delta_{j_0}\propto k^2$ and $k^{5/2}$ for $k_{\!_C}\lesssim \mathcal{H}$ and $k_{\!_C}\gtrsim \mathcal{H}$ respectively, reminiscent of the magnetic field spectrum. After inflation, parametric resonance can amplify the charge fluctuations in the $m^2|\varphi|^2$ model, provided we are in the broad resonance regime $g_{\!_A}|\bar{\varphi}|\gtrsim m$ (i.e. $k_{\!_C}\gtrsim \mathcal{H}$ with a spectrum $\propto k^{5/2}$ on superhorizon scales). Note that unlike the magnetic field case, we see some production of charge fluctuations on superhorizon scales at the end of inflation even when $m\gtrsim g_{\!_A}|\bar{\varphi}|$. This fluctuation production is related to the oscillation of the inflaton, but cannot be analysed with simple Floquet analysis since it occurs on superhorizon scales where expansion plays a significant role. 

If we imagine that the gauge field in question is the electromagnetic field and put $g_{\!_A}=2e$ one can compare the charge fluctuations to the observational bounds from vorticity and cosmological magnetic fields on the electrical charge asymmetry of the universe \cite{Caprini:2003gz}, namely that $\Delta_{j_0}/(en_{\!_B})\lesssim 10^{-26}$ on co-moving scales corresponding to $10^2\, \text{kpc}$ today, where $n_{\!_B}\sim 1.5\times 10^{-10}\,T^3$ is the number of baryons and $T$ is the photon temperature. For $k_{\!_C}<\mathcal{H}$, we can assume the universe reheats immediately after the end of inflation and the charge fluctuation is transferred to a charge asymmetry in the ordinary matter.\footnote{Note that at the end of inflation this charge would correspond to longitudinal electric fields of enormous strength. This may lead to Schwinger pair production, which can annihilate the charge on this scales. For $k_{\!_C}=\mathcal{H}$, $e\Delta_{E^L}\sim10^{-2}\,\text{GeV}\gg m_{\text{electron}}$. However, this estimate is not reliable, since the vev of the Standard Model Higgs is largely uncertain at the end of inflation. It is not clear what the masses of the fermions are.} The charge asymmetry will then redshift as $a^3\Delta_{j_0}=\text{const}$, akin to number density. Then the fractional charge asymmetry
${\Delta_{j_0}}/{en_{\!_B}}\sim 10^9\times\left({k}/{a_{\text{rh}}T_{\text{rh}}}\right)^2\left(g_{\!_A}{|\varphi_{\text{inf}}|}/{T_{\text{rh}}}\right)\sim 10^{-36}\,$,
where we used $|\varphi_{\text{inf}}|\sim \mpl$, $k=10\,\text{Mpc}^{-1}$, $T_{\text{end}}\sim10^{16}\,\text{GeV}$ and $a_{\rm end}T_{\rm end}\sim T_0=2.5\times 10^{-4}\rm eV$.\footnote{We caution the reader that this is a very rough estimate.} For the broad resonance regime, $k_{\!_C}\gtrsim \mathcal{H}$, the amplitude of the charge fluctuations can be parametrically amplified up to $10^3-10^4$. However, there is also a suppression factor which dominates on the scales of interest. Explicitly, for $k_{\!_C}=10^3\,\mathcal{H}$ discussed in Fig. \ref{fig:Ps}, if we assume that the universe reheats immediately after back-reaction has taken place ($2.5$ e-folds after the end of inflation) and then the universe expands for another $57.5$ e-folds, the suppression factor is roughly $\sim\sqrt{\exp(-2.5)k/k_{\!_C}}$, implying $\Delta_{j_0}/(en_{\!_B})\approx10^{-45}$.

Again, we recall that in the $m^2|\vp|^2$ model, broad resonance does not happen for $k_{\!_C}\lesssim \mathcal{H}$. For steeper power-law potentials, parametric resonance is efficient for light gauge fields and the excess of charged particles per baryon can go up to $10^{-32}$. We once again caution the reader that the calculations here are meant to be schematic.

\subsection{Metric perturbations}

The production of gauge fields may affect the primordial metric perturbations, giving rise to (for example) non-gaussianities and gravitational waves. Since the gravitational production of gauge fields during inflation is not significant, they are not expected to give rise to any interesting signals. However, if the post-inflationary dynamics include parametric amplification of fields then interesting signatures are possible. For example, light scalar fields might develop a non-zero vev across the observable universe today along with perturbations around this vev within each Hubble patch at the time of preheating. The details of the following non-linear stage might depend on the vev in each separate universe. For instance, in \cite{Chambers:2007se,Chambers:2008gu,Bond:2009xx} it was shown that as the inflaton decays to a light $\chi$ field in $\phi^2\chi^2$ models, the back-reaction depends on the initial vev $\bar{\chi}_{\text{i}}$ within the Hubble patch. This in turn affects the expansion history within the Hubble patch and therefore may lead to non-gaussianities. $\bar{\chi}_{\text{i}}$ also affects the gravitational waves produced during the non-linear evolution within the separate patches. This may lead to low-multiple corrections to the stochastic gravitational wave background \cite{Bethke:2013aba,Bethke:2013vca}.

For $k_{\!_C}\lesssim\mathcal{H}$ one can show that the gauge field sector has a light scalar with a scale invariant power spectrum - it is $\delta\vp^1$ in the Coulomb gauge, see eq. (\ref{eq:GaugeTransfCoulomb}) and Appendix \ref{sec:dS}. This implies that $\delta\vp^1$ develops a non-zero vev across the sky with deviations from it within each Hubble patch at the time of preheating. Its back-reaction should lead to similar effects to the $\vp^2\chi^2$ models, since the resonance structure is similar. Note that for the $m^2|\vp|^2$ model we have been considering, backreaction cannot happen because resonance is inefficient if $k_{\!_C}\lesssim \mathcal{H}$ (which was necessary for developing a vev for $\delta\varphi^1$ during inflation). The absence of resonance for $k_{\!_C}\lesssim\mathcal{H}$ is a feature of $m^2|\vp|^2$ model. For steeper potentials (e.g $\lambda|\varphi|^4$), broad resonance can occur for $k_{\!_C}\lesssim\mathcal{H}$.

\section{Conclusions}
\label{sec:conclusions}

In this paper, we carried out a self-contained analysis of particle production during and after inflation in models with charged scalars coupled to Abelian and non-Abelian fields. We calculated the power spectra of the produced gauge fields in the regime where the equations of motion could be linearized. We also provided a prescription for setting up initial conditions for lattice simulations to further evolve the fields nonlinearly. 

To make our treatment self-contained, we provided the necessary equations for linear perturbations (including metric fluctuations) in terms of gauge invariant variables, and provided a dictionary to relate these variables to those in some common gauges. This should allow results to be translated between gauges quite easily. For pedagogical purposes, we carried out the initial analysis for the Abelian model. In the later half of the paper we provided the explicit generalization to the non-Abelian cases.

We carried out the quantization and evolution of the perturbations during inflation in terms of gauge invariant variables. After substituting the constraint equations into the original action, the quantization and subsequent evolution was carried out in a straightforward manner. Gauge invariant variables have the advantage of automatically dealing with the correct number of degrees of freedom; we never having to worry about spurious gauge modes. We numerically calculated the electric and magnetic field power spectra at the end of inflation for a simple $m^2|\varphi|^2$ inflation (see Fig. \ref{fig:PsEndOfInflation}). We provided an understanding of the shape of the spectra (blue tilt and broken power law behavior -- see eqns. \eqref{eq:subhCompton} and \eqref{eq:suphCompton}) in terms of two scales: the co-moving Compton wavenumber of the gauge fields $k_{\!_C}\equiv a|\bar{\varphi}|g_{\!_A}/2$ ($\bar{\varphi}$ is the vev of the inflaton condensate during inflation, $g_{\!_A}$ is the coupling strength and $a$ is the scalefactor) and the conformal Hubble scale $\mathcal{H}$. We found that the transverse modes are always dominant over the longitudinal ones on sub-Compton scales, $k\gtrsim k_{\!_C}$. For gauge fields with $k_{\!_C}\gtrsim\mathcal{H}$ the longitudinal modes are as energetic as the transverse ones on super-Compton scales, $k\lesssim k_{\!_C}$, whereas when $k_{\!_C}\lesssim\mathcal{H}$ the longitudinal modes dominate on super-Compton scales. The longitudinal mode of the electric field is directly proportional to charge fluctuations, and hence charge fluctuations are also generated during inflation. While our numerical calculations assumed a particular model, these results are expected to hold more generally for potentials where the inflaton is rolling slowly. As a further check, we carried out a semi-analytic analysis in de Sitter universe and confirmed the shapes of the electric and magnetic field spectra.

While gauge invariant variables are suited for calculations during inflation, they become ill-defined when the inflaton starts oscillating after inflation. After inflation, the Coulomb gauge turns out to be a particularly well-suited for preheating studies. We carried out an explicit numerical calculation of resonant gauge field production during preheating and provided power spectra of the corresponding electric and magnetic fields at the end of preheating (see Fig. \ref{fig:Ps}). For an analytic understanding of the main features of the spectra, we carried out a Floquet analysis of the instabilities with the gauge constraints included (see Fig. \ref{fig:2dFloq_m2Phi2}). In both calculations, we found that the longitudinal modes are as prone to resonant excitation during preheating as the transverse ones. The characteristic wavenumber range of the instability is $k\lesssim m(g_{\!_A}\mpl/m)^{1/2}$, whereas the growth rate of fluctuations is determined by $g_{\!_A} \mpl/m$. We also estimated that the gauge fields back-react on the scalar condensate within about $10$ oscillations in the case of a $m^2|\varphi|^2$ potential and for  couplings, $g_{\!_A}\gsim10^{-3}$ (see Fig. \ref{fig:BackReactionCriterion}). For $g_{\!_A}\lesssim10^{-4}$ gauge fields are not produced significantly enough to back-react. 

If back-reaction becomes important, the occupation numbers in the gauge fields are usually very high.  Further evolution, which is usually highly nonlinear, can be investigated by numerical lattice simulations within the classical approximation. We provided a scheme for initializing such simulations with the gauge constraints being violated only at second order in the perturbations. We also provide a work-around enabling the gauge constraints to be satisfied more precisely at the expense of insignificant (second order in perturbations) modification to the spectrum of the gauge fields. Our prescription for initial conditions is not restricted to a particular gauge; we provide an explicit example of setting up such conditions in the temporal gauge which is commonly used for simulations (see Section \ref{sec:App}). The derived power spectra as well as the initial conditions prescription might be useful for lattice simulations of Abelian and non-Abelian fields at the end of inflation (for example, \cite{Rajantie:2000fd,Deskins2013,Adshead2015,Dufaux2010,Figueroa2015,GarciaBellido2003,DiazGil2007,DiazGil2008,Enqvist:2015sua,Figueroa:2016ojl}).

We considered two non-Abelian models. In the $SU(2)$ model we showed that to linear order in perturbations, the problem conveniently reduces to three decoupled identical replicas of the Abelian model. We then extended the local group to $SU(2)\times U(1)$, with the Electroweak sector of the Standard Model of Particle Physics in mind. Again, to linear order in the perturbations, the problem splits into three copies of the Abelian model. There are two identical copies, describing the evolution of the massive $W$ bosons, and a similar one for the massive $Z$ boson, the only difference being the coupling constants. The massless photon $\mathcal{A}$ remains decoupled from the other sectors of the theory. The framework describing the longitudinal and transverse modes in the Abelian model can be straightforwardly applied to the $W$ and $Z$ sectors, including the scheme for initialising lattice simulations. We provided Floquet charts to capture the instability of these fields during preheating, assuming the scalar condensate to be characterised by $\lambda(\boldsymbol{\vp}^{\dagger}\boldsymbol{\vp})^2$ self-interaction (see Fig. \ref{fig:2dFloq_lambdaPhi4}). In this case, we again confirm that the longitudinal and transverse mode can be excited at a comparable level.

Our analysis allowed us to estimate the magnetic field and charge fluctuations from inflation and preheating in a self-consistent manner. We found that the charged inflaton produces magnetic field that cannot exceed $\sim10^{-50}\,\text{G}$ on $1\,\text{Mpc}$ scales today (consistent with \cite{Finelli:2000sh}) for $m^2|\vp|^2$ potential. Parametric resonance in models with steeper potentials, e.g. $\lambda|\vp|^4$ may boost the magnetic field by a factor of $10^5$, which is still not enough to seed an efficient galactic dynamo mechanism. The charge fluctuations are shown to be also below the observational bounds, if the inflaton is assumed to be electrically charged. The excess charged particles per baryon are found to be at most $\lesssim 10^{-36}$ on $0.1\,\text{Mpc}$ scales for $m^2|\vp|^2$ potential. Preheating in models with steeper potentials, e.g. $\lambda|\vp|^4$ may amplify this by a factor of $10^4$. This is well within the observational bound of $\lesssim 10^{-26}$ \cite{Caprini:2003gz}. The possibility of an electrically charged inflaton is not ruled out.

Our analysis also reveals that the asymuthal degree of freedom of the charged inflaton develops a scale-invariant power spectrum if the gauge fields are lighter than the Hubble scale. This gives us the possibility of generating non-Gaussianities and gravitational waves \cite{Chambers:2007se,Chambers:2008gu,Bond:2009xx,Bethke:2013aba,Bethke:2013vca,Figueroa:2016ojl} from the non-linear stage of reheating which we leave for future work.

\begin{acknowledgments}
We thank Anthony Challinor for useful conversations.
\end{acknowledgments}

\appendix

\section{Gauge field perturbations in de Sitter space}
\label{sec:dS}

Let us try to understand the electric and magnetic field power spectra at the end of inflation using some simplified semi-analytic analysis. For this purpose, we approximate the space-time to be de Sitter, $H=\text{const}$, and the inflaton to roll slow enough (so that $d\ln \rho/d\ln a \ll 1$), or more precisely, $\rho\rightarrow\text{const}$. The equations governing the evolution of the mode functions are then given by
\Beq
\label{eq:deSitterEoM}
\partial^2_\tau u_k^{T\pm}+\left(k^2+k_{\!_C}^2\right)u_k^{T\pm}=0\,,\\
\partial^2_\tau u_k^{L}-\frac{2}{\tau}\frac{\partial_\tau u_k^{L}}{1+(k_{\!_C}/k)^2}+\left(k^2+k_{\!_C}^2\right)u_k^{L}=0\,,
\Eeq
where $k_{\!_C}=ag_{\!_A}\bar{\rho}/2=-g_{\!_A}\bar{\rho}/2H\tau$. Note that $k_{\!_C}$ depends on $\tau$ and that $\tau$ is negative during inflation.\footnote{The equations look a lot simpler, and are easier to analyze by doing the following change of variables: $x = -k\tau$ and $\alpha = k_{\!_C}/\mathcal{H}$. Doing this makes $x=-k\tau$ the only independent variable, with $\alpha$ acting as a time and scale independent parameter. We do not use $x$ and $\alpha$ in the presentation to avoid introducing too many new variables.} 

The analytic solutions for the transverse modes of constant mass in de Sitter space-time are known, e.g. cf. \cite{Davis:2000zp}. Using the WKB initial conditions, \eqref{eq:uWKB}, we have
\Beq
\label{eq:uTHankel}
u^{T\pm}_k(\tau)=\frac{\sqrt{-k\tau}}{(2\pi)^{3/2}}\sqrt{\frac{\pi}{4k}}H^{(1)}_z(-k\tau)\exp\left(iz\frac{\pi}{2}+i\frac{\pi}{4}\right)\,,
\Eeq
where $H^{(1)}_z(-k\tau)$ is the Hankel function of first kind, of order $z=\sqrt{({1}/{4})-(k_{\!_C}/\mathcal{H})^2}$. Now the distinction between the $k_{\!_C}\ll \mathcal{H}$ and $k_{\!_C}\gg \mathcal{H}$ regimes discussed in Section~\ref{sec:Inflps} becomes a bit more evident. For $k_{\!_C}/\mathcal{H}>1/2$, $z$ is purely imaginary. For $k_{\!_C}/\mathcal{H}<1/2$, $z$ is purely real, and cannot exceed $(1/2)$. The asymptotes of the Hankel functions depend on $z$ and its complex phase.

\begin{figure*}[t] 
   \centering
    \hbox{\hspace{-26ex}\includegraphics[width=10.7in]{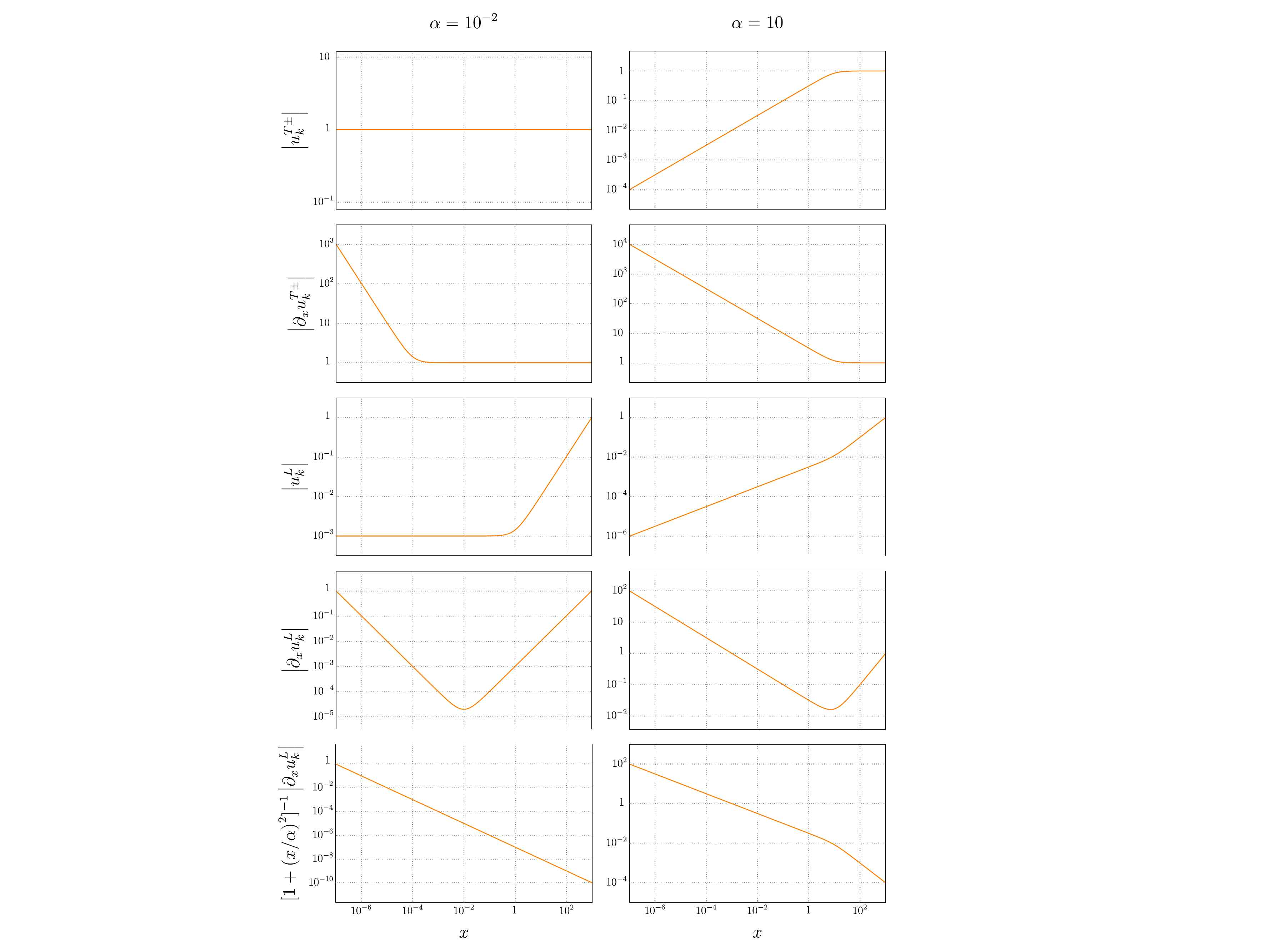}}
   \caption{Numerical solutions to eq.~\eqref{eq:deSitterEoM}, i.e. the field modes in de Sitter space-time with the inflaton assumed to be constant. Fields are evolved backwards in {\it x}, from $x_\text{in}=(-k\tau)_\text{in}\gg\alpha=k_{\!_C}/\mathcal{H}=g_{\!_A}\bar{\rho}/(2H)$.}   \label{fig:ModesInflation}
\end{figure*}

We shall now explain the double power-laws observed in the magnetic and transverse electric fields. We will first focus on the $k_{\!_C}/\mathcal{H}>\frac{1}{2}$, i.e. $z^2<0$ regime.
At early enough times $k\gg k_{\!_C}$ (recall that $k_{\!_C}$ gets smaller at earlier times), the transverse mode functions $u^{T\pm}_k\sim e^{-ik\tau}/\sqrt{k}$ and $\partial_\tau u^{T\pm}_k\sim\sqrt{k}e^{-ik\tau}$. This can be verified analytically or just by solving the equations of motion, see Fig.~\ref{fig:ModesInflation}. Similarly at late enough times when  $k\ll k_{\!_C}$ it is easy to verify from the analytic solution in eq. \eqref{eq:uTHankel} and Fig.~$\ref{fig:ModesInflation}$ that $u^{T\pm}_k\sim1/\sqrt{k_{\!_C}}$ and $\partial_\tau u^{T\pm}_k\sim\sqrt{k_{\!_C}}$. Recalling that $a^4\Delta^2_{B^{T\pm}}\sim k^5|u^{T\pm}_k|^2$ and $a^4\Delta^2_{E^{T\pm}}\sim k^3|\partial_\tau u^{T\pm}_k|^2$ one finds the familiar scalings
\Beq
\Delta^2_{B^{T\pm}}&\approx\frac{H^4}{4\pi^2}\times\begin{cases}
                     \left(k/\mathcal{H}\right)^{\!4}(k/k_{\!_C})\,, & \text{if } k\ll k_{\!_C}\,,\\
                     \left(k/\mathcal{H}\right)^{\!4}\,,        & \text{if } k\gg k_{\!_C}\,,
                   \end{cases}\\
\Delta^2_{E^{T\pm}}&\approx \frac{H^4}{4\pi^2}\times\begin{cases}
                     \left(k/\mathcal{H}\right)^{\!3}(k_{\!_C}/\mathcal{H})\,, & \text{if } k\ll k_{\!_C}\,,\\
                     \left(k/\mathcal{H}\right)^{\!4}\,,        & \text{if } k\gg k_{\!_C}\,.
                   \end{cases}
\Eeq
We note two things. Firstly, the break in the double power-laws occurs where it did in the plots based on the numerical calculation for $V(\rho)=m^2\rho^2/2$ evaluated at the end of inflation provided in Fig. \ref{fig:PsEndOfInflation} (cf. eq. \eqref{eq:subhCompton}).  Secondly, the difference in the $k_{\!_C}$ dependence implies that at late times ($k\ll k_{\!_C}$) there will be more power stored in the form of transverse electric field and less in the form of magnetic field. The product of the transverse electric and magnetic power spectra, however, is unchanged. 

We repeat the same procedure for the magnetic and transverse electric fields in the weak coupling regime, $k_{\!_C}/\mathcal{H}<\frac{1}{2}$, i.e. $\frac{1}{4}>z^2>0$. Initially, when $k\gg k_{\!_C}$, $u^{T\pm}_k\sim e^{-ik\tau}/\sqrt{k}$ and $\partial_\tau u^{T\pm}_k\sim \sqrt{k}e^{-ik\tau}$. Later on, $k\ll k_{\!_C}$, $u^{T\pm}_k\sim 1/\sqrt{k}$. That is why in this coupling regime the magnetic power spectrum is described by a single power-law. The late time behaviour of the $\partial_\tau u^{T\pm}_k$ is slightly more intriguing. For $k_{\!_C}^2/\mathcal{H}\ll k\ll k_{\!_C}$, $\partial_\tau u^{T\pm}_k\sim \sqrt{k}$. However, when $k\ll k_{\!_C}^2/\mathcal{H}$, $\partial_\tau u^{T\pm}_k\sim k_{\!_C}/\sqrt{k}$, implying a double power-law for the transverse electric field, at a scale defined by the square of $k_{\!_C}$. The actual power spectra again agree with what we found in Section~\ref{sec:Inflps} for the $m^2\rho^2/2$ inflation (cf. eq. \eqref{eq:suphCompton})
\Beq
\Delta^2_{B^{T\pm}}&\approx \frac{H^4}{4\pi^2}\left(\frac{k}{\mathcal{H}}\right)^{\!4}\,, \\
\Delta^2_{E^{T\pm}}&\approx \frac{H^4}{4\pi^2}\times\begin{cases}
                     \left(k/\mathcal{H}\right)^{\!2}\left(k_{\!_C}/\mathcal{H}\right)^{\!2}\,, & \text{if $k\ll k_{\!_C}^2/\mathcal{H}$}\,,\\
                     \left(k/\mathcal{H}\right)^{\!4}\,,        & \text{if $k\gg k_{\!_C}^2/\mathcal{H}$}\,.
                   \end{cases}
\Eeq

The longitudinal mode is harder to approach analytically. It can be solved analytically only for $k\gg k_{\!_C}$ and $k\ll k_{\!_C}$. In the former (subhorison) limit the equation of motion reduces to
\Beq
\label{eq:LsubdS}
\partial^2_\tau u^L_k-\frac{2}{\tau}\partial_\tau u^L_k+k^2u^L_k=0\,.
\Eeq
The exact solution is given by
\Beq
\label{eq:deSitterLongSubHorSol}
u^L_k(\tau)&=\frac{1}{(2\pi)^{3/2}\sqrt{2k}}\left(\frac{k}{k_{\!_C}}-\frac{i}{k_{\!_C}\tau}\right)\exp(-ik\tau)\,,\\
\partial_\tau u^L_k(\tau)&=\frac{1}{(2\pi)^{3/2}\sqrt{2k}}\left(-\frac{ik^2}{k_{\!_C}}\right)\exp(-ik\tau)\,,
\Eeq
where we have used the WKB initial conditions, cf. eq. \eqref{eq:uWKB}. Note $u^L_k$ has two sorts of harmonic terms: ones of constant amplitude and ones that are linear in $-k\tau$ (recall $k_{\!_C}=-g_{\!_A}\bar{\rho}/2H\tau$ in de Sitter space-time). However, $\partial_\tau u^L_k$ has only terms of the latter kind\footnote{This is independent on the initial conditions and holds for the general solution of eq. \eqref{eq:LsubdS}.} - this will be important when discussing the weak coupling power spectrum of $\Delta^2_{E^L}$ on super-Hubble scales.

In the superhorison limit, $k\ll k_{\!_C}$, the longitudinal mode is governed by
\Beq
\partial^2_\tau u^L_k-\frac{2k^2}{\tau k_{\!_C}^2}\partial_\tau u^L_k+k_{\!_C}^2u^L_k=0\,.
\Eeq
This equation has a general solution in terms of hypergeometric functions which is not very illuminative and we shall not give here. It is more straightforward just to solve the full equation of motion eq.~\eqref{eq:deSitterEoM} for $u^L_k$ and from its evolution to infer $\Delta^2_{E^L}$. We did the same thing with the transverse modes.

We again start with the strong coupling limit, $k_{\!_C}/\mathcal{H}\gg\frac{1}{2}$. At early times, $k \gg k_{\!_C}$, $u^{L}_k\sim e^{-ik\tau}\sqrt{k}/k_{\!_C}$ and $\partial_\tau u^{L}_k\sim e^{-ik\tau}k^{3/2}/k_{\!_C}$, as shown in Fig.~\ref{fig:ModesInflation}. See why the two are similar in eq.~\eqref{eq:deSitterLongSubHorSol}. At late times, $k\ll k_{\!_C}$, $u^{L}_k\sim 1/\sqrt{k_{\!_C}}$ and $\partial_\tau u^{L}_k\sim\sqrt{k_{\!_C}}$. Rewriting the longitudinal electric field power spectrum as $\Delta^2_{E^L}\sim k^3|\partial_\tau u^L_k|^2/[1+(k/k_{\!_C})^2]^2$, we recover the familiar {\it k}-scalings (cf. eq. \eqref{eq:subhCompton})
\Beq
\Delta^2_{E^L}&\approx\frac{H^4}{4\pi^2}\times\begin{cases}
                     \left(k/\mathcal{H}\right)^{\!3}k_{\!_C}/\mathcal{H}\,, & \text{if $k\ll k_{\!_C}$}\,,\\
                     \left(k/\mathcal{H}\right)^{\!2}\left(k_{\!_C}/\mathcal{H}\right)^{\!2}\,, & \text{if $k\gg k_{\!_C}$}\,.
                   \end{cases}
\Eeq

The final case we consider is the weak coupling regime of the longitudinal modes, $k_{\!_C}/\mathcal{H}\ll\frac{1}{2}$. At the beginning, when $k\gg\mathcal{H}$, $u^{L}_k\sim e^{-ik\tau}\sqrt{k}/k_{\!_C}$. However, after the {\it k}-mode crosses out the Hubble horizon, but is still shorter than the Compton wavelength, i.e. $k_{\!_C}\ll k\ll\mathcal{H}$, $u^{L}_k\sim1/\sqrt{k}$. This transition upon Hubble horizon exit for sub-Compton modes is not observed for $\partial_\tau u^{L}_k$. Instead for all $k\gg k_{\!_C}$, $\partial_\tau u^{L}_k\sim e^{-ik\tau}k^{3/2}/k_{\!_C}$, cf. eq.~\eqref{eq:deSitterLongSubHorSol}. Later on when $k\ll k_{\!_C}$, $u^{L}_k\sim1/\sqrt{k}$ and $\partial_\tau u^{L}_k\sim k_{\!_C}/\sqrt{k}$. The power spectrum of the longitudinal electric field than becomes
\Beq
\Delta^2_{E^L}&\approx\frac{H^4}{4\pi^2}\times\begin{cases}
                     \left(k/\mathcal{H}\right)^{\!2}\left(k_{\!_C}/\mathcal{H}\right)^{\!2}\,, & \text{if $k\ll k_{\!_C}$}\,,\\
                     \left(k/\mathcal{H}\right)^{\!2}\left(k_{\!_C}/\mathcal{H}\right)^{\!2}\,, & \text{if $k\gg k_{\!_C}$}\,.
                   \end{cases}
\Eeq
Although, the {\it k}-scaling of $\Delta^2_{E^L}$ in the weak coupling regime (cf. eq. \eqref{eq:suphCompton}) is accounted for by our calculations in de Sitter space-time, the power excess on super-Hubble scales seen in Fig.~\ref{fig:PsEndOfInflation} is not. This requires a more general consideration in which the time dependences of {\it H} and $\rho$ are included.

\section{The Abelian model in Coulomb gauge}
\label{sec:LSCG}
In Coulomb gauge $\partial_i A^i= 0$, the background variables we consider are $(\bar{\vp}^0,\bar{\vp}^1)=(\bar{\vp}^0(\tau),0)$ (using the $U(1)$ gauge symmetry). The linearised perturbations in Fourier space are $\delta\tilde{\vp}^0_\bk,\delta\vp^1_\bk,A_{0\bk},A^{T\pm}_\bk$. Note that we have not chosen a particular space-time slicing, i.e. we are working in diffeomorphism invariant variables. E.g. $\delta\tilde{\vp}^0$ is given by $\delta\tilde{\rho}$ from eq.~\eqref{eq:diffinv}, with $\bar{\rho}\rightarrow\bar{\vp}^0$. The corresponding equations of motion are as follows.

For the background dynamics, the equations of motion are
\Beq
\label{eq:EoMbckgrndCoulomb}
&\partial^2_\tau\bar{\vp}^0+2\mathcal{H}\partial_\tau\bar{\vp}^0+a^2\partial_{\bar{\vp}^0}V(\bar{\vp}^0)=0\,,\\
\Eeq
and equations of motion for the perturbations in real and imaginary parts of $\vp$ are given by:
\Beq
\label{eq:EoMdphiCoulomb}
&\partial_{\tau}^2\delta\tilde{\vp}^0_{\bk}+2\mathcal{H}\partial_{\tau}\delta\tilde{\vp}^0_{\bk}+k^2\delta\tilde{\vp}^0_{\bk}+a^2\partial^2_{\bar{\vp}^0} V(\bar{\vp}^0)\delta\tilde{\vp}^0_{\bk}\\
&\qquad+\frac{2}{\mpl^2}\left[\left(\mathcal{H}\partial_{\tau}\bar{\vp}^0+\frac{a^2}{2}\partial_{\bar{\vp}^0} V(\bar{\vp}^0)\right)\frac{\delta\tilde{\vp}^0_{\bk}\partial_{\tau}^2\bar{\vp}^0-\partial_{\tau}\bar{\vp}^0\left(\partial_{\tau}\delta\tilde{\vp}^0_{\bk}+\mathcal{H}\delta\tilde{\vp}^0_{\bk}\right)  }{\partial_{\tau}\mathcal{H}-\mathcal{H}^2+k^2} - \left(\partial_{\tau}\bar{\vp}^0\right)^{\!2}\delta\tilde{\vp}^0_{\bk} \right]=0\,,\\
&\partial^2_{\tau}\delta{\vp}^1_\bk+2\left[\frac{\mathcal{H}-\frac{\partial_\tau\bar{\vp}^0}{\bar{\vp}^0}\left(\frac{g_{\!_A}\bar{\vp}^0a}{2k}\right)^{\!\!2}}{1+\left(\frac{g_{\!_A}\bar{\vp}^0a}{2k}\right)^{\!\!2}}\right]\partial_\tau\delta\vp^1_\bk\\
&\qquad+\left\{\frac{a^2}{\bar{\vp}^0}\partial_{\bar{\vp}^0}V(\bar{\vp}^0)+2\frac{\left(\frac{g_{\!_A}\partial_\tau\bar{\vp}^0a}{2k}\right)^{\!\!2}+\mathcal{H}\left(\frac{g_{\!_A}a}{2k}\right)^{\!\!2}\bar{\vp}^0\partial_{\tau}\bar{\vp}^0}{1+\left(\frac{g_{\!_A}\bar{\vp}^0a}{2k}\right)^{\!\!2}}+k^2+\left(\frac{g_{\!_A}\bar{\vp}^0a}{2}\right)^{\!\!2}\right\}\delta\vp^1_\bk=0\,.
\Eeq
The equation of motion for the transverse modes in the gauge field are
\Beq
\label{eq:EoMATCoulomb}
&\partial^2_\tau A^{T\pm}_\bk+\left[k^2+\left(\frac{g_{\!_A}\bar{\vp}^0a}{2}\right)^{\!\!2}\right]A^{T\pm}_\bk=0\,,
\Eeq
and finally, the constraint equation yields
\Beq
\label{eq:EoMA0Coulomb}
&A_{0\bk}=\frac{g_{\!_A}}{2}\frac{\left[\delta \vp_{\bk}^1\partial_{\tau}\bar{\vp}^{0}-\bar{\vp}^0\partial_{\tau}\delta\vp_{\bk}^1\right]}{\left(\frac{k}{a}\right)^{\!2}+\left(\frac{g_{\!_A} \bar{\vp}^0}{2}\right)^{\!\!2}}\,.\\
\Eeq

\bibliography{bibLozanovAmin16v3}
\bibliographystyle{utphys}

\end{document}